\documentclass[fleqn,usenatbib]{mnras}
\usepackage{newtxtext,newtxmath}

\usepackage[T1]{fontenc}
\usepackage{ae,aecompl}
\usepackage[labelformat=empty]{subfig}

\usepackage{verbatim}
\usepackage{graphicx}	
\usepackage{amsmath}	

\usepackage{soul}

    \title[Deep learning, galaxy structure \& quenching]{A deep learning approach to test the small-scale
galaxy morphology and its relationship with star formation activity in
hydrodynamical simulations}

\author[L. Zanisi ]{
Lorenzo Zanisi  $^{1,2}$ $^{*}$ 
Marc Huertas-Company$^{3,4}$, 
Fran\c{c}ois Lanusse$^{5}$,
Connor Bottrell$^{6}$, \newauthor
Annalisa Pillepich,$^{7}$
Dylan Nelson,$^{8,9}$
Vicente Rodriguez-Gomez,$^{10}$
Francesco Shankar$^{1}$, \newauthor
Lars Hernquist$^{11}$,
Avishai Dekel$^{12}$,
Berta Margalef-Bentabol$^{13}$,
Mark Vogelsberger$^{14}$, \newauthor
Joel Primack$^{15}$
\\
$^{1}$  Department of Physics and Astronomy, University of Southampton, Highfield, SO17 1BJ, UK\\
$^{2}$ DISCnet centre for Doctoral Training, University of Southampton, Highfield, SO17 1BJ, UK\\
$^{3}$  Instituto de Astrof\'isica de Canarias (IAC); Departamento de Astrof\'isica, Universidad de La Laguna (ULL), E-38200, La Laguna, Spain\\ 
$^{4}$ LERMA, Observatoire de Paris, CNRS, PSL, Universit\'e Paris Diderot, France\\
$^{5}$ Berkeley Center for Cosmological Physics, Department of Physics, University of California, Berkeley, California, USA\\
$^{6}$ Department of Physics and Astronomy, University of Victoria, Victoria, British Columbia V8P 1A1, Canada\\
$^{7}$ Max-Planck-Institut f{\"u}r Astronomie, K{\"o}nigstuhl 17, D-69117 Heidelberg, Germany\\
$^{8}$ Max-Planck-Institut f{\"u}r Astrophysik, Karl-Schwarzschild-Str. 1, 85741 Garching, Germany\\
$^{9}$ Universit{\"a}t Heidelberg, Zentrum f{\"u}r Astronomie, Institut f{\"u}r theoretische Astrophysik, Albert-Ueberle-Str. 2, 69120 Heidelberg, Germany\\
$^{10}$ Instituto de Radioastronom\'ia y Astrof\'isica, Universidad Nacional Aut\'onoma de M\'exico, Apdo. Postal 72-3, 58089 Morelia, Mexico\\
$^{11}$ 
Institute for Theory and Computation
, Harvard-Smithsonian Center for Astrophysics, 60 Garden Street, Cambridge, MA 02138, USA\\
$^{12}$ Racah Institute of Physics, The Hebrew University, Jerusalem 91904, Israel\\
$^{13}$ Department of Physics and Astronomy, University of Pennsylvania, Philadelphia, PA 19104, USA\\
$^{14}$ Department of Physics, Massachusetts Institute of Technology
, Cambridge, MA 02139, USA, \\
$^{15}$ Physics Department, University of California, Santa Cruz, Santa Cruz, CA 95064, USA\\
$^{*}$ Email: L.Zanisi@soton.ac.uk
}

\date{Accepted for publication in MNRAS}

\pubyear{2019}
\begin{document}
\label{firstpage}
\pagerange{\pageref{firstpage}--\pageref{lastpage}}
\maketitle


\begin{abstract}
Hydrodynamical simulations of galaxy formation and evolution attempt to fully model the physics that shapes galaxies. The agreement between the morphology of simulated and real galaxies, and the way the morphological types are distributed across galaxy scaling relations are important probes of our knowledge of galaxy formation physics. Here we propose an unsupervised deep learning approach to perform a stringent test of the fine morphological structure of galaxies coming from the Illustris and IllustrisTNG (TNG100 and TNG50) simulations against observations from a subsample of the Sloan Digital Sky Survey. Our framework is based on PixelCNN, an autoregressive model for image generation with an explicit likelihood. We adopt a strategy that combines the output of two PixelCNN networks in a metric that isolates the small-scale morphological details of galaxies from the sky background. We are able to \emph{quantitatively} identify the improvements of IllustrisTNG, particularly in the high-resolution TNG50 run, over the original Illustris. However, we find that the fine details of galaxy structure are still different between observed and simulated galaxies. This difference is mostly driven by small, more spheroidal, and quenched galaxies which are globally less accurate regardless of resolution and which have experienced little improvement between the three simulations explored. We speculate that this disagreement, that is less severe for quenched disky galaxies,  may stem from a still too coarse numerical resolution, which struggles to properly capture the inner, dense regions of quenched spheroidal galaxies. 
\end{abstract}
\section{Introduction}
\label{sec:introduction}
In the recent years, cosmological hydrodynamical simulations of galaxy formation and evolution have reached unprecedented accuracy. Early efforts (e.g. \citealt{Croft+08}, \citealt{Crain+09}, \citealt{Schaye+10_OWLS}, \citealt{Nuza+10}, \citealt{Dimatteo+12}) have paved the way to state-of-the art simulations  (\citealt{Vogelsberger+14},\citealt{Schaye+15_EAGLE}, \citealt{Dubois+14_Horizon},\citealt{Dave+19_SIMBA}, \citealt{Pillepich+18_TNGdescription}), which broadly agree with a number of observations.

The resemblance of the simulated galaxies to real ones may be considered an important hallmark of the quality of simulations and hence a crucial assessment of our knowledge of the relevant physical processes implemented therein. Indeed, one of the major successes of simulations is the ability to produce galaxies with a wide variety of morphologies, something that was not possible until only a few years ago \citep{Vogelsberger+14_nature, Schaye+15_EAGLE, Dubois+14_Horizon}. Furthermore, most simulations are now able to generate galaxies whose physical properties are in the ballpark of observations, such as the galaxy stellar mass function \citep{Furlong+15,Pillepich+18}, the size-mass relation ($R_e-M_{\rm star}$) and its evolution \citep{Genel+18,Furlong+17}, the color bimodality \citep{Trayford+17,Nelson+18_results} and the star formation activity (\citealt{Donnari+19}, though in this last instance some tensions may still be present, especially at $z\gtrsim 1$). 

A key challenge for simulations is to try to reproduce the well known correlation between galaxy morphology and star formation activity (e.g., \citealt{Eales+17}), and how it propagates onto the galaxy scaling relations, which are observed to be different in the two cases (e.g., \citealt{Shen+03},\citealt{Wuyts+11},\citealt{Bell+12_SFR_morpho_structure}) . As shown in some works \citep{Huertas-Company+16}, this goes beyond the simplistic late type-early type dichotomy, since these two broad morphological types include a wide variety of subclasses.

Assessing the level of agreement between the morphologies of the full populations of observed and simulated galaxies is a hard task, due to the intrinsic complexity of galaxy shapes. The approach followed by some authors  (\citealt{Snyder+15_morpho_SF}, \citealt{Bottrell+17_bulgeIllustris}, \citealt{Bottrell+17a}, \citealt{Rodriguez-Gomez+19}, \citealt{Bignone+19}, \citealt{Baes+20}) consisted in making use of integrated, parametric and nonparametric quantities as diagnostics (such as the popular $C-A-S-G-M_{20}$ statistics (e.g., \citealt{Abraham+94}, \citealt{Conselice03_CAS}, \citealt{Lotz+04_G_M20}), with the aim of describing galaxy morphologies with only a few numbers. However, such an approach may still not grasp the full complexity of a galaxy image. In fact, although technically all the pixels of a galaxy image are used to retrieve these quantities, their choice may be incomplete (i.e. the $C-A-S-G-M_{20}$ spatial diagnostics may in principle be extended, see for instance \citealt{Freeman+13_MID}, \citealt{Wen+14}, \citealt{Pawlik+16}, \citealt{Rodriguez-Gomez+19}), and, for this reason, limited in power (i.e. the similarity of these statistics between observed and simulated galaxies, although informative, is no guarantee of the overall quality of simulated galaxy properties). The key point is that all the precious information contained in the pixels of an image may not be fully accessible with standard techniques, which may be a major shortcoming when comparing the morphologies of observed and simulated galaxies. Moreover, the different statistics provide separate pieces of information, while it would be desirable to assess the quality of a simulation using a single-valued metric. A recent attempt to generalise over the (non) parametric techniques outlined above has been carried out in \citet{Huertas-Company+19} where a supervised deep learning framework was devised to classify the morphology of simulated galaxies. Using Bayesian Neural Networks, \citet{Huertas-Company+19} were able to identify galaxies in the simulation for which the network would produce a high variance in the output label -  a sign that the the network struggled to assign a clear morphology to some objects (mainly small galaxies), which therefore may not be very realistic.

\vspace{1em}

We build upon the work of \citet{Huertas-Company+19} by introducing a fully unsupervised framework to compare the morphologies of simulated galaxies with observations.
Comparing images coming from different datasets is a task that in Machine Learning is known as \emph{Out of Distribution} detection (OoD). The high-level idea is to have a deep learning model that learns the details of a dataset and condenses it in a single-valued function, the likelihood, which can be used as a metric to assess candidate OoD images. Deep Generative Models (DGMs) have been proposed in the literature to perform this kind of assessment. In short, a DGM trained on a given dataset computes the likelihood for each of the in-distribution images (i.e. images that come from the same distribution of the training set) as well as for all the candidate OoD images (i.e. data not seen by the Network at training time that may or may not come from the same distribution of the training set). A comparison between the likelihood distributions of both datasets will reveal whether the candidate OoD sample agrees with the training set \citep{Bishop1994}. 

However, the reliabilty of the likelihood of DGMs for OoD detection tasks has been questioned in the literature. In particular, \citet{serra+19} found that the likelihood a DGM computed for a test image is a function of the image complexity with contributions from both the background and the subject. Moreover, it has also been found that the image background can have a significant confounding effect on the likelihood that the network computes. For example, \citet{Ren+19_LLR} showed that the likelihood correlates with the number of pixels that have a value of zero. By combining the likelihood of two DGMs trained on datasets that share a similar background, Ren et al. showed that  the contribution of the subject of the image may be isolated. Here we take a step forward and try to overcome the issues highlighted in \citet{Ren+19_LLR} and \citet{serra+19} by combining the likelihood of two DGMs in a way that factors out both the contribution of the background and that marginalizes over the trivial properties of galaxy light profiles. 

\vspace{1em} 

In this paper we propose the use of PixelCNN, an autoregressive DGM, as a novel tool to compare the morphology of simulated and observed galaxies. PixelCNNs  (\citealt{vandenOord+16b}, \citealt{vandenOord+16a}) explicitly learn the probability distribution of the pixel values of images coming from a given dataset in an autoregressive fashion (i.e. the value of each pixel is conditioned to that of previously processed pixels). The appeal of PixelCNN is that it features an explicit, tractable likelihood with probabilistic meaning. Other deep learning frameworks, such as Generative Adversarial Networks \citep{Goodfellow+14} and Variational Autoencoders \citep{Kingma&Welling2014} are less suited for OoD tasks, as their likelihood is not tractable. Nevertheless, GANs have been proposed in \citet{Margalef-Bentabol+20} to perform OoD tasks based on an \emph{anomaly score}. We will discuss  how our approach compares to theirs more in detail in the remainder of this paper.

\vspace{1em}
In this proof-of-concept work, our aim is to  \emph{quantitatively} assess the fidelity of the stellar morphologies of galaxies produced by  the Illustris and IllustrisTNG simulations by comparing them with available observations.  We further explore whether an increase in resolution may be able to lead to an even better agreement between the morphology of simulated and observed galaxies by exploiting the higher resolution offered by a realization of IllustrisTNG in a smaller cosmological box, TNG50, and how this depends on star formation activity. The novelty of this work is in that we devise a methodology which is sensitive to the relationship between the fine morphological structure and the global properties of the galaxies' light profile, which is a very stringent test for simulations.

\vspace{1em}
The outline of the paper is as follows. We describe the Sloan Digital Sky Survey observations and the fully realistic mock observations of the Illustris and IllustrisTNG simulations in Section \ref{sec:data}. In Section \ref{sec:PixelCNN} we present our deep learning model and in Section \ref{sec:strategy} we outline the strategy with which we compare images of galaxies coming from simulations and observations. This is done \emph{quantitatively} according to a metric which is a combination of the output of two neural networks. In Section \ref{sec:main_result} we show that our framework is able to recognize an improvement in IllustrisTNG compared to Illustris. Such an improvement is still not enough to achieve a full agreement with SDSS observations. In Section \ref{sec:quiescent_not_reproduced_LLR} we show that there has been a continual advance in the realism of simulated galaxy morphologies overall. However, this is less true for quiescent galaxies, particularly the spheroidal ones, which do not compare well to observations, even at the enhanced numerical resolution provided by TNG50. In Section \ref{sec:scalingrelations} we show how the quality of simulated galaxies varies across scaling relations, and conclude that quiescent small and/or high S\'ersic index galaxies are the ones that are most problematic. Indeed, in Section \ref{sec:interpretation} we show that most of the discrepancy between simulations and observations for this galaxy population comes from the galaxy inner regions.
In Section \ref{sec:discussion}  we discuss how similar findings have started to emerge in the literature along with some potential caveats in common with our study, as well as the potential reasons of this disagreement. Finally, in Section \ref{sec:conclusions} we give a concise summary of our findings and discuss future applications of our framework.

\section{Data}
\label{sec:data}

\subsection{Simulations}
\label{sect:simulations}
We make use of the Illustris Simulation (\citealt{Vogelsberger+14_nature},\citealt{Vogelsberger+14}, \citealt{Genel+14}, \citealt{Sijacki+14}) and its successor IllustrisTNG (\citealt{Pillepich+18}, \citealt{Nelson+18_results}, \citealt{Nelson+18_datarelease}, \citealt{Marinacci+18}, \citealt{Springel+18}, \citealt{Naiman+18}).
Illustris and IllustrisTNG are hydrodynamical cosmological simulations , run with the \textit{AREPO} moving-mesh code \citep{Springel2010}. The Illustris simulation has proved capable of reproducing several observables, but presented several shortcomings as summarized in \citet{Nelson+15}. In the IllustrisTNG simulation significant changes have been made with respect to the Illustris framework. These include the modelling of magnetic fields \citep{Pakmor+11,Pakmor+13,Pakmor+14}, the substitution of the \emph{bubble mode} AGN feedback at low accretion rates \citep{Sijacki+07} with a kinetic AGN feedback \citep{Weinberger+17}, a modification of the implementation of galaxy-wide winds and updated mass yields from star particles (see \citealt{Pillepich+18_TNGdescription} for a detailed summary). 

The IllustrisTNG simulation was run with identical subgrid physics in three cosmological volumes of progressively larger sizes and with comparatively lower resolution. 
 Here we compare the run of Illustris, which is performed in a cubic box of about 100 comoving Mpc a side to the IllustrisTNG framework implemented in a box of the same size, TNG100, and one of about 50 Mpc a side, TNG50 \citep{Pillepich+19_TNG50, Nelson+19_TNG50presentation}. We use in all three cases the highest resolution realizations of each volume, which correspond to baryonic mass resolutions of $\sim 10^6 M_\odot$ for TNG100 and Illustris, whereas TNG50 has a fifteen times higher mass resolution of $\sim 8\times10^4 M_\odot$, more comparable to zoom-in simulations.

 From the simulations we select galaxies with $M_{\rm star}>10^{9.5}M_\odot$ at $z=0.0485$\footnote{snapshot 95 for IllustrisTNG and 131 for Illustris.}, for a total of $\sim 12,500 $ galaxies for Illustris and TNG100 and $\sim1,700$ objects for TNG50.  The images are processed with a joint use of the radiative transfer code \texttt{SKIRT} (\citealt{Baes+11_SKIRT}, \citealt{Camps&Baes2015}), the nebular modelling code \texttt{MAPPINGS-III} \citep{Groves+08}  and the \citet{Bruzual&Charlot2003} \texttt{GALAXEV} stellar population synthesis code. The methodology is described in detail in \citet{Rodriguez-Gomez+19}. Briefly, each stellar particle in either simulation (which represents a coeval stellar population) is modelled with \texttt{GALAXEV} for stellar particles older than 10 Myr, while younger stellar particles are treated as a starbursting population with \texttt{MAPPINGS-III}. To model dust, it is assumed that the diffuse dust content of each galaxy is traced by the star-forming gas, that the dust-to-metal mass ratio is constant and equal to 0.3 \citep{Camps+16}, and that dust is a mix of graphite grains, silicate grains, and polycyclic aromatic hydrocarbons \citep{Zubko+04_dustmodel}. Full dust-inclusive radiative transfer is run only if the fraction of star forming gas exceeds 1\% of the total baryonic mass. The simulated galaxies are mock-observed in the SDSS $r-$band at $z=0.0485$ along a random line of sight with the pixel scale of the SDSS telescope ($\approx 0.396 ''/\text{pix}$). Full observational realism is included as described in Section~\ref{sec:realsim}.
 
 The structural properties of the mock-observed simulated galaxies, such as the effective radius $R_e$ and the S\'ersic index $n_{ser}$, are obtained with \textsc{statmorph} \citep{Rodriguez-Gomez+19}. \textsc{statmorph} is a Python package\footnote{Available at \url{https://statmorph.readthedocs.io/en/latest/}} for calculating non-parametric morphological diagnostics of galaxy images, as well as fitting 2D Sérsic profiles. The stellar mass of galaxies is computed as the mass of all the bound stellar particles within 30kpc from the galaxy center, while the star formation rates (SFR) are computed within twice the half-mass radius of each galaxy. Other stellar mass and SFR definitions have been discussed in \citet{Pillepich+18} and \citet{Donnari+19}.

\subsection{Observations}
\label{sec:observations}
\begin{figure}
    \centering
 \includegraphics[height=0.45\textwidth, clip=True]{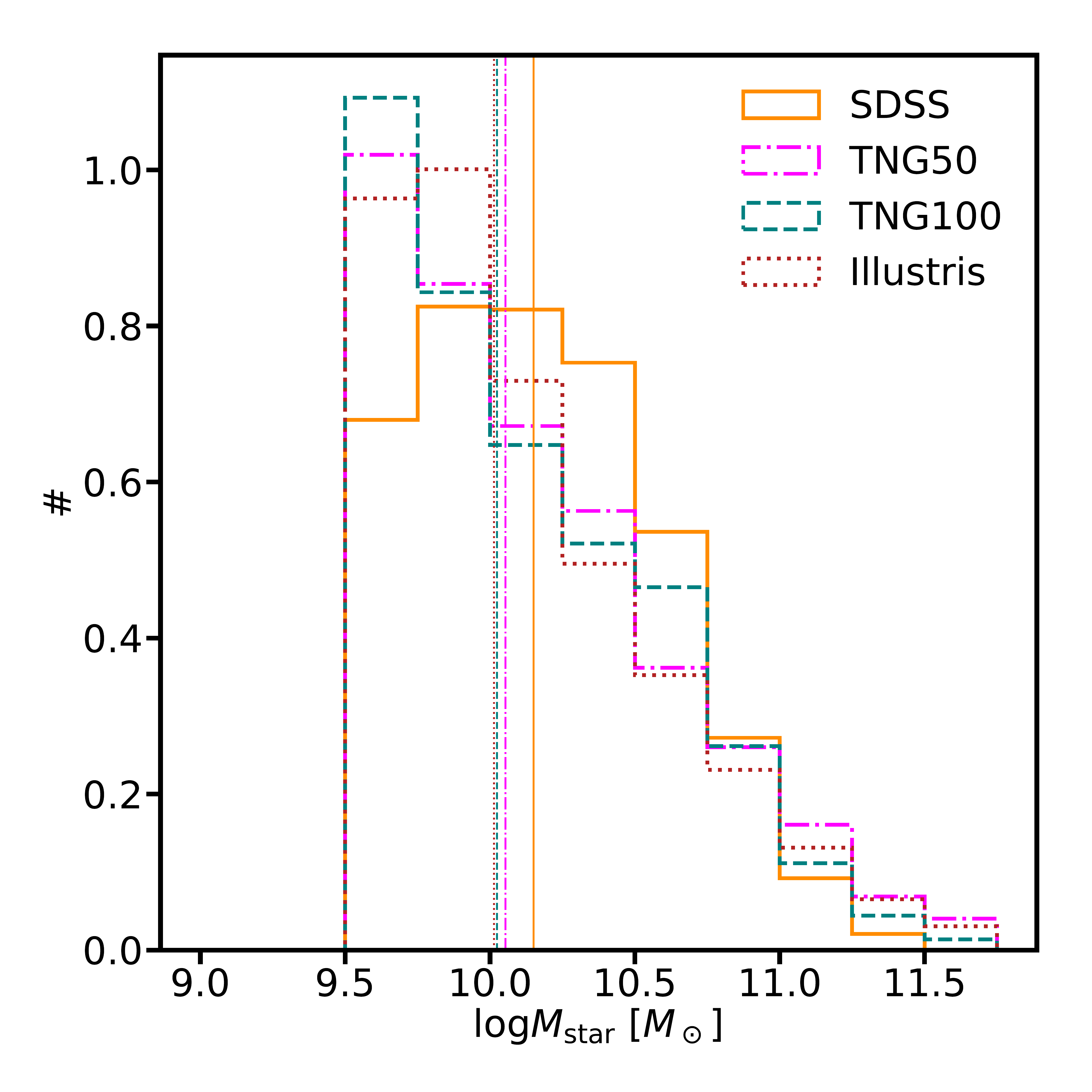} 
   \caption{The normalized stellar mass distributions for SDSS (solid orange line), TNG50 (dot-dashed magenta line), TNG100 (teal dashed line) and Illustris (red dotted line). The vertical lines indicate the median mass of each distribution. It can be seen that SDSS is incomplete at $M_{\rm star} \lesssim 10^{10}M_\odot$, but overall the mass distributions are similar. }
    \label{fig:SMF}
\end{figure}

In the following we will use the SDSS DR7 \citep{Abazajian+09}  spectroscopic sample \citep{Strauss+02}. We use the \citet{Meert+15}  catalogues for galaxies in this sample, totaling 670,722 galaxies. The Meert et al. morphology catalogues have specifically been shown to offer improved profile fits to the sample's brightest galaxies compared to previous catalogues (e.g. \citealt{Simard+11}) - owing to a highly robust sky-subtraction algorithm. The galaxy stellar masses  are computed adopting S\'ersic photometric fits and the mass-to-light ratio $M_{\rm star}$/L by \citet{Mendel+14}. Although the spectral energy distribution of galaxies contains information which is critical to understand the physical processes that regulate galaxy formation, in this exploratory work  we choose to adopt only single band images (specifically $r$-band). We plan to expand our work to multi-band photometry in the future. 

We also match the \citet{Meert+15} photometric catalog with measurements of star formation rate (SFR) from \citet{Brinchmann+04} and with the group catalogs of \citet{Yang+07, Yang+12_groups}, which will allow us to identify satellite and central galaxies in our observed galaxy sample.

We use the images of SDSS galaxies that have a stellar mass $M_{\rm star}>10^{9.5}M_\odot$ as our training sample. An important issue that must be dealt with when choosing the training sample is that of the redshift evolution of the angular diameter distance driven by cosmology. Indeed, the pixel physical scale\footnote{i.e. kpc/pix} is a strong function of redshift, which means that the training sample must be chosen so that the average pixel scale is as close as possible to the pixel scale at the redshift of the snapshot that we use for the simulations (i.e. $z\sim0.0485$, see Section \ref{sect:simulations}). Hence, we also limit the redshift range of the SDSS training sample to $0.033<z<0.055$,  which gives a median pixel scale only 7\% larger than the pixel scale at $z=0.0485$. This redshift cut leaves us with $\approx$44000 galaxies in SDSS, of which we use $\approx$ 32000 for training and $\approx$ 12000 for testing. We note that in principle with the pixel scale of the SDSS camera (i.e. 0.396"/pix) the minimum physical scales probed at z$\sim$0.0485 would be around $\sim$0.3 kpc. However, when the SDSS PSF ($\gtrsim 1$"$\sim$3-4 pixels) is accounted for the smallest scales to which we are sensitive are around $\sim 1$ kpc. Such low resolution is still enough for some trends to arise, as shown in the following Sections.

In Figure \ref{fig:SMF} we compare the stellar mass distribution of SDSS with that of the simulations. The slightly higher median mass of SDSS compared to Illustris and IllustrisTNG results from the incompleteness of observations below $M_{\rm star}\lesssim10^{10} M_\odot$. Indeed, we checked that the distributions have a very similar median value if only galaxies above that mass are considered. In the remainder of this paper, we will break down our results above and below the completeness threshold.

 \subsection{Galaxy archetypes}
 \label{sec:archetypes}
 Our methodology implies training a second DGM on a simplified version of the same galaxies used to train the first one. In other words, we would like to have a second dataset where the global properties of SDSS (such as brightness, size, ellipticity and light concentration) are retained, but where more complex features, such as the spiral arms of a disk galaxy, are ignored. This can be constructed by using the S\'ersic fits of the SDSS galaxies described in the previous subsection. We produce these images using \texttt{GalSim} \citep{galsim} and the values of the best-fitting $r-$band S\'ersic parameters provided in \citet{Meert+15}. We include full observational realism as detailed in the next Section.

 \subsection{Observational realism}
 \label{sec:realsim}
 Both the simulations and the best-fitting S\'ersic images are idealized objects. Therefore, for a fair comparison with observation we add the same kind of observational effects that are found in SDSS, that is, the presence of a noisy sky background and interlopers, as well as the convolution with the SDSS Point Spread Function.  \citet{Bottrell+17a} and \citet{Bottrell+17_bulgeIllustris} presented \texttt{RealSim}, an algorithm that enables such procedure.  Briefly, with \texttt{RealSim} it is possible to place a galaxy from a given simulation in a real SDSS field. The mock galaxy is convolved with the Point Spread Function of that particular field; the effects of shot noise and cosmological surface brightness dimming are also included. For more details about \textit{RealSim}, we refer the reader to the original papers. \citet{Bottrell+19} have shown that the including the correct level of realism in mock observations is crucial when using neural networks for classification tasks.  

\subsection{Volume effects}
\label{sec:volume}
Given that the cosmological volume spanned by the TNG100 and Illustris simulations is more than 8 times larger than that of TNG50, one thing we must worry about is cosmic variance. Indeed, \citet{Genel+14} showed that the statistics of galaxy populations may vary quite substantially in sub-boxes of 25/h$\approx$35 Mpc a side in the Illustris simulation. Therefore, it is very much possible that the volume probed by TNG50 results in a biased galaxy population. 

The way we address this issue in the following is by creating several realizations of SDSS, TNG100 and Illustris of the same sample size of TNG50, and then use the mean and variance of the bootstrapped distributions where possible.

In principle, cosmic variance could also affect the comparison between SDSS and simulations. However, the test set of SDSS that will be used in the following shares a very similar sample size with Illustris and TNG100. While this is not strictly a measure of the volume spanned by SDSS, we can reasonably assume that a similar sample size should enable a meaningful comparison between observations and those two simulations, since they have similar stellar mass distributions.

\section{Methods}
\subsection{PixelCNN}
\label{sec:PixelCNN}
PixelCNN (\citealt{vandenOord+16b,vandenOord+16a}) is an autoregressive generative model with an explicit likelihood. Given an image $X$, the likelihood of PixelCNN is ``autoregressive" in the sense that the likelihood a given pixel is assigned is conditioned on all the previous pixels of the image (which sometimes are collectively called "context"), so that
\begin{equation}
\label{eq:likelihood}
p_{\theta}(X) = \prod\limits_{i=1}^{N^2} p_{\theta}(X_i|X_{1...i-1}),
\end{equation}
where N is the pixel width/height of the square cutout.
Here $p_{\theta}(X_i|X_{1...i-1})$ is the probability distribution function of pixel i evaluated at $X_i$ and conditioned on all the previous $X_{1...i-1}$ pixels, and $\theta$ are the weights of the network. It is worth stressing that eq. \ref{eq:likelihood} models \emph{explicitly} the likelihood of the training sample. In the following we will use the negative log-likelihood, which is less prone to floating point limitations,
\begin{equation}
    \label{eq:loglikelihood}
    \mathcal{L} \equiv -\log{p_{\theta}(X)} = -\sum_{i=1}^{N^2} \log{p_{\theta}(X_i|X_{1...i-1})}
\end{equation}
The ansatz of eq. \ref{eq:likelihood} imposes the choice of an ordering for the pixels. We follow a prescription according to which the image is scanned from top left to bottom right, row by row. This is a standard implementation of PixelCNN that takes advantage of the way convolutions are typically implemented in deep learning frameworks. The autoregressive nature of PixelCNN is achieved by means of a particular type of convolutions that mask the pixels to the right and bottom of the current pixel, so that the network is forced to learn the relationship between each pixel and the previous context only. This is fully detailed in \citet{vandenOord+16b,vandenOord+16a}).\\

We here adopt the \texttt{PixelCNN++} architecture proposed by \citet{Salimans+17}\footnote{Available at \href{https://github.com/openai/pixel-cnn}{https://github.com/openai/pixel-cnn}} interfaced with a higher level \texttt{Tensorflow} API\footnote{Available at \href{https://github.com/pmelchior/scarlet-pixelcnn}{https://github.com/pmelchior/scarlet-pixelcnn}}. Briefly, Salimans et al. adopt a fully convolutional autoencoder-like architecture, with three downsampling and three upsampling stages respectively, where downsampling and upsampling are implemented using strided convolutions \footnote{Transposed convolutions in the case of upsampling.}. Each stage consists of an adjustable number of Gated Resnet layers (\citealt{vandenOord+16b}, \citealt{He+15_ResNet}), which entail zero-padding convolutions to preserve dimensionality. Stages in the downsampling and upsampling  parts of the network with the same dimensionality are connected with shortcut connections as in \citet{Ronneberger+15_Unet}, to ensure that part of the information lost in the downsampling is efficiently recovered. We refer the reader to \citet{Salimans+17} and \citet{vandenOord+16a, vandenOord+16b} for further details of the implementation.

\vspace{1 em}

Obviously, not all images will have the same likelihood. Rather, PixelCNN maps a distribution of images into a distribution of likelihoods. This feature is in principle extremely powerful, since it allows to collapse the complexity that characterizes images into a single-valued function. However, as briefly mentioned in the Introduction and as fully explained in the following, the likelihood alone may not be a good proxy for the quality of an image. Rather, the combination of the likelihood from two independent models may give a better estimate for it. Therefore, we train two PixelCNNs models:
\begin{itemize}
    \item $p_{\theta_{\rm SDSS}}$, a network trained on the SDSS sample described in Section \ref{sec:observations}
    \item $p_{\theta_{\rm sersic}}$, a model trained on the best S\'ersic fits with the added observational realism as described in Section \ref{sec:archetypes}.
\end{itemize}

\subsubsection{Training}
\label{sec:training}
The images which originally were of size of 128x128 pixels, are augmented 10 times with random rotations and then cropped to 64x64 and degraded to reach the size of 32x32 pixel\footnote{We use the publicly available \texttt{scipy} library.} in order to meet memory and time constraints. 

To train PixelCNN we use 32000 galaxies randomly extracted from our SDSS sample, corresponding to the $75\%$ of the dataset. We also trained a second PixelCNN on the best r-band S\'ersic fits of the same SDSS galaxies. The likelihood distributions in the two cases are shown in Figure \ref{fig:traintest}.

One complication that astronomical images suffer compared to standard applications which use \texttt{png} images is that in the latter case the range of values that a pixel can take is limited (i.e. from 0 to 255), while this does not apply to the astronomical standard where the value of each pixel of a \texttt{fits} image is a flux and hence it is not bounded in principle. Here we use \texttt{fits} images that can be downloaded from the SDSS server and the Illustris and IllustrisTNG websites, as detailed in the "Data Availability" Section. Therefore eq. \ref{eq:likelihood} should be interpreted as the product of the conditional probability distribution functions evaluated at $X_i$, rather than the probability mass. To ensure the stability of training, we reduce the dynamical range of pixel values by dividing each image by 1000 and subsequently applying the \texttt{arcsinh} function. We further impose a hard upper limit of 1 to the rescaled flux per pixel. The choice of this threshold involves a trade-off between training convergence and information lost in the small-scale details of the images. With our choice of 1 as an upper limit, we do not see any trends between the metric that we use in this paper (see next Section) and the fraction of pixels that are above the chosen threshold, which is less than 1.5\% for the vast majority of the images in our samples.

\subsection{Strategy and the LLR metric}
\label{sec:strategy}
The likelihood of generative models such as PixelCNN has been proposed as a tool to compare different datasets on the grounds that the likelihood distribution of a candidate OoD dataset should peak at lower values \citep{Bishop1994}. However, the interpretation of the likelihood is not an easy task, as discussed in the following. 

Firstly, the background of an image is thought to play an important role in determining the likelihood of a given sample \citep{Ren+19_LLR}. This is because the log-likelihood is an additive quantity, and therefore all the pixels will contribute to it, including those where the subject (i.e., the galaxy in our case) is not present.  To factor out the undesired contribution of the background \citet{Ren+19_LLR} proposed the use of two DGMs, where the second network is trained on a dataset that has similar background statistics to the training set of the first. In our case, we have the networks $p_{\theta_{\rm SDSS}}$ and $p_{\theta_{\rm sersic}}$ which both are trained to learn a similar sky background by construction.
The likelihood of a test image $X_{test}$ evaluated by both models can be decomposed simply in the roughly independent contributions of the background pixels $X_{\rm background}$ and pixels of the subject, $X_{\rm subject}$, 
\begin{equation}
\label{eq:LLR_1}
p_{\theta_i}(X_{test})=  p_{\theta_i}(X_{\rm background})p_{\theta_i}(X_{\rm subject})   \end{equation}
with $i=SDSS, sersic$. Then the log-likelihood ratio (LLR)
\begin{align}
    \label{eq:LLR_def}
    LLR &= log\Bigl \{\frac{ p_{\theta_{\rm SDSS}}(X_{test})}{p_{\theta_{\rm sersic}}(X_{test})} \Bigr \} \\ &=  log \Bigl \{\frac{ p_{\theta_{\rm SDSS}}(X_{\rm background})p_{\theta_{\rm SDSS}}(X_{\rm subject})}{p_{\theta_{\rm sersic}}(X_{\rm background})p_{\theta_{\rm sersic}}(X_{\rm subject})} \Bigr \}
\end{align}
should not depend on the background pixels, since both models capture the background equally well. 

Secondly, the complexity of an example image (both background and subject) has been found to anticorrelate with the likelihood \citep{serra+19} (see also Appendix \ref{app:robustness}). However we are interested in the complexity of the galaxy only, $X_{\rm subject}$, and irrespective of its global features such as brightness, size, ellipticity and S\'ersic index, $X_{\rm  global}$. Indeed, the expression in eq. \ref{eq:LLR_def} does not only help isolating the galaxy from the background, but it also provides information about the small-scale morphological details, $X_{\rm details}$. In fact, the contribution of the subject of the image $X_{\rm subject}$ can be decomposed in the contributions from  $X_{\rm  global}$ and $X_{\rm details}$ using the theorem of compound probability,
\begin{align}
    \label{eq:subject_details}
    p_{\theta_i}(X_{\rm subject})  & =p_{\theta_i}(X_{\rm details}, X_{\rm  global}) \\
    &=p_{\theta_i}(X_{\rm details}|X_{\rm  global})p_{\theta_i}(X_{\rm  global})
\end{align}
where we have accounted for the dependence of certain morphological features from global properties in the term $p_{\theta_i}(X_{\rm details}|X_{\rm  global})$ (e.g., spiral galaxies, which have very distinctive features, also tend to be larger than spheroids as shown by a vast body of literature - see for example \citealt{ Shen+03, Bernardi+14,Lange+16, Zanisi+20_sizefunct}). The log-likelihood ratio, LLR (where only the contribution of $X_{\rm subject}$ remains, see eq. \ref{eq:LLR_def}), is now
\begin{align}
    \label{eq:LLR_details_smooth}
    LLR &= log\Bigl \{\frac{ p_{\theta_{\rm SDSS}}(X_{\rm subject})}{p_{\theta_{\rm sersic}}(X_{\rm subject})} \Bigr \} \nonumber \\ &=  \log{ \Bigl \{\frac{ p_{\theta_{\rm SDSS}}(X_{\rm details}| X_{\rm  global}) p_{\theta_{\rm SDSS}}(X_{\rm  global})}{p_{\theta_{\rm sersic}}(X_{\rm  global})} \Bigr \}}
\end{align}
where we have used the fact that the best S\'ersic fits are featureless and so a model trained on them will only learn about $X_{\rm  global}$. If $p_{\theta_{\rm SDSS}}$ and $p_{\theta_{\rm sersic}}$ are able to learn the global features equally well, then the only contribution left to the LLR is
\begin{equation}
    \label{eq:LLR_final}
LLR \approx
log \bigl \{ p_{\theta_{\rm SDSS}}(X_{\rm details}| X_{\rm  global}) \bigr\}.
\end{equation}
Therefore our LLR should be able to capture only the relationship between the fine morphological details and the global properties, without the contribution from the latter alone.

\subsubsection{The LLR is informative of the agreement between simulations and observations}

A key property of the LLR is that it serves as a metric to assess which of two competing models gives a better fit to the data. In our case our models are two PixelCNNs which are trained on $r-$band images of SDSS galaxies as well as their best-fitting S\'ersic images (
$p_{\theta_{\rm SDSS}}$ and $p_{\theta_{\rm sersic}}$ respectively).
Suppose that our samples $X_{test,j}$ are extracted from a test distribution $q$, i.e $X_{test}\sim q$. For us, $X_{test,j}$ represents a single image from one of the simulations used in this work, and $q$ is the collection of all these images.

The expected value of the LLR reads
\begin{align}
    \mathbb{E}_{x\sim q}[LLR] 
    & \equiv  \sum_{j=1}^M log\Bigl \{\frac{ p_{\theta_{\rm SDSS}}(X_{test,j})}{p_{\theta_{\rm sersic}}(X_{test,j})} \Bigr \} q(X_{test,j})  \\
    & =  \sum_{j=1}^M \Bigl\{ \log{\Bigl [ \frac{ q(X_{test,j})}{p_{\theta_{\rm sersic}}(X_{test,j})} \Bigr ]} q(X_{test,j})  \\ & \qquad -  \log{\Bigl [ \frac{q(X_{test,j})}{p_{\theta_{\rm SDSS}}(X_{test,j})}\Bigr ]} q(X_{test,j})\Bigr \}  \\
    & = D_{KL}(q || p_{\theta_{\rm sersic}}) - D_{KL}(q || p_{\theta_{\rm SDSS}}),
    \label{eq:DKL}
\end{align}
where the second equation is obtained by dividing and multiplying the argument of the logarithm by $q(X_{test,j})$. Here $D_{KL}(f ||g) = \sum_{i=1}^N [\log{f(x_i)/g(x_i)}] f(x_i)$ is the Kullback-Leibler divergence, which is a way to quantify the distance between two distributions. Thus, if $\mathbb{E}_{x\sim q}[LLR]>0$, then $D_{KL}(q || p_{\theta_{\rm sersic}}) > D_{KL}(q || p_{\theta_{\rm SDSS}})$, that is, the distance of $q$ from the $p_{\theta_{\rm sersic}}$ model is larger than that from the $p_{\theta_{\rm SDSS}}$ model, and therefore $q$ is closer to the distribution of SDSS galaxy images. Hence, Eq. \ref{eq:DKL} leads us to conclude that \emph{the larger the expected value of the LLR, the more similar $q$ is to $p_{\theta_{\rm SDSS}}$}. A clear indication of our mathematical derivation is that SDSS should have the highest mean LLR (i.e., $q \equiv p_{\theta_{\rm SDSS}}$). Conversely, the collection of galaxies coming from a given simulation (i.e. $q \neq p_{\theta_{\rm SDSS}}$) should ideally have a mean LLR that is as close as possible to that of SDSS, but it is predicted that the condition $\langle LLR \rangle\leq \langle LLR_{SDSS} \rangle$ should hold. More formally, we can \emph{quantify} how much simulations depart from SDSS by computing the difference between the mean LLR of simulated galaxies and that of SDSS, $\Delta \langle LLR \rangle \equiv \langle LLR \rangle - \langle LLR_{SDSS} \rangle$. Since $\mathbb{E}_{x\sim q}[LLR]$ is highest for observations by construction 
then the largest value that $\Delta \langle LLR \rangle $ can assume is zero. To make it abundantly clear, this means that the closer the $\Delta \langle LLR \rangle$ is to zero, the more consistent a data set is with SDSS. A simulation for which $\Delta \langle LLR \rangle=0$ perfectly reproduces the observed galaxy morphologies. We stress again that the level of agreement between simulations and data is independent of both the sky background and global morphology with this metric, and depends only on the small-scale structural details of simulated galaxies (see eq. \ref{eq:LLR_final}). We also emphasize that in this study we are limited by the relatively low resolution of SDSS images, which is mimicked in the mock observations of Illustris and IllustrisTNG galaxies. In principle the same identical framework may be applied to higher-resolution imaging. 

\vspace{1em}

The framework outlined above applies if all the global parameters are the same, i.e. for galaxy samples with reasonably compatible global scaling relations, which is roughly true in our case (but see Section \ref{sec:summary}). On the other hand, should the simulated galaxy population be extremely biased, our methodology would not be applicable. For example, ad absurdum, let's take the case of an hypothetical cosmological simulation that produces only a single, perfectly realistic galaxy, or multiple identical copies thereof. The galaxy population in this simulation, as a whole, is clearly not realistic, since real galaxies span a range of properties. However, the $LLR$ distribution of the simulated sample would be a delta-Dirac function centered at a high value of $LLR$, resulting in a very high, or even positive, $\Delta \langle LLR \rangle$. It is clear that such value of the $\Delta \langle LLR \rangle$ does not indicate a good agreement between the small-scale morphology of the \emph{population} of simulated and real galaxies. We will explore other methodologies that will allow to circumvent this specific limitation of our approach in future work.

\vspace{1em}

Other techniques to compare distributions, such as the popular Kolmogorov-Smirnov (KS) test, are available in the literature. However, we found that the KS test is not sensitive enough to describe the difference between the LLR distributions of observed and simulated galaxies. Indeed, the p-value of a KS test under the null hypothesis that the LLR distributions of SDSS and each simulation are identical is always zero - perhaps not surprisingly, since the distributions that we will present in the following are substantially different. A p-value of zero in all cases prevents us from achieving one of our main aims, that is quantifying the improvement between the various simulations. Therefore, in the following we will use the LLR as a metric to compare observations and simulations. We discuss the robustness of this approach compared to using the likelihood of the $p_{{\theta}_{SDSS}}$ model only in Appendix \ref{app:robustness}. 

\begin{figure}
    \centering
 
    \includegraphics[width=0.5\textwidth, clip=True]{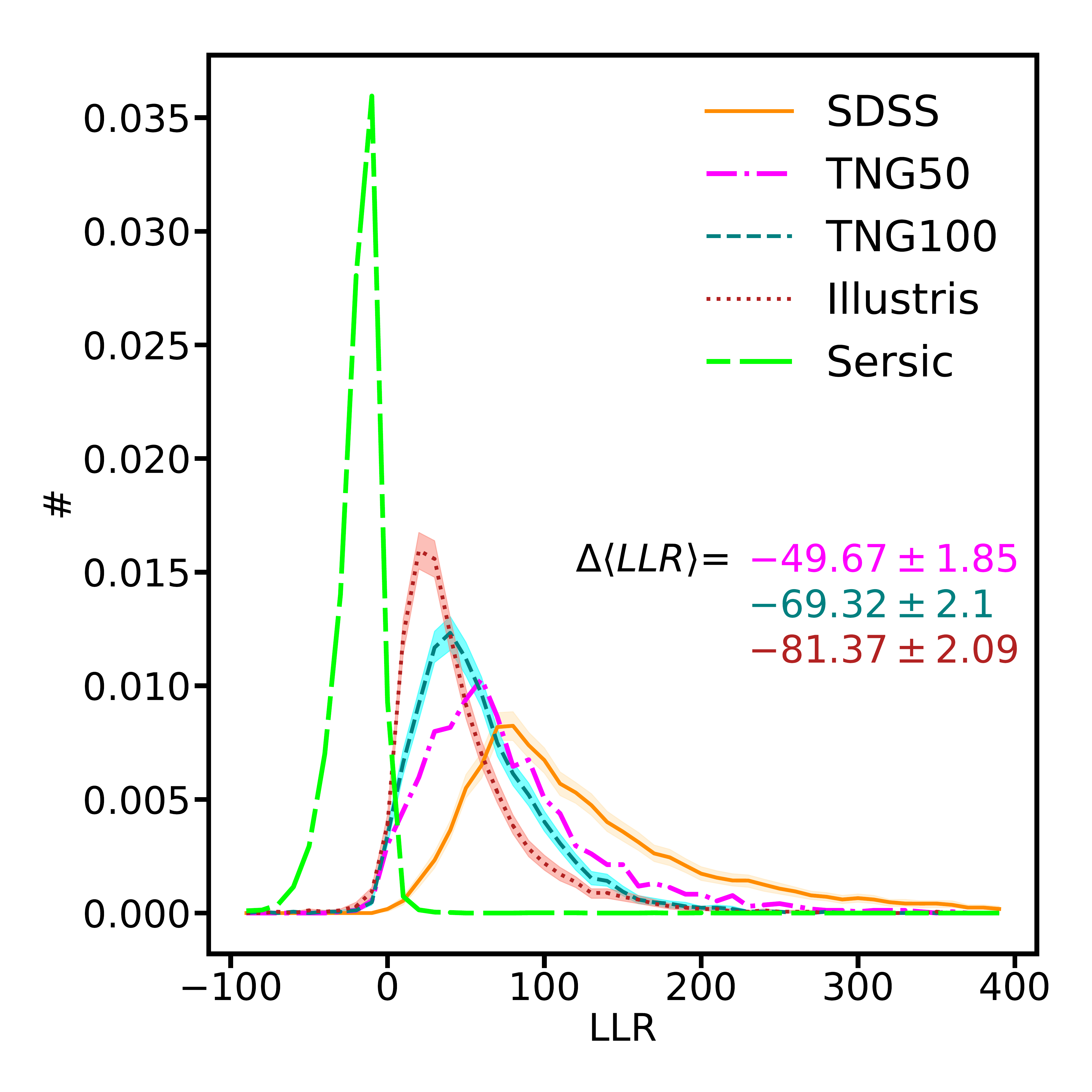}
    \caption{The log-likelihood ratio (LLR) distributions of SDSS (orange solid line), TNG50 (magenta dot-dashed line), TNG100 (dashed line),  Illustris (red dotted line) and the best S\'ersic fits (green long dashed line), for galaxies with $M_{\rm star}>10^{9.5}M_\odot$. The shaded regions show the 1 sigma confidence level obtained by bootstrapping SDSS, TNG100 and Illustris 1000 times to the same sample size of TNG50. The $\Delta \langle LLR \rangle$ for each simulation is also reported, inclusive of the 1$\sigma$ confidence interval resulting from the bootstrapping. 
    The higher the value of the $\Delta \langle LLR \rangle$, the more similar a dataset is to SDSS. Therefore, TNG50 is the simulation that best reproduces the morphology of SDSS galaxies, followed by TNG100 and Illustris. }
    \label{fig:LLR}
    
\end{figure}

\section{PixelCNN can distinguish simulations and observations}
\label{sec:main_result}

The LLR distributions for  Illustris, TNG100, TNG50 and the test sets of SDSS and their best-fitting S\'ersic profiles are shown in  Figure \ref{fig:LLR}, which constitutes the main result of our paper. The first consideration to emphasize is that the SDSS test set is the one with highest LLR, while the best S\'ersic fits of SDSS galaxies have a negative LLR. This confirms the findings outlined at the end of the previous Section: a higher LLR is a signature that a dataset is better represented by SDSS observations and, conversely,  the smaller the LLR the more the dataset is similar to featureless S\'ersic profiles. 

With this in mind, we now bring the reader's attention to a very clear trend: the distribution of SDSS peaks at the highest LLR followed, in order, by TNG50, TNG100 and Illustris. This results in values of $\Delta \langle LLR \rangle$ of -49.67$\pm$1.85, -69.32$\pm$1.93 and -81.37$\pm$2.09.  According to our framework, this means that Illustris is the simulation that gives the worst performance of the three. The IllustrisTNG implementation markedly improves over Illustris, with TNG50 being the closest to SDSS. We recall that Illustris and the two IllustrisTNG simulations differ in the implementation of the physics that shapes galaxies while their resolution is comparable. Therefore, we must conclude that the physical modelling implemented in IllustrisTNG is able to generate more realistic galaxies compared to the original Illustris model. Moreover, TNG50 features a factor of ~2.5 spatial resolution compared to the other two simulations used here. We then conclude that the improvement in resolution in TNG50 leads to further agreement with observations. 

It is noteworthy, however, that even the newest generation of simulations, although remarkably more accurate compared to earlier efforts, still struggles to reproduce the small-scale morphological details of SDSS observed galaxies, down to scales of $\approx1$ kpc (see Section \ref{sec:training}). 

\section{The small-scale stellar morphology of quiescent galaxies is not well reproduced by simulations}
\label{sec:quiescent_not_reproduced_LLR}
\begin{figure*}
    \centering
    \subfloat[]{\includegraphics[width=0.9 \textwidth, clip=True]{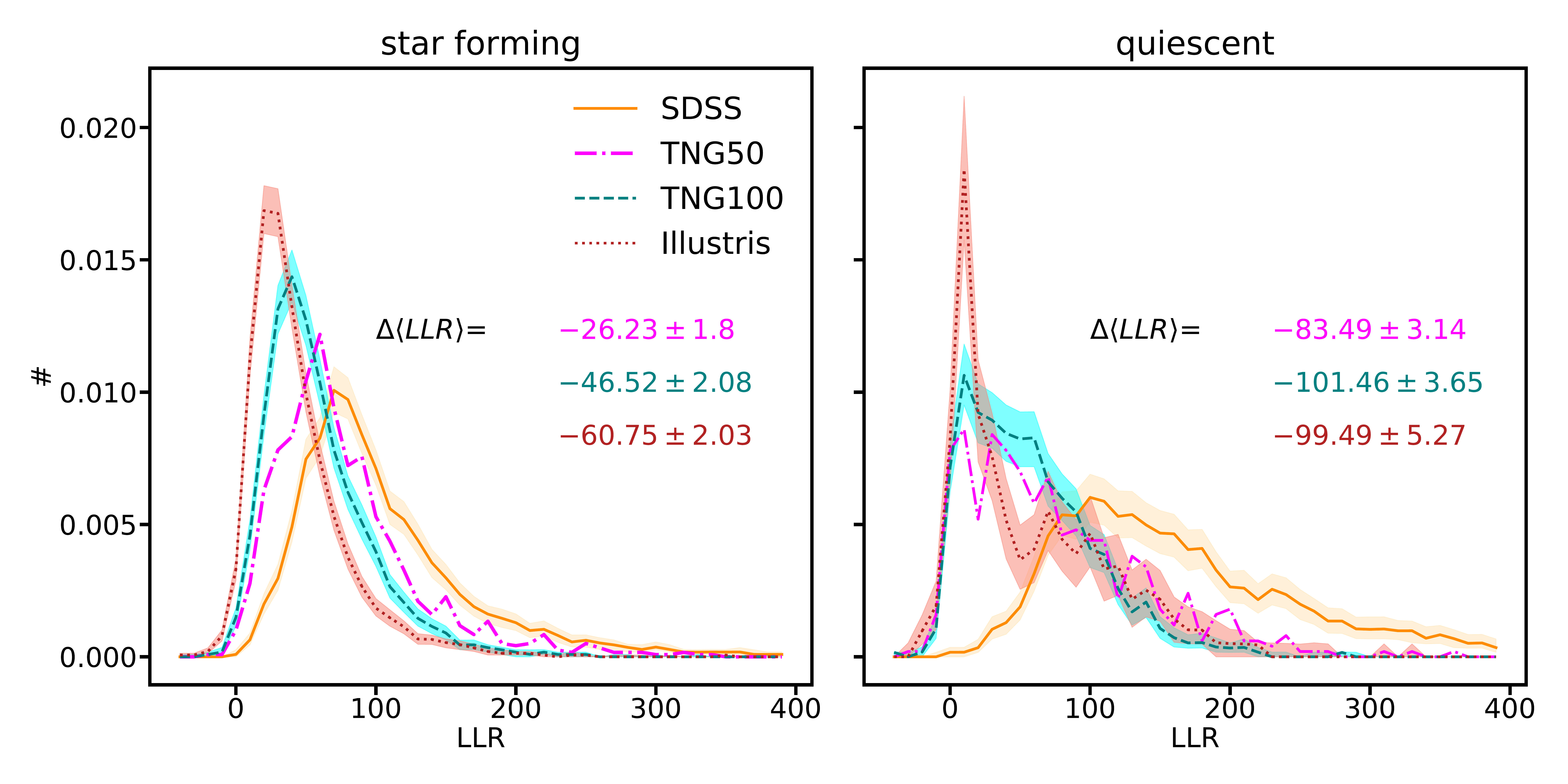}}\\
    \vspace{-1 em}
    \subfloat[]{\includegraphics[height=0.3\textwidth, clip=True]{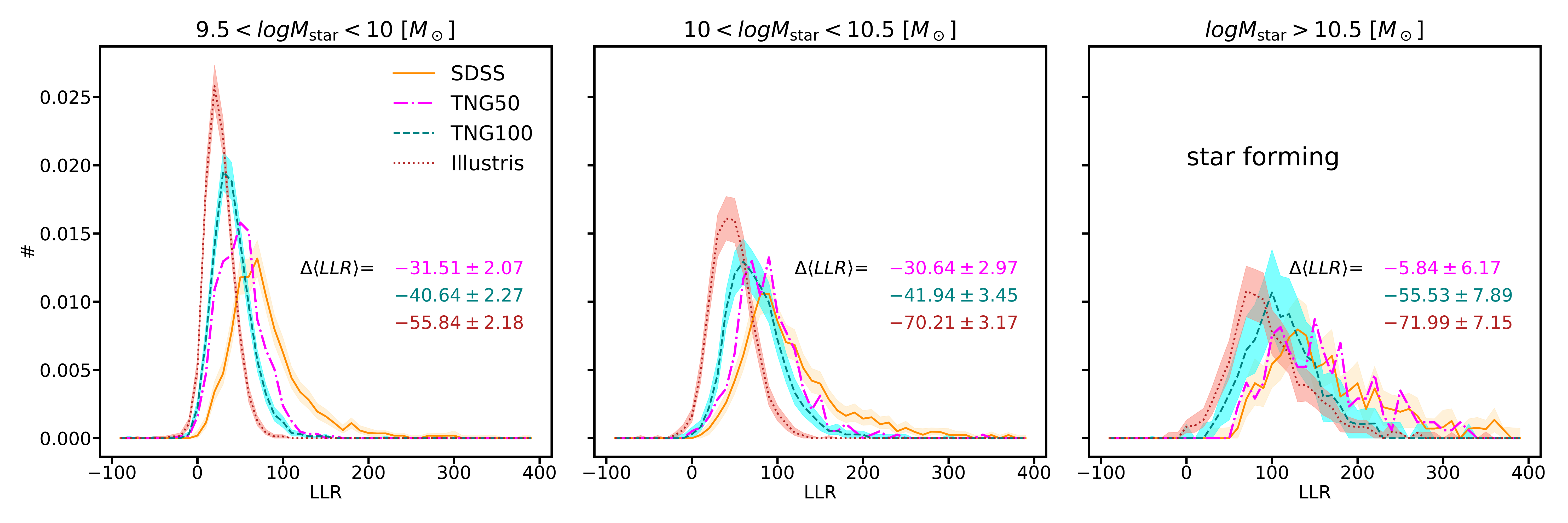} }\\
    \vspace{-1 em}
    \subfloat[]{\includegraphics[height=0.3\textwidth, clip=True]{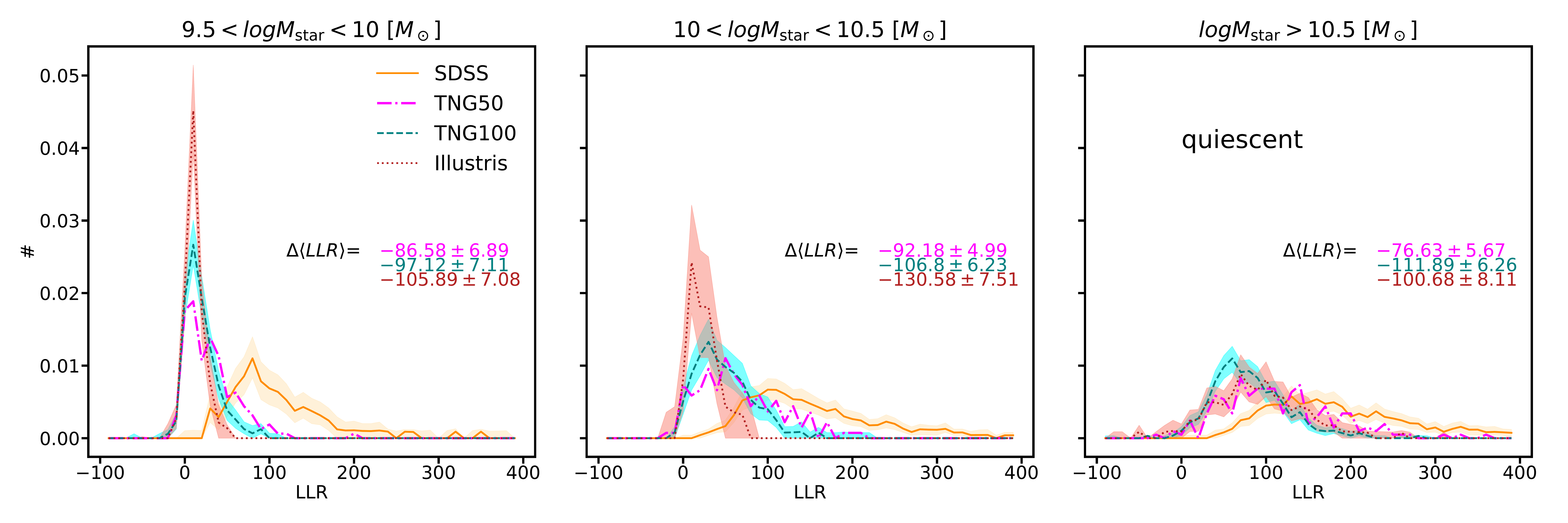} } 
    \caption{\emph{Upper row:} The log-likelihood ratio (LLR) distribution of star forming (left) and quiescent (right) galaxies for SDSS (orange), TNG50 (magenta), TNG100 (teal) and Illustris (red). \emph{Middle row:} The LLR distributions of star forming galaxies in three bins of galaxy stellar mass. \emph{Bottom row:} The LLR distributions of quiescent galaxies in three bins of stellar mass. Colors and line styles in the Middle and Bottom rows are as in Upper row.  The shaded regions show the 1 sigma confidence level obtained by bootstrapping SDSS, TNG100 and Illustris 100 times to the same sample size of TNG50.
    For star forming galaxies the $\Delta \langle LLR \rangle$
    is the lowest for TNG50, followed by TNG100 and Illustris, indicating that TNG50 is the simulation that best models star forming galaxies.
    Instead, all simulations struggle to accurately model quiescent galaxies, for which the $\Delta \langle  LLR \rangle$ remains low in all cases. These trends are robust across the stellar mass bins considered.  }
    \label{fig:LLR_SFQ}
\end{figure*}

%

\subsection{Star forming galaxies vs quiescent galaxies}
We have seen how the LLR provides a useful metric to evaluate the quality of galaxy images produced by simulations which is aware of the morphological details of galaxy structure. Based on this, in the previous Section we have also demonstrated that the latest generation of simulations of galaxy formation still struggles to produce realistic samples, despite a marked improvement compared to earlier work.  So, why is it that simulations are yet to reproduce in order to make realistic-looking galaxies? 

Here we try to answer this question by raising one issue that has been broadly debated in the literature, that is, the effectiveness of the implementations of the subgrid physics that regulates star formation and quenching. In the following we will advocate that most of the discrepancy between observations and simulation stems from an imperfect relationship between star-formation activity and small-scale morphological features.

To do this, we here exploit the power of our LLR framework, which prescribes that the higher the mean value of the LLR distribution of a dataset, the better it resembles observations. The LLR distributions for star forming ($\log{\rm{sSFR/yr}^{-1}}>-11 $) and quiescent ($\log{\rm{sSFR/yr}^{-1}}<-11 $) galaxies in our simulations and SDSS are shown in the upper panel of Figure \ref{fig:LLR_SFQ}. These distributions have been obtained by resampling SDSS, Illustris and TNG100 with the same sample size of TNG50 similarly to Figure \ref{fig:LLR}. 

The left top panel of Figure \ref{fig:LLR_SFQ} shows how the mean of the LLR distribution for simulated star forming galaxies is the closest to SDSS for TNG50, followed by TNG100 and with Illustris being the furthest away from it. The higher LLR of TNG100 with respect to Illustris is suggestive that the improved physical model for galaxy formation adopted in the IllustrisTNG framework is overall an improvement compared to the original Illustris implementation \citep{Pillepich+18_TNGdescription}. Furthermore, the unprecedented agreement with observations reached by TNG50 star forming galaxies is also a sign that a higher resolution is key to effectively model star formation. We note, however, that all simulated data sets are still inconsistent at the 1 sigma level with SDSS.

On the other hand, it can be seen that the improvement noted for star forming galaxies does not seem to propagate to quiescent galaxies as well (top right panel of Figure \ref{fig:LLR_SFQ}). We start by noting that in this case Illustris galaxies show a tail of high LLR that is consistent with IllustrisTNG at the 1 sigma level. However the large variance suggests that this tail is very scarcely populated, whereas the very small variance found for the spike at low LLR is indicative that the bulk of the population of Illustris quiescent galaxies lies there, i.e. they are very far from reproducing SDSS. Yet, while there seems to be an overall improvement from the original Illustris framework to IllustrisTNG, the better resolution offered by TNG50 over TNG100 does not appear to significantly modify the overall LLR distribution for quiescent galaxies. Indeed, the distributions of the two IllustrisTNG volumes are practically consistent but at very high LLR, where the probability density of TNG50 is slightly higher.

In short, what the top of Figure \ref{fig:LLR_SFQ} is telling us is that there has been a clear amelioration from the Illustris to the IllustrisTNG framework, in both the modelling of star formation regulation and of quenching, yet IllustrisTNG still produces small-scale stellar morphological details which differ from those in SDSS, and especially so for quiescent galaxies. Most importantly, while the higher resolution featured by TNG50 generates a sizeable improvement in the morphology of star forming galaxies, this is not the case for quiescent galaxies. This is suggestive that the physics which couples small-scale stellar morphological details to star-formation quenching in the IllustrisTNG simulations warrants improvement, or that an even higher resolution is needed to accurately model the processes that lead to quiescence.

\subsection{Mass dependence}
\label{sec:LLR_mass_SFQ}
The physical mechanisms that quench star formation in a galaxy are thought to depend on stellar mass. At low masses, it is generally accepted that quiescence mainly occurs in satellite galaxies due to environmental processes (e.g., \citealt{Peng+10}) -- this behaviour naturally emerging also in IllustrisTNG \citep{Joshi+20_disksTNG,Donnari+20_satellites_AGN_preprocessing}. Conversely, a plethora of possible mechanisms has been identified for quenching higher-mass galaxies (see Introduction). In IllustrisTNG, AGN feedback is responsible for halting star formation in galaxies with $M_{\rm star}\gtrsim 10^{10.5} M_\odot$ \citep[e.g.][]{Weinberger+17,Zinger+20}, regardless of whether they are centrals or satellites (\textcolor{blue}{Donnari et al. 2020, submitted}). Therefore, we break down the upper panel of Figure \ref{fig:LLR_SFQ} in the following bins of galaxy stellar mass: $10^{9.5}<M_{\rm star}/M_\odot<10^{10}$,$10^{10}<M_{\rm star}/M_\odot<10^{10.5}$ and $M_{\rm star}>10^{10.5}M_\odot$. The choice of these bins is not casual. Indeed, the lowest mass bin is where SDSS is incomplete and so the comparison between the datasets should be taken with a grain of salt. The other two bins are chosen to be around a mass scale that is thought to be key in galaxy formation, namely $M_{\rm star}\approx 3$x$10^{10}M_\odot \approx 10^{10.5}M_\odot$ (e.g., \citealt{Cappellari2016_review}). In IllustrisTNG that is roughly the mass scale where the AGN feedback mode switches from \emph{thermal} to \emph{kinetic} (\citealt{Weinberger+17}, \citealt{Terrazas+20}). We note that the prescriptions for AGN feedback have been significantly changed from Illustris to IllustrisTNG (see Section \ref{sect:simulations} and \citealt{Pillepich+18_TNGdescription}), and that it has been argued that AGN feedback plays a role in establishing the morphology of massive galaxies (e.g., \citealt{Dubois+16_MorphoAGN},\citealt{Genel+15_angomomMorphoAGN}).



For star forming galaxies, we can see that the trend of the top panel of Figure \ref{fig:LLR_SFQ} persists across all masses: star forming galaxies are best reproduced by TNG50, followed by TNG100 and Illustris, from the least massive to the most massive galaxies. In particular, it is noteworthy that the $\Delta \langle LLR \rangle$ of massive star forming galaxies in TNG50 is consistent with zero at the 1$\sigma$ level, meaning that these galaxies are reproduced extremely well by TNG50.

Quiescent galaxies, instead, feature a significantly worse, i.e. lower, $\Delta \langle LLR \rangle$ consistently across all masses and for all simulations. We note that the higher resolution of TNG50 improves only marginally on TNG100 and Illustris in the lower mass bins, but it is more significant for massive galaxies. This evidence suggests that environmental quenching in all simulations always produces galaxy morphologies that differ from those of SDSS, with a weak dependence on resolution. For massive galaxies ($M_{\rm star}\gtrsim 10^{10.5}M_\odot$), we note that for TNG100 the $\Delta \langle LLR \rangle$ is lower than for Illustris (although they are consistent at the 1$\sigma$ level), while TNG50 improves on both. The fact that quenched galaxies in Illustris and TNG100 have a similar performance is puzzling. In fact, the distinct implementations of AGN feedback in the two simulations may be expected to generate different levels of agreement with SDSS. We speculate below on the possible reasons for this somewhat unexpected result.

One possibility is that the exact implementation of AGN feedback does not significantly affect morphology at the resolution of Illustris and TNG100, at least at the redshift probed here, $z\approx0.05$. It could be possible that AGN feedback may have an impact on morphology at higher redshift, but then major mergers substantially change the morphology of quiescent galaxies (e.g. \citealt{Rodriguez-Gomez+17_mergers_spin}, \citealt{Clauwens+18}, \citealt{Martin+18_morpho_HorizonAGN}, \citealt{Tacchella+19}), at which point the small-scale collisionless dynamics of the stars in the merger remnant depends on numerical resolution. This argument would be favoured by the fact that major mergers are observed to occur with similar rates in Illustris \citep{Rodriguez-gomez+16} and TNG100 \citep{Huertas-Company+19} for massive galaxies. In the pictures outlined above, the better match of TNG50 with SDSS could simply be due to an improved resolution, but not necessarily a better physical model for AGN feedback.

We also note that some tensions in the relationship between size, color and non-parametric morphological estimators between observations and both Illustris and, to a lesser extent, TNG100 massive galaxies was already highlighted in \citet{Rodriguez-Gomez+19}. The variety of non-parametric morphological indicators adopted in \citet{Rodriguez-Gomez+19} provide separate pieces of information compared to our unsupervised LLR strategy, which generalises over human-biased non-parametric approaches and summarises the small-scale details in a single value. Moreover, we note that some higher-level details in the galaxy structure may be lost with the higher pixel scale (i.e. lower resolution, $0.396$ arcsec/pix) of SDSS, which we adopt here, compared to the Pan-STARRS observations used in \citet{Rodriguez-Gomez+19}, where a lower pixel scale  (i.e., higher resolution, $0.25$ arcsec/pix) is available. Therefore, a direct comparison between our results and those presented in \citet{Rodriguez-Gomez+19} is significantly non trivial. This is true in general, and it applies in particular to the case of massive galaxies which we discussed here.

\subsection{Does environment matter?}
\label{sec:environment}
\begin{figure*}
    \centering
    \subfloat[\label{fig:LLR_SFQ_cen}]{{\includegraphics[width=0.7\textwidth]{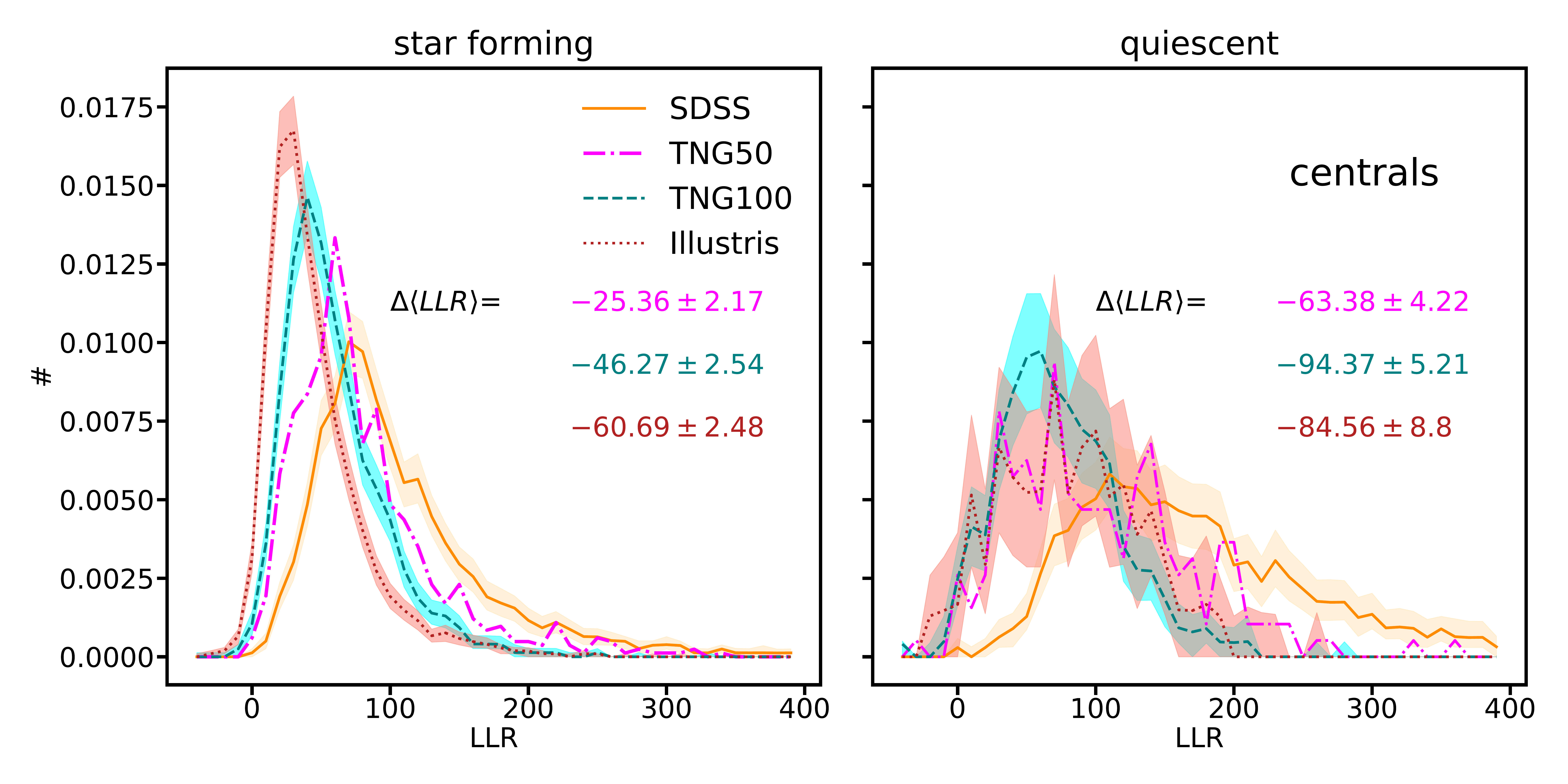} }}\\
    \subfloat[\label{fig:LLR_SFQ_sat}]{{\includegraphics[width=0.7\textwidth]{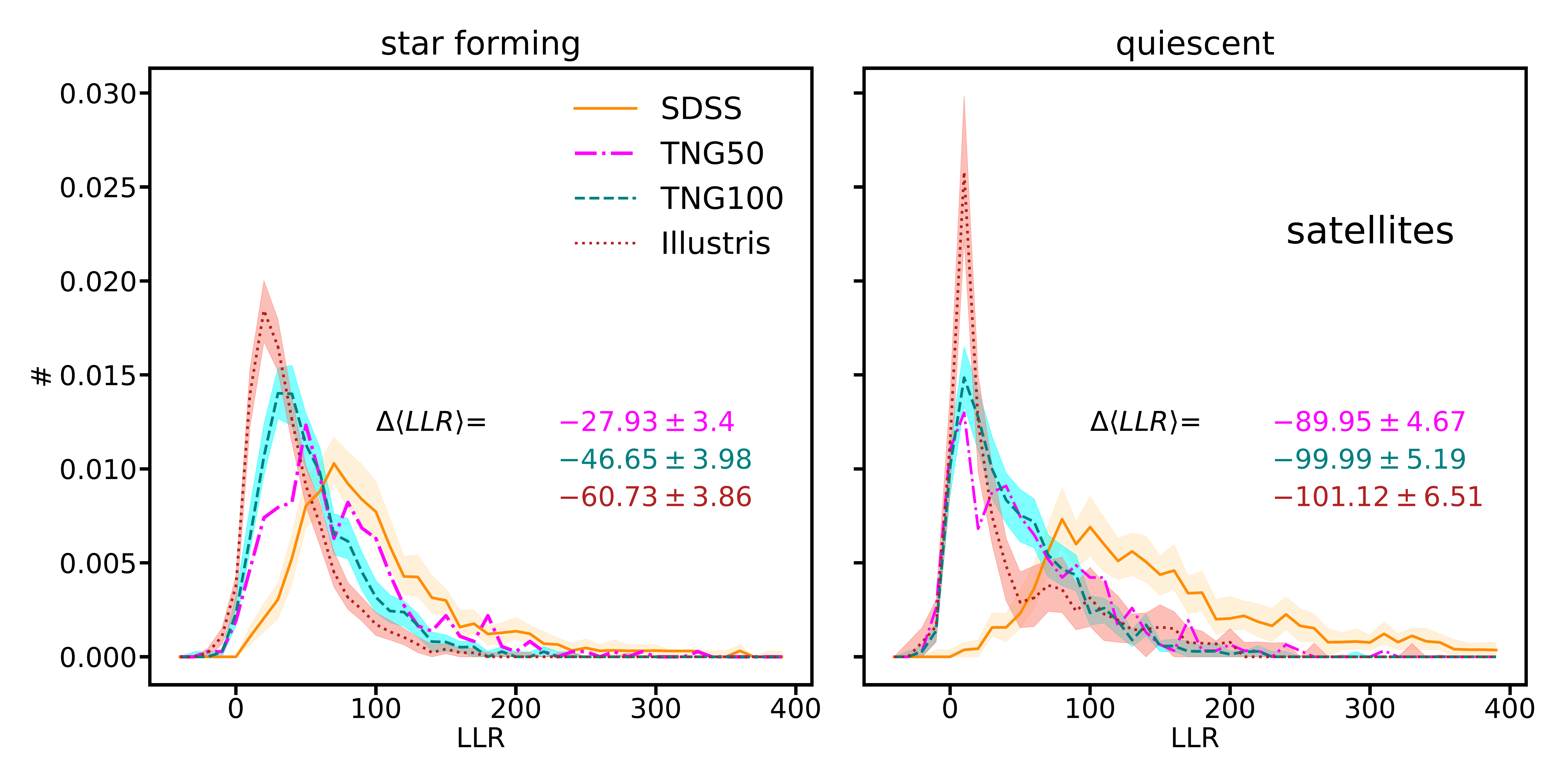} }}
    \caption{The log-likelihood ratio (LLR) distributions of quiescent and star forming galaxies for centrals (top) and satellites (bottom). The LLR distributions of star forming centrals and satellites follow the same trends highlighted in Figure \ref{fig:LLR_SFQ}. Contrary to star forming galaxies, quiescent galaxies display a lower $\Delta \langle LLR \rangle$ both for centrals and satellites: this indicates that quiescent galaxies are not well reproduced in simulations regardless of the quenching mechanism (environmental quenching for low mass satellites, AGN for centrals and massive satellites).}
    \label{fig:LLR_SFQ_censat}
\end{figure*}

We have shown in the previous Sections that while the modelling of star forming galaxies has continuously improved in the years, quenched galaxies still seem inaccurate according to our deep learning framework. These results have been presented for the full galaxy population of our data sets, as well as for three stellar mass cuts. In particular, we have discussed the connection between stellar mass, quiescence and the environment. Here, we test the link with environment more explicitly.

We show the LLR distributions of quiescent and star forming central and satellite galaxies in Figure \ref{fig:LLR_SFQ_censat}. Let's start by comparing the trends for star forming galaxies. It is clear that in this case both satellites and centrals markedly improve from Illustris to TNG100, and from the latter to TNG50, as was shown in Figure \ref{fig:LLR_SFQ} for the full population. It is also interesting to note that the $\Delta \langle LLR \rangle$ for star forming centrals and satellites are almost identical for all simulations. In fact, this is a trend that we observe also for the quenched population: by comparing the $\Delta \langle LLR \rangle$ quoted in the right column of Figure \ref{fig:LLR_SFQ_censat} for central and satellite quiescent galaxies, we observe that similarly low values are achieved. The only exception is for TNG50, where central quiescent galaxies feature a significantly higher $\Delta \langle LLR \rangle$ compared to quiescent satellites. Since the population of quenched quiescent galaxies is dominated by massive galaxies in IllustrisTNG, we refer the reader to the discussion at the end of the previous Section for a speculative explanation of this behaviour.

In summary, Figure \ref{fig:LLR_SFQ_censat} suggests that the different processes that quench central and satellite galaxies result in a similar disagreement with observations. This in turn suggests that the main culprit for the disagreement is not necessarily to be searched in {\it the way} gas is removed and star formation halted (e.g. via ram-pressure stripping vs gas expulsion via BH feedback in the TNG runs) but rather on how the stellar light distribution is realized in the numerical models in the case of quenched galaxies. We note, however, that the relatively small volumes probed by the IllustrisTNG and Illustris simulations, as well as SDSS, implies that the statistic of cluster-sized dark matter haloes, which host most of the satellite galaxies, is subject to significant cosmic variance: there are about 6 clusters with $M_{\rm halo}>10^{13.5} M_\odot$ in SDSS and TNG50, while TNG100 features more than 50 of them. Since environmental quenching is found to be a rather steep function of host halo mass in both observations (e.g., \citealt{Davies+19}) and simulations (e.g., Donnari et al. in prep), as well as of individual cluster assembly history \citep{Joshi+20_disksTNG}, we caution that the disagreement found for satellite galaxies should be taken with a grain of salt.


\section{The realism of simulated galaxies across scaling relations and the role of quenching}
\label{sec:scalingrelations}
\begin{figure*}
    \centering
    \subfloat[]{\label{fig:sizemass}\includegraphics[height=0.3\textwidth]{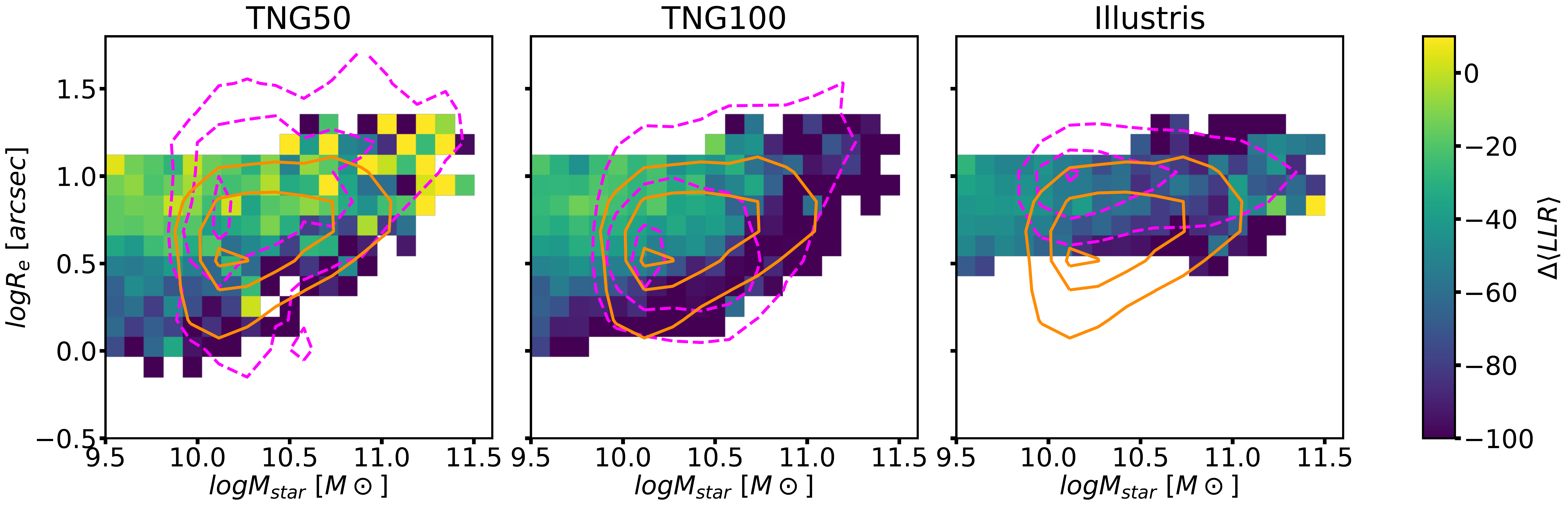} }\\
    \subfloat[]{\label{fig:sSFR}\includegraphics[height=0.3\textwidth]{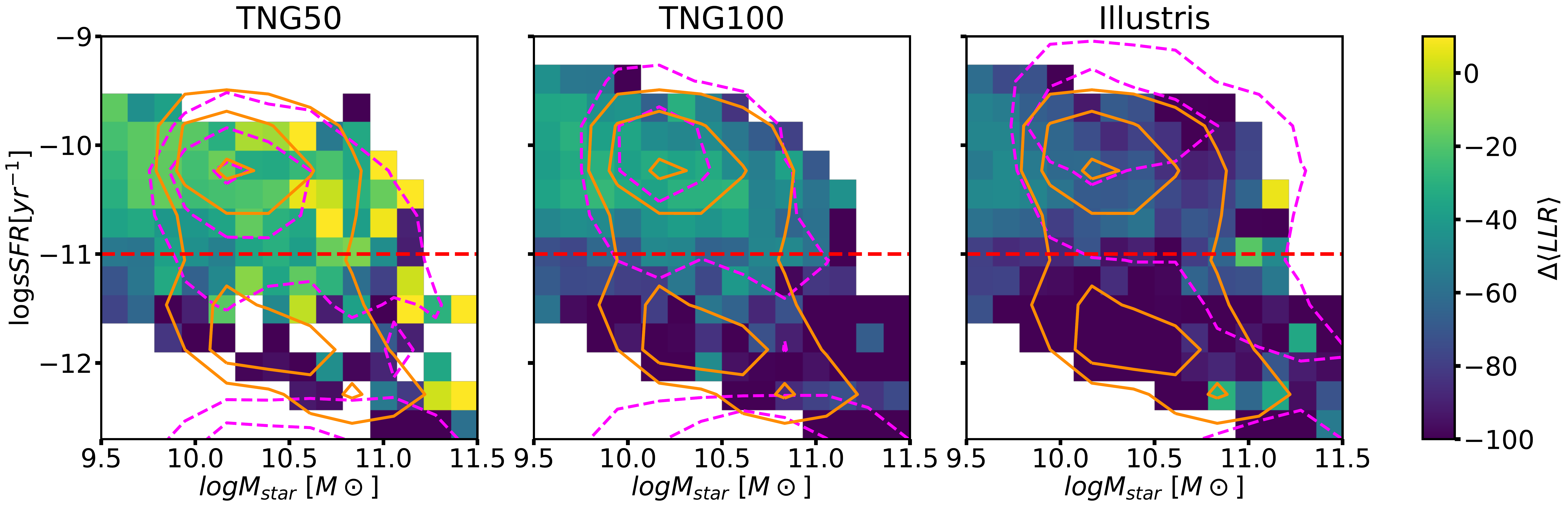}}\\
    \subfloat[]{\label{fig:n_R}\includegraphics[height=0.3\textwidth]{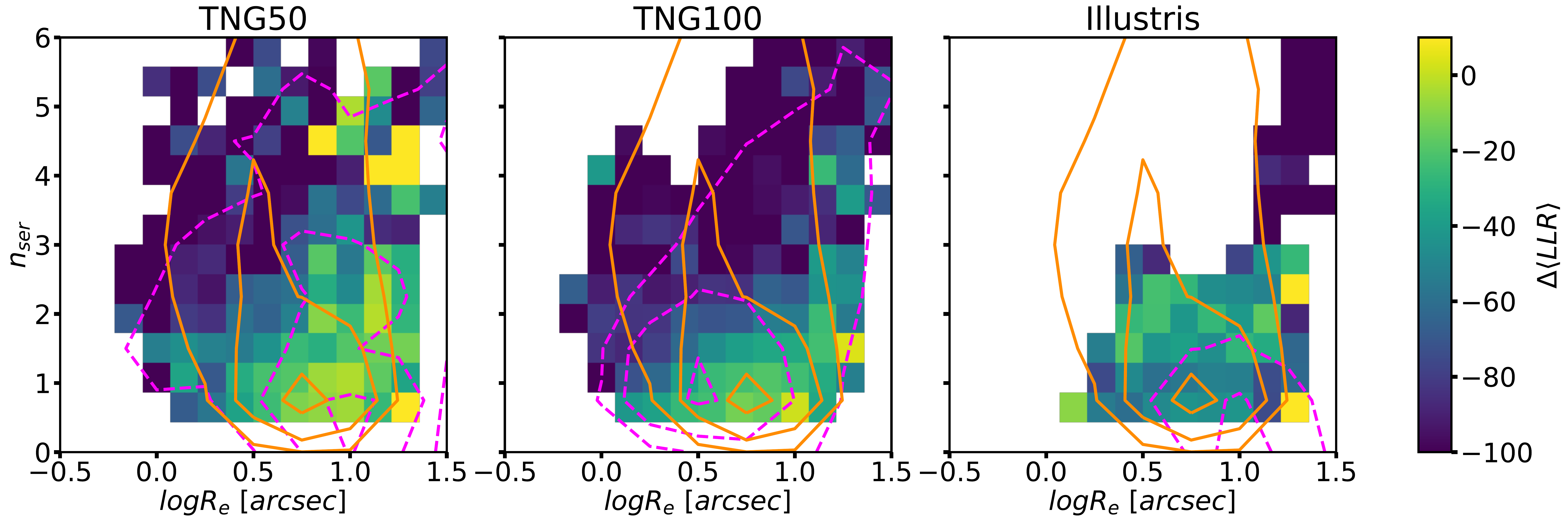} } 
    \caption{The size-mass relation (top panel),    $sSFR-M_{\rm star}$ relation(middle panel) and the S\'ersic index-size relation   (bottom panel) for the three simulations studied in this work as labelled. The color code is the difference between the mean LLR of each simulation and the mean LLR of SDSS at each point of the scaling relations. A brighter color indicates a better agreement with SDSS. In the middle panel we also show with a red dashed line the sSFR threshold that defines star forming ($\log{sSFR/yr^{-1}}\gtrsim -11$) and quiescent($\log{sSFR/yr^{-1}}\lesssim -11$) galaxies. We also impose a strict lower limit on the sSFR at $\log{sSFR/yr^{-1}}=-12.5$. We show with orange solid contours the 10th, 50th and 90th percentiles of the 2D distributions for SDSS galaxies for galaxies above the mass completeness threshold of $ M_{\rm star} \approx 10^{10} M_\odot$. Contours for the same mass cut are also shown with magenta dashed lines for simulations, which are in the ballpark of the observed scaling relations (especially so for TNG50, less so for Illustris). It can be seen that quenched, concentrated, small galaxies are the ones with the lowest $\Delta \langle LLR \rangle$, and so their fine stellar morphology substantially disagrees with observations.}
    \label{fig:scalingrelations}
\end{figure*}

The neural networks that we use here are aware of galaxy structure only by design and are not trained with any direct information about star formation activity. Yet, we have just shown that the morphologies of quiescent and simulated galaxies are not reproduced equally well by simulations according to our deep learning framework. The reason for this behaviour must therefore be investigated more thoroughly.

One way to address this issue is to explore the quality of simulations across galaxy scaling relations. More specifically, we study how the average LLR of simulated galaxies deviates from the average LLR of SDSS galaxies at each point on the planes defined by scaling relations. Thus, in this case the  $\Delta \langle LLR \rangle$ gives an indication of how realistic simulated galaxies are in a given region of the planes defined by galaxy properties. Note that this kind of analysis is possible only because simulations are in the ballpark of observations, at least at the redshift of interest. Yet some data points for simulations still lie outside of the manifold, and therefore we exclude them in the following. To make this abundantly clear, the blank space in the following Figures may mean either that SDSS observations or simulated galaxies are not present in that region of the manifold. Nevertheless, we show contours in each panel for the distributions of SDSS (orange solid curves) and the simulated (magenta dashed curves) galaxies to give an idea of how the different samples populate the depicted planes.

As an example, we take three scaling relations that have been widely studied in the literature: the $R_e-M_{\rm star}$ relation (size-mass relation, e.g. \citealt{Shankar+10_sizefunct},\citealt{Bernardi+14}, \citealt{Lange+15}, \citealt{Zanisi+20_sizefunct}), the $n_{ser}-R_e$ relation (S\'ersic index-size relation e.g. \citealt{Trujillo+01_sersic}, \citealt{Ravikumar+05_sersic}) and  the $sSFR-M_{\rm star}$ relation (specific star formation rate-stellar mass relation, e.g. \citealt{Salim+07}, \citealt{Elbaz+11}).
These are shown for each simulation in Figure \ref{fig:scalingrelations}, and are color coded by the  $\Delta \langle LLR \rangle$. We discuss each of these relations separately at first, and we then propose an interpretation.

In the size-mass relations of both IllustrisTNG simulations there is a clear gradient in  $\Delta \langle LLR \rangle$, where at fixed stellar mass larger galaxies deviate the least from SDSS and smaller ones are progressively less realistic. Instead, this behaviour is not present in Illustris, due to the well-known lack of small galaxies in this simulation at low redshift \citep{Snyder+15_morpho_SF}. Interestingly,  massive galaxies seem to be better reproduced in Illustris compared to TNG100. As Illustris massive galaxies are typically more star forming than TNG100 galaxies \citep{Donnari+20_quenchedFract_mocks}, and since star forming galaxies are on average better reproduced, then the $\Delta \langle LLR \rangle$ is likely biased high for Illustris massive galaxies when no cut on star formation activity is made. We will discuss trends for star forming and quenched galaxies separately later in this Section.


The $\rm{sSFR}-M_{\rm star}$ relations reveal that star forming galaxies notably improve from Illustris to TNG100 and from the latter to TNG50. In particular, it is worth noticing that massive star forming galaxies seem to be slightly more accurate than less massive ones. On the contrary, it can be seen that on average passive galaxies differ the most from SDSS.  Lastly, it is also worth reminding the reader of the well-known uncertainties in retrieving SFR from the observed optical colours only (e.g. \citealt{Donnari+19}, \citealt{Eales+17,Eales+18}), which could affect dramatically the distribution of SDSS observations for  $\log{\rm{sSFR/yr}^{-1}}\lesssim -11$  and hence this kind of region-wise comparison with simulations. We will address this point in the following.

Finally, the bottom panels of Figure \ref{fig:scalingrelations} show the $n_{ser}-R_e$ relations for the three simulations. Although the trends are somewhat less obvious is this case, a close inspection of the figure reveals a few interesting details. First of all, Illustris is not able to produce galaxies with medium-to-low sizes and high S\'ersic indices, as already noted by \citet{Bottrell+17_bulgeIllustris}. While this is something that is reproduced in TNG100, we note that high S\'ersic index galaxies tend to differ the most from their SDSS counterpart. In TNG50, instead, we clearly see that high mass galaxies with a high $n_{ser}$ are much better in agreement with SDSS.

\begin{figure*}
    \centering
    \subfloat[]{\includegraphics[height=0.3\textwidth]{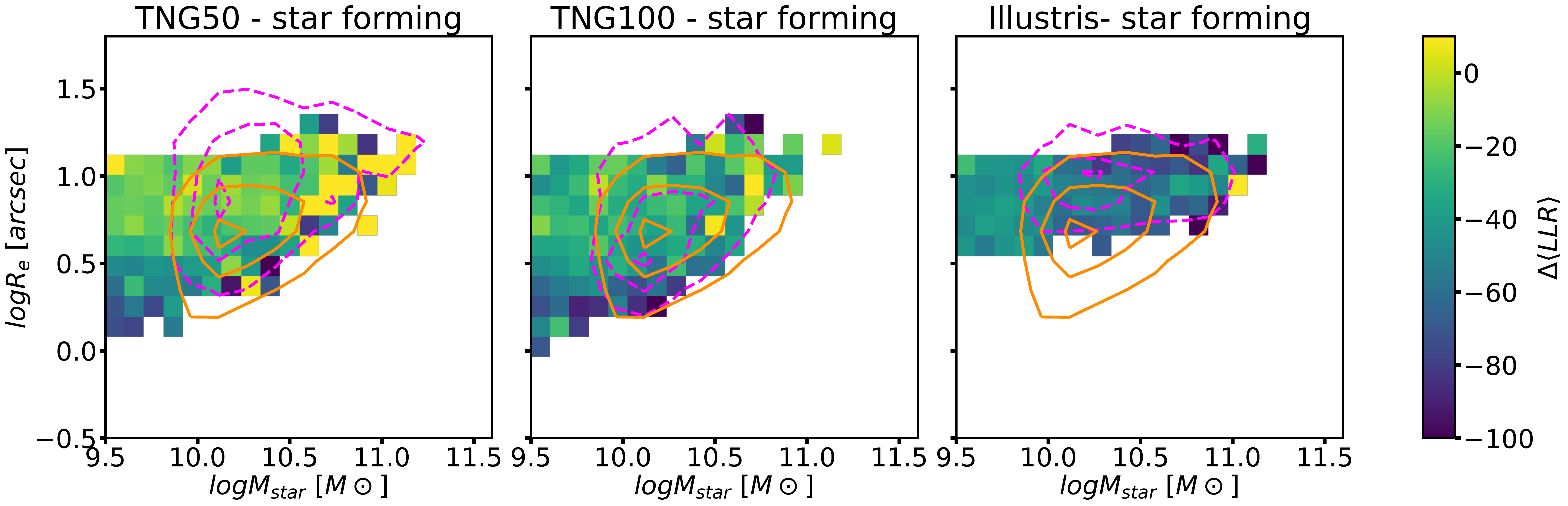} }\\
    \subfloat[]{\includegraphics[height=0.3\textwidth]{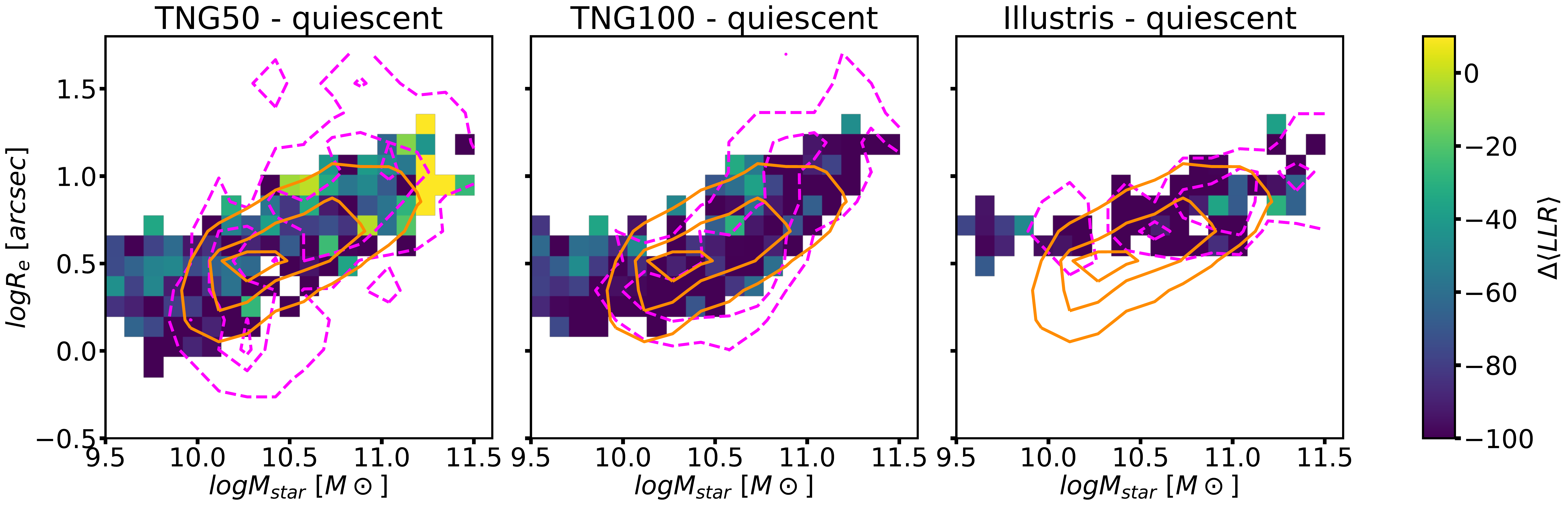} } \\
    \caption{The size-mass relation for star forming (top row) and quiescent galaxies (bottom row). The color code is the same as in Figure \ref{fig:scalingrelations}. Here TNG100 and Illustris have been randomly sampled to the same sample size of TNG50. See \textbf{online supporting material} for other realizations of the sampling.  We also show with orange solid contours the 10th, 50th and 90th percentiles of the 2D distributions for SDSS galaxies for galaxies above the mass completeness threshold of $ M_{\rm star} \approx 10^{10} M_\odot$. Contours for the same mass cut are also shown with magenta dashed lines for simulations. It can be seen that quiescent galaxies are in general less well reproduced, with the exception of massive quenched TNG50 galaxies. Note that although the morphology of star forming galaxies is better reproduced by simulations, smaller simulated star forming galaxies feature a lower $\Delta \langle LLR \rangle$ compared to larger ones at fixed stellar mass.}
    \label{fig:sizemass_SFQ}
\end{figure*}
\begin{figure*}
    \centering
    \subfloat[]{\includegraphics[height=0.3\textwidth]{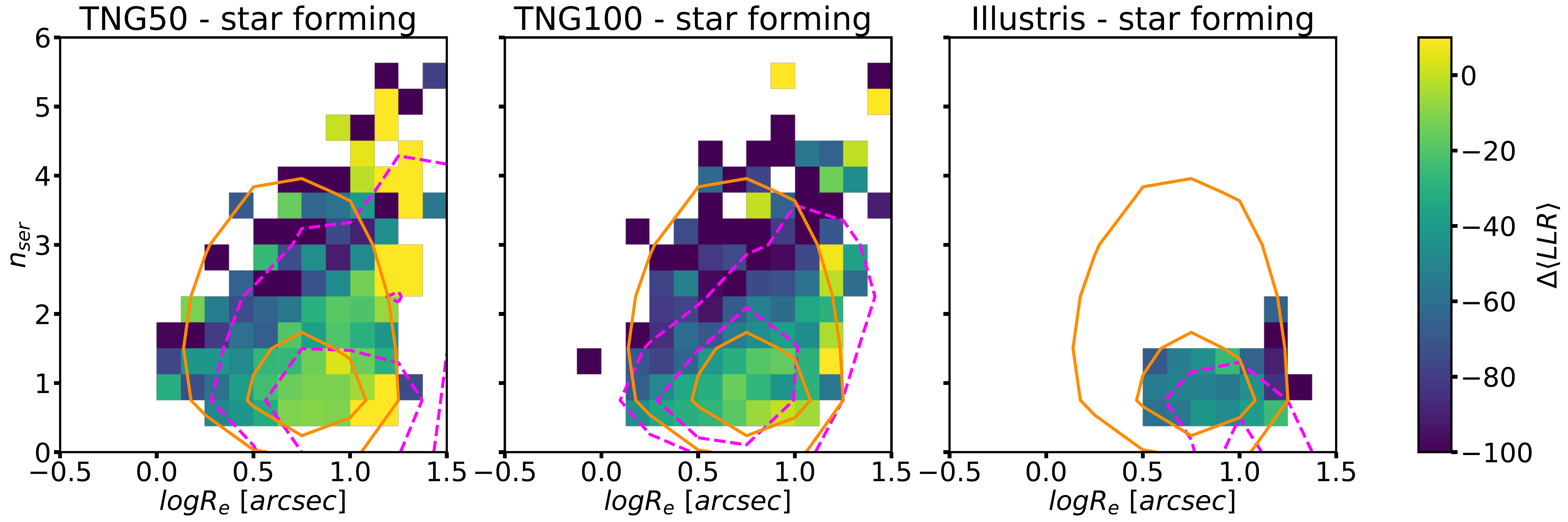} }\\
    \subfloat[]{\includegraphics[height=0.3\textwidth]{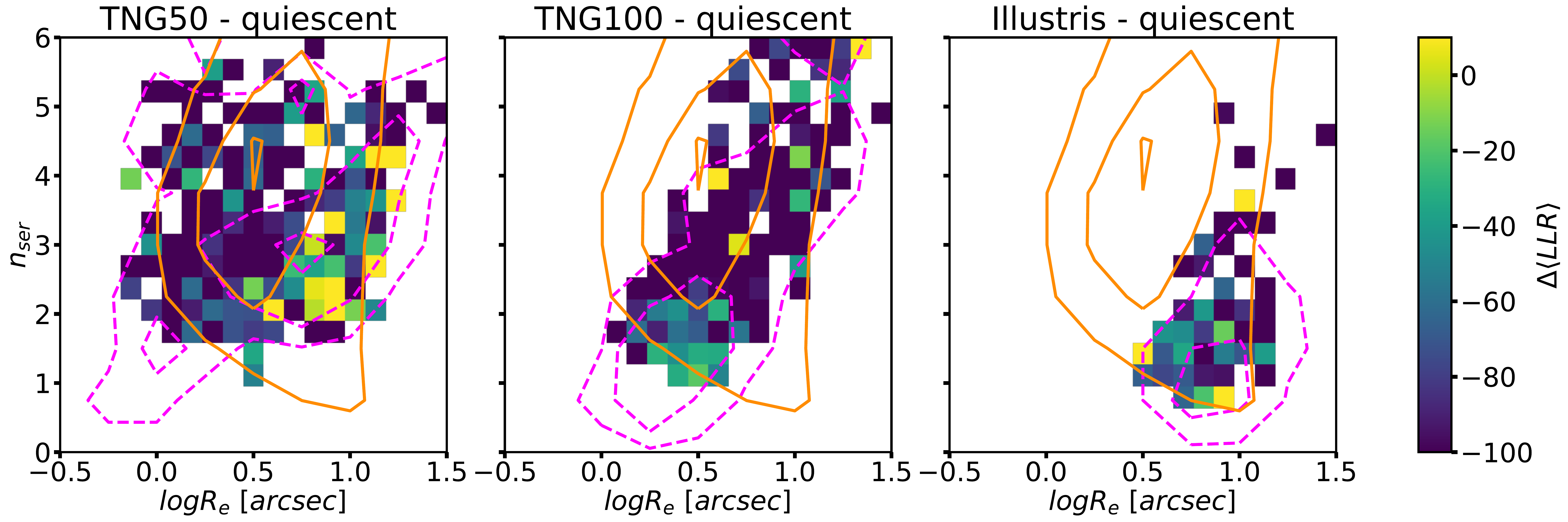} } \\
    \caption{The S\'ersic index-size relation for star forming (top row) and quiescent galaxies (bottom row). The color code is the same as in Figure \ref{fig:scalingrelations}. Here TNG100 and Illustris have been randomly sample to the same sample size of TNG50. See \textbf{online supporting material} for other realizations of the sampling.  We also show with orange solid contours the 10th, 50th and 90th percentiles of the 2D distributions for SDSS galaxies for galaxies above the mass completeness threshold of $ M_{\rm star} \approx 10^{10} M_\odot$. Contours for the same mass cut are also shown with magenta dashed lines for simulations. Note the absence of small, high-S\'ersic index galaxies in Illustris and (although to a less extent) TNG100. Also not that this population is instead present in the higher resolution TNG50. Moreover, it is worth observing that large galaxies with a medium-to-high S\'ersic index are better reproduced in TNG50, both in the quiescent and star forming populations, compared to TNG100.}
    \label{fig:n_R_SFQ}
\end{figure*}

To summarize our findings, simulations seem to still struggle at reproducing the small-scale stellar structural features of galaxies that are more concentrated and smaller in size, at fixed stellar mass. We wish to further explore how the quality of simulated galaxies across the scaling relations studied here depends on star formation activity. Therefore, we split our data sets in star forming and quiescent galaxies, as done in the previous Section. Note that by binning in sSFR above and below $\log{\rm{sSFR/yr}^{-1}}\approx-11 $, we alleviate the issue about the reliability of the comparison between the sSFR inferred from observations and simulations.


There is an important caveat to mention before we proceed. As discussed already in Section \ref{sec:data}, not all galaxy populations may be statistically well represented in the volume of TNG50, which is more than 8 times smaller compared to the other simulations. This would explain, for instance, the fact that in TNG50 the quiescent region of the $sSFR-M_{\rm star}$ relation seems to be less densely populated in Figure \ref{fig:scalingrelations}. We alleviate this issue in the following by showing random realizations of TNG100 and Illustris of the same sample size of the smaller IllustrisTNG volume, as done previously. In the \textbf{online supplementary material} we show the same Figures for different random samples to show that our interpretation is robust.

 Figure \ref{fig:sizemass_SFQ} shows the well-known trend where on average star forming and quiescent galaxies lie above and below the mean of the size-mass relation at fixed stellar mass respectively for the IllustrisTNG simulations \citep{Genel+18}. This trend agrees with observations and it
is something that is not seen in Illustris. Indeed the absence of this differential size-mass relation in Illustris was raised as a cause of concern by \citet{Bottrell+17_bulgeIllustris}. We note that the overall too-large sizes of Illustris galaxies, independent of color/SFR, was taken into account for the TNG model calibration \citep{Pillepich+18_TNGdescription}. However, it is clear from Figure \ref{fig:sizemass_SFQ} that quiescent galaxies in both IllustrisTNG volumes have a consistently lower  $\Delta \langle LLR \rangle$ value compared to star forming galaxies, with the exception of massive quiescent galaxies in TNG50.

The $n_{ser}-R_e$ relations for star forming and quiescent galaxies are shown in Figure \ref{fig:n_R_SFQ}. In this Figure we see that for star forming galaxies there is a definite improvement from TNG100 to TNG50, especially for large, high-$n_{ser}$ galaxies. In the original Illustris simulation very few star forming extended, high S\'ersic index galaxies even exist. For quiescent galaxies, the improvement is less marked from Illustris to TNG100. However when comparing the latter to TNG50 quiescent galaxies, we do see hints that extended galaxies with $3 \lesssim n_{ser} \lesssim 4$ are better reproduced in the smaller IllustrisTNG volume. Interestingly, we also see that TNG50 is able to produce compact, highly concentrated galaxies, which however still differ more from SDSS galaxies in terms of their small-scale stellar morphological details..

In summary, the variation of the quality of simulations across galaxy structural scaling relations, as quantified by the  $\Delta \langle LLR \rangle$, seems to support the idea that simulations do not generate realistic small-scale features in the stellar morphology of quenched galaxies, particularly those small in size and/or highly concentrated.
This holds true even in the IllustrisTNG simulations, where the bimodality of structural scaling relations is broadly reproduced.

\section{Interpreting the LLR}
\label{sec:interpretation}

\begin{figure*}
    \centering
    \subfloat[]{\includegraphics[height=0.4\textwidth]{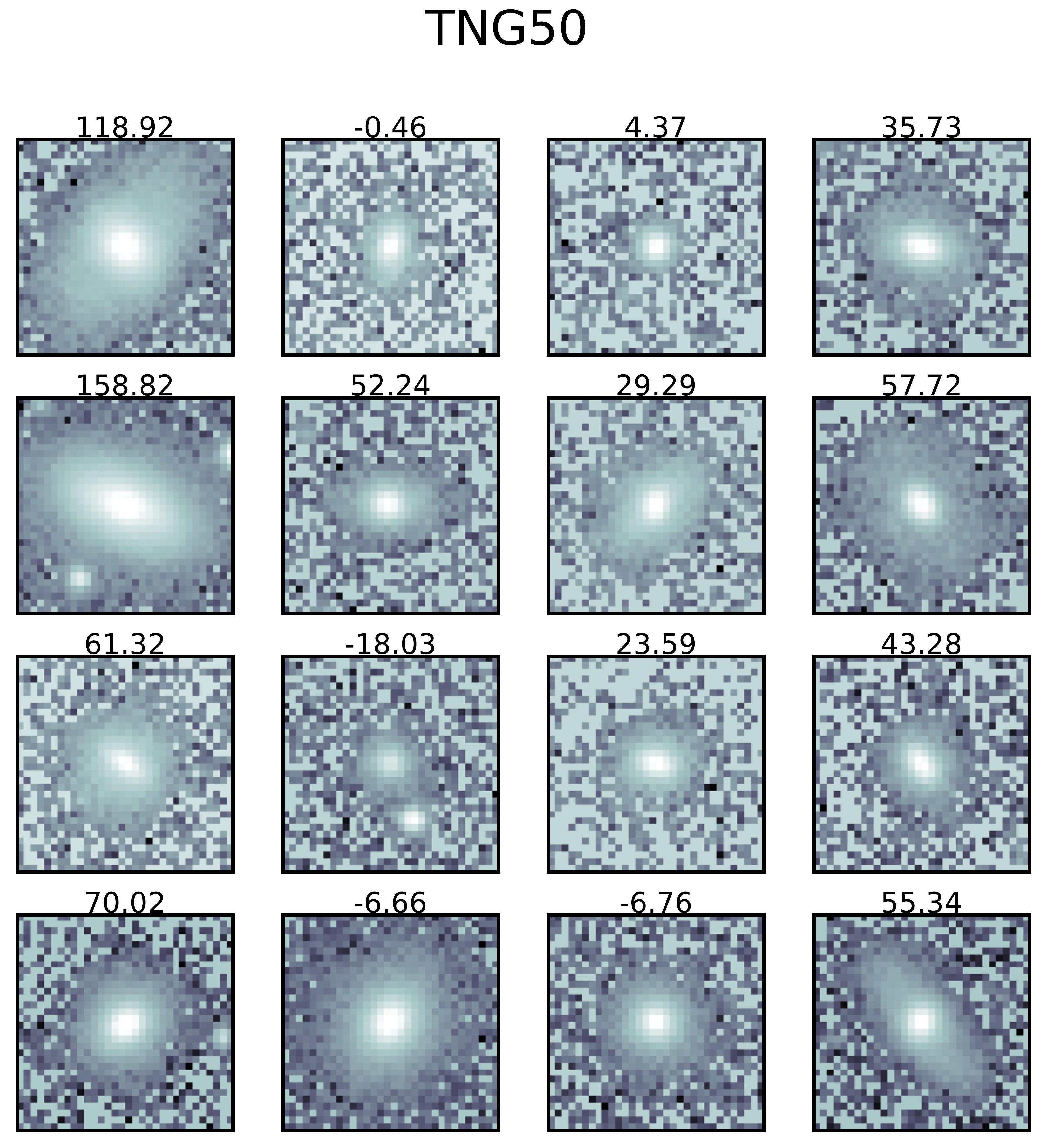} }
    \subfloat[]{\includegraphics[height=0.4\textwidth]{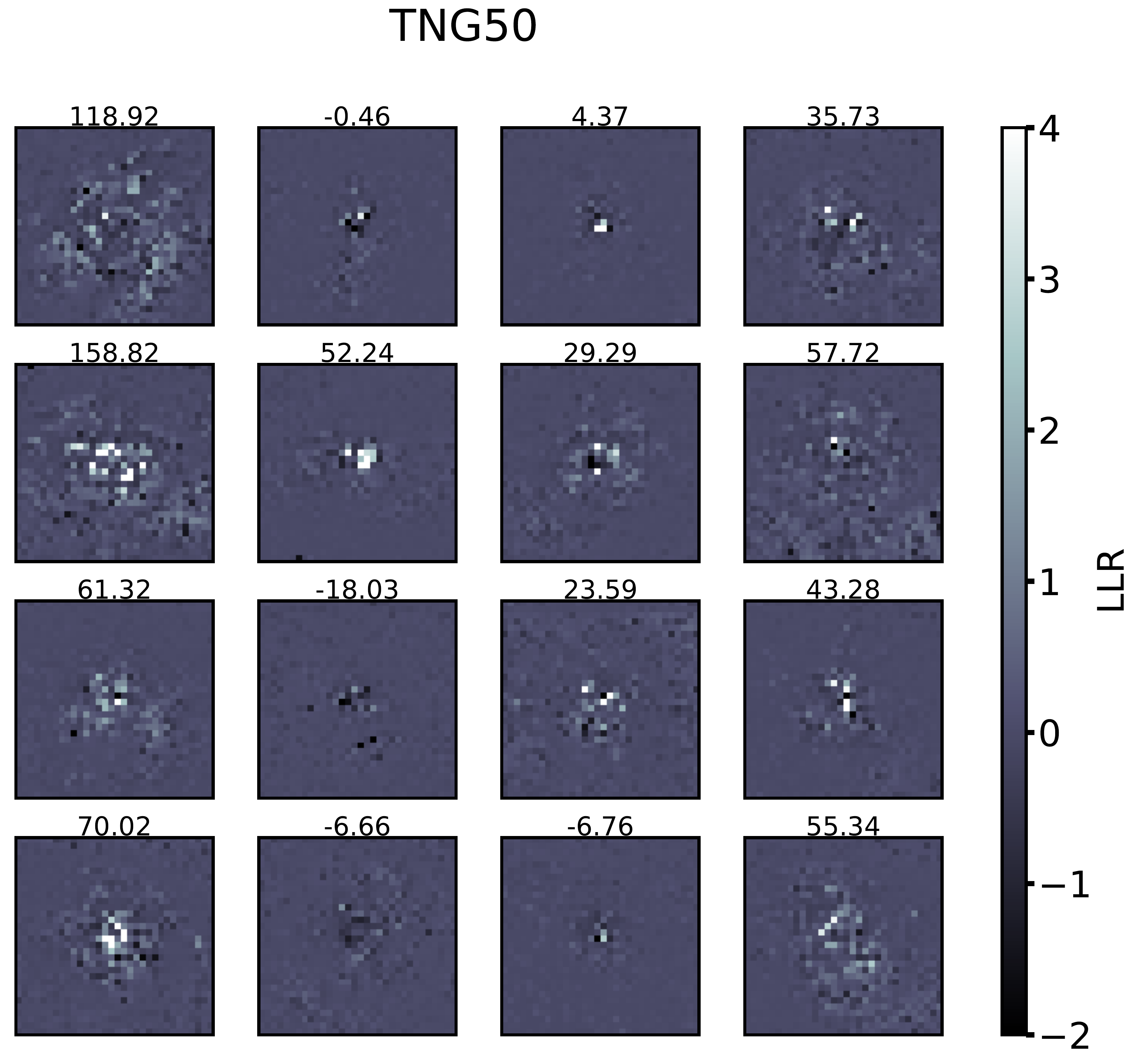} }\\
    \subfloat[]{\includegraphics[height=0.4\textwidth]{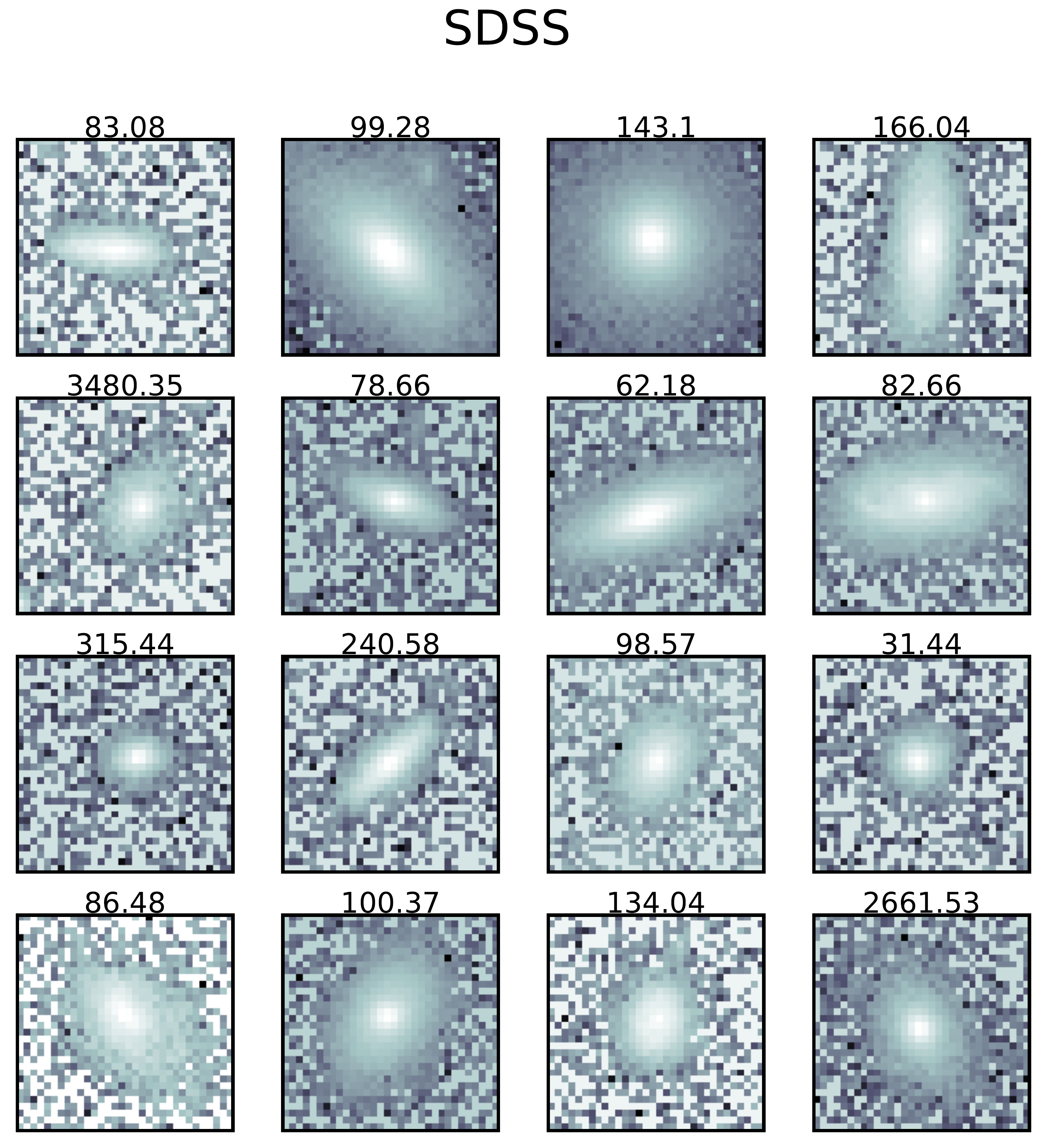} }
    \subfloat[]{\includegraphics[height=0.4\textwidth]{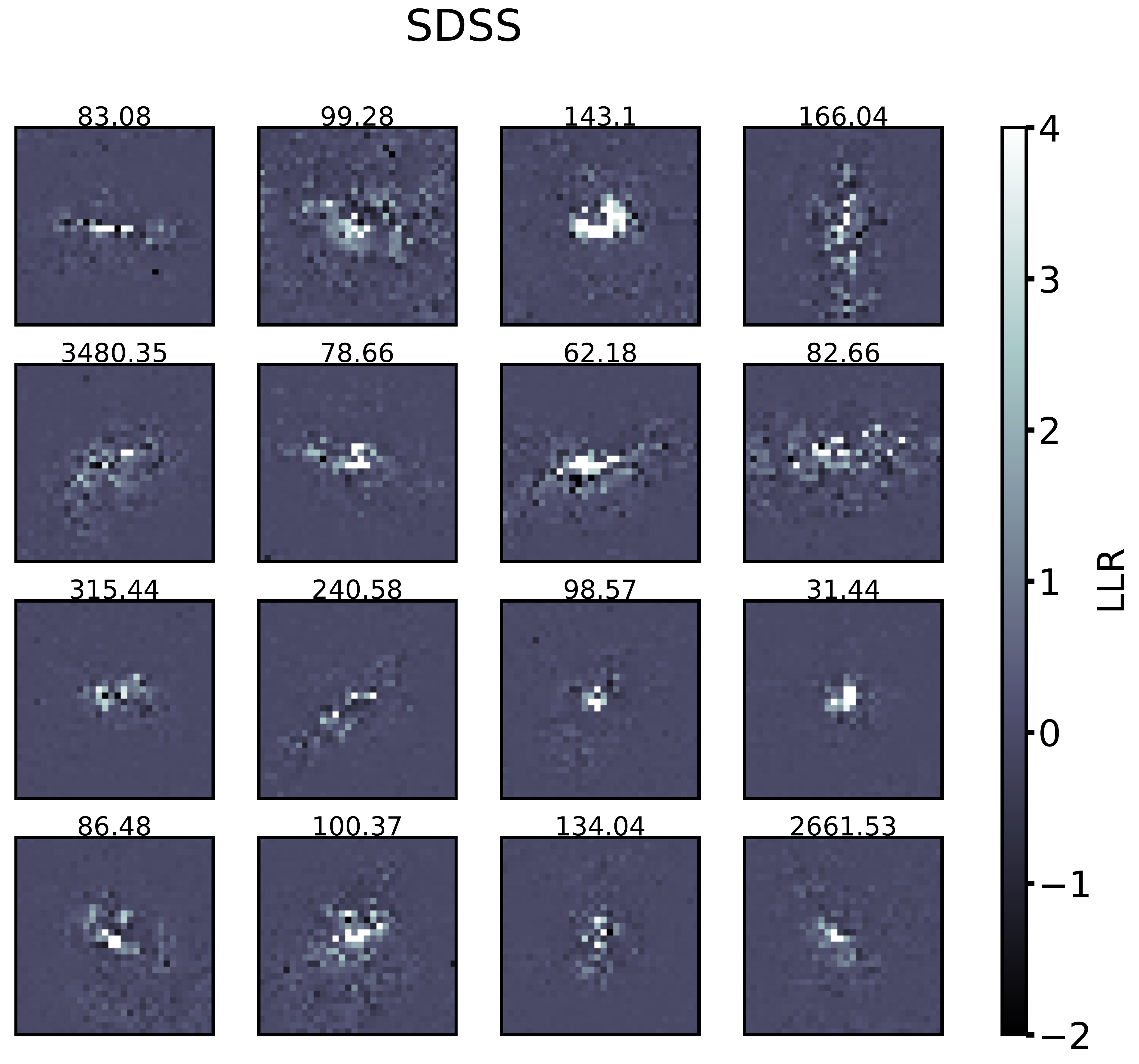} }\\
    
    \caption{Thumbnails of TNG50 (top left) and SDSS (bottom left) quenched galaxies with $R_e<3$'', $n_{ser}>4$. The top right and bottom right panels show the pixel-wise contributions to the LLR for the TNG50 and SDSS galaxies respectively. Each panel is labelled with its value of the LLR. The color scale in the right column is identical for all the panels. Note that we have manually limited the color scale to values from -2 to 4 for practical reasons, but pixels can assume also higher and lower values. For instance, if the contribution of a given pixel is 100, it will saturate to the color corresponding to the value of 4.} It can be seen that the central regions of TNG50 galaxies are much less prominent in the LLR maps compared to SDSS, despite the thumbnails of real and simulated galaxies look fairly similar. This indicates a failure in the simulation to properly capture the densest regions of quenched galaxies.
    \label{fig:LLR_interpret}
\end{figure*}

The interpretability of the outcome of deep learning studies is always problematic. Substantial progress has been made in the case of convolutional neural networks (CNNs) applied to classification tasks with techniques such as GradCam \citep{gradCAM} and Saliency Maps \citep{SaliencyMaps}. These algorithms provide a way to visualize the regions of an image a CNN mostly focuses on to output a certain prediction. Saliency maps have been already applied in galaxy morphology classification \citep{Huertas-Company+19}, and GradCam in merger stage identification \citep{Snyder+20_mergerclassification}.
Unfortunately, by nature, these techniques cannot be applied to generative models.
PixelCNN, however, has the very amenable feature that the likelihood is constructed pixel by pixel, and so the LLR, which is the ratio of the likelihood of two PixelCNN networks. Therefore, it is possible to identify which pixels contribute the most to the LLR, and therefore infer what the networks believe a realistic galaxy looks like.

As an example, we focus here on a population of galaxies which is poorly reproduced in simulations, that is, small, concentrated quiescent galaxies (see previous Section). A sample of this population for SDSS and TNG50 is shown in Figure \ref{fig:LLR_interpret} (left column), along with the pixel-wise contributions to the LLR (``LLR maps", right column). The colour scale of the LLR maps saturates at values of 4 and -2 for practical reasons, but the LLR contribution of individual pixels can be much higher or lower.  First and foremost, we note that it is impossible for the human eye to observe any difference between SDSS galaxies and simulated ones. Admittedly, it is also not obvious to identify clear patterns in the behaviour of the pixel-wise contribution to the LLR. At a closer look, however, it can be seen that the central regions of SDSS galaxies contribute much more to the LLR compared to TNG50. This means that the simulated galaxies are most inaccurate in the central parts, where instead the differences with a smooth S\'ersic model are more pronounced for SDSS. Indeed, for some simulated galaxies the LLR map is almost featureless,  a sign that there is not much difference between the simulated galaxy and a Sersic model (e.g. the two central galaxies in the bottom row of the top right panel in Figure \ref{fig:LLR_interpret}), despite the fact that the galaxies themselves (the two central galaxies in the bottom row in the top left panel of Figure \ref{fig:LLR_interpret}) look reasonably realistic. We also note that the deviation from a S\'ersic profile may occur at different levels in different parts of a galaxy with a non-trivial spatial distribution. This is because, while the galaxy light distribution may not display any interesting feature at a visual inspection, the interplay between the likelihoods of the two networks will determine the complex behaviour observed in the LLR maps.

Given the behaviour of the LLR, it is entirely possible that the light profiles of simulated galaxies differ substantially from SDSS. This may be because the resolution elements are still too coarse to properly capture the inner regions of the light distribution, as discussed in Sections \ref{sec:convergence} and \ref{sec:shortcomings}.

\section{Related work, caveats and discussion}
\label{sec:discussion}
In this study we used deep generative neural networks to perform a quantification of the extent to which the morphologies of galaxies produced in simulations of galaxy formation agree with observations. We compare our framework with other works, in which either more classical techniques or other deep learning methods were used, bearing in mind that a full assessment of their relative performance is out of the scope of this paper. In Section \ref{sec:caveats} we also discuss a caveat that these works share with the present paper.

\subsection{Non-parametric morphologies}
\label{sec:nonparam_discussion}
One way to study the details of galaxy morphology that go beyond the simple S\'ersic index is to use the model-independent \emph{non-parametric} morphologies \citep{Conselice03_CAS,Lotz+04_G_M20}. These provide a quite flexible framework based on the way light is distributed across the galaxy and have been used, amongst other applications, in automated classification tasks \citep{Huertas-Company+11} and merger identification \citep{lotz+11}. The use of these moment-based approaches has been proposed in some studies to attempt a meaningful comparison between the morphology of real and simulated galaxies \citep{Snyder+15_morpho_SF,Rodriguez-Gomez+19, Bignone+19}. \citet{Snyder+15_morpho_SF} found a good agreement between the non-parametric morphologies of Illustris galaxies and SDSS observations, also across scaling relations. That being said, in \citet{Rodriguez-Gomez+19} it was also shown that in fact TNG100 much better reproduces observed PanSTARRS morphologies compared to the original Illustris implementation. This is also the case for the EAGLE simulation, as shown with similar techniques in \citet{Bignone+19}. Although a direct, quantitative comparison with non-parametric approaches is not possible, our deep learning-based analysis qualitatively agrees with the findings of \citet{Rodriguez-Gomez+19}, as shown in Figure \ref{fig:LLR}. We further proved that the improved resolution provided by TNG50 is key to reproducing star forming galaxies, while quiescent galaxies appear to be the most dissimilar from SDSS for both IllustrisTNG realization. The lack of small, quiescent, bulge-dominated galaxies of Illustris was identified in \citet{Snyder+15_morpho_SF} and \citet{Bottrell+17_bulgeIllustris}, but the dependence of galaxy morphology on star formation activity for TNG100 is something that was not addressed explicitly in \citet{Rodriguez-Gomez+19}. Nevertheless, \citet{Rodriguez-Gomez+19} argued that the correlations between galaxy morphology, size and color in TNG100 is in tension with PanSTARRS observations, which qualitatively agrees with our results.

\subsection{Other deep learning frameworks}
\label{sec:other_deep_L}
In  \citet{Huertas-Company+19} a CNN was trained on images from \citet{Nair+10} \footnote{Where galaxies were assigned labels in the form of TType by means of eyeball classification by the authors.} to perform a supervised classification of galaxy morphology and it was then applied to both SDSS and the IllustrisTNG simulation. \citet{Huertas-Company+19} found a remarkable agreement between the morphological scaling relations of observed and simulated galaxies. However the fully supervised approach taken in \citet{Huertas-Company+19}  works under the non-trivial assumption that the training (SDSS) and the test (IllustrisTNG) data come from the same underlying distribution. This is a \emph{critical assumption}, since it is not known a priori whether simulations agree with observations. In fact, a test image will always be assigned a predicted class by the CNN, sometimes with high confidence, even though it looks nothing like any of the images in the training set. In \citet{Huertas-Company+19} this issue was addressed by using Monte Carlo dropout, which is equivalent to Bayesian Neural Networks \citep{Gal2016Dropout}. Monte Carlo dropout consists in making repeated label predictions for any given image, each time randomly setting to zero a number of weights in the CNN. This technique allowed to select objects for which the network finds a high variance in the output label, that is to identify galaxies in IllustrisTNG which do not look realistic. Interestingly, it was found that for compact TNG100 galaxies, the prediction uncertainty was the highest, something which qualitatively agrees with our finding that those galaxies are not well reproduced in simulations.

\vspace{1em}

More recently, other unsupervised approaches based on generative models, like ours, have been proposed to compare simulations and observations. In \citet{Margalef-Bentabol+20} a Generative Adversarial Network (GAN,\citealt{Goodfellow+14}) was used for the first time with the aim of comparing CANDELS high-redshift observations \citep{Koekemoer+11,Grogin+11} with galaxies produced by the Horizon-AGN simulation \citep{Dubois+14_Horizon}. As done in this work, \citet{Margalef-Bentabol+20} treated the problem as an OoD detection task. However, while we adopt a generative model with an explicit likelihood for this purpose, in the case of a GAN the likelihood is not explicit. Therefore, \citet{Margalef-Bentabol+20} resorted to the \emph{anomaly score}, a single-valued metric that measures how well a trained GAN can reproduce a test image. Objects with a higher anomaly score are considered outliers. Moreover, a difference in the \emph{distribution} of anomaly scores of a test set compared to that of the training sample is interpreted as a sign that the two populations differ as a whole. Using an anomaly score-based comparison between CANDELS observations and the Horizon-AGN simulation, Margalef-Bentabol et al. concluded that the two populations differ statistically. Again, they also report the highest anomaly score for spheroidal, small, high-S\'ersic index galaxies. This is in agreement with our results at low redshift.

\subsection{A note on synthetic images}
\label{sec:caveats}
\begin{figure}
    \centering
    \includegraphics[width=0.5\textwidth]{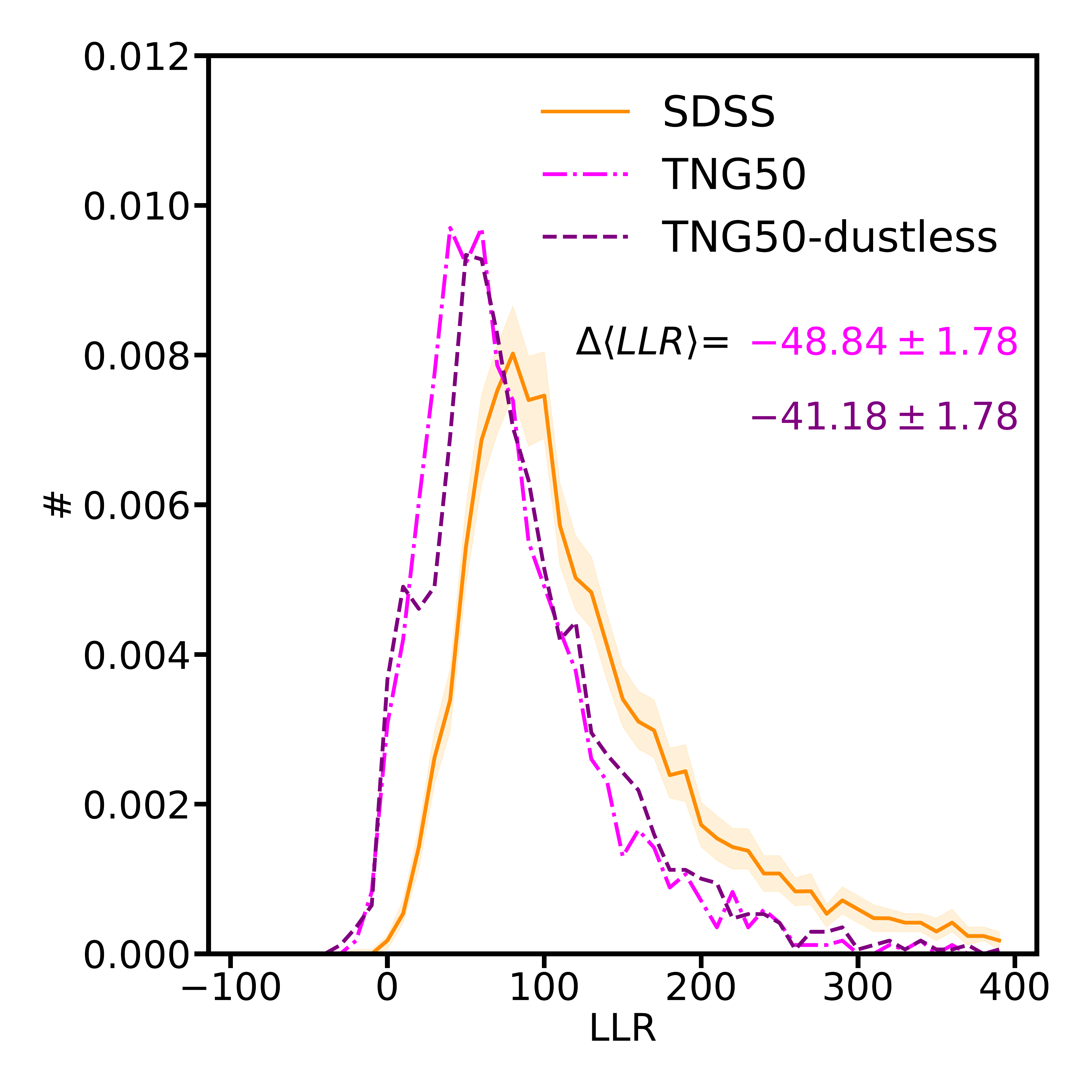}
    \caption{The LLR distribution of TNG50 for the case of dust-inclusive radiative transfer, which we used throughout this work (magenta dot-dashed line), and the case where dust was not modelled (i.e., only stellar light contributes to galaxy morphology, with no dust absorption or emission, purple dashed line). Not including dust results in a better performance for the simulation. This Figure highlights the challenge faced by dust radiative transfer models. The main results presented in this paper remain valid as dust was  included in all the simulations in the same way.} 
    \label{fig:dust}
\end{figure}
The generation of galaxy images from simulations comes with a number of crucial assumptions that may significantly affect the comparison with observations. For example, the fluxes measured from synthetic images strongly depend on
the  assumed  stellar  initial mass function (IMF),  the assumed stellar population synthesis model and the adopted  model  for  dust  effects, such as obscuration and scattering. Different implementations can potentially generate substantial variance in the resulting galaxy morphology. All the simulations that we use in this work have been processed identically, and therefore any uncertainty in the image generation process is propagated in the same way across simulations. Moreover, we stress once again that the mock images of observed galaxies have been convolved with real SDSS PSF and feature a realistic sky background that includes interlopers and the known sources of noise. Therefore we believe that any difference between real and mock observations stems from the galaxy in the center of the cutouts.

A major uncertainty comes from the fact that dust is not explicitly traced in the simulations used here (see \citealt{McKinnon+18_cosmoSim_Dust} for a simulation where this is done), and hence important assumptions must be made for dust production in star forming regions and in the interstellar medium , as shown in detail in \citet{Trayford+17}. The uncertainty in dust modelling results in different dust geometries, and hence varied obscuration patterns \citep{Rodriguez-Gomez+19}. Dust is ubiquitous in star forming galaxies (e.g. \citealt{Galliano+18_review_dust}). The discrepancy that we find between simulated and real star forming galaxies, as quantified by the LLR in Figure \ref{fig:LLR_SFQ}, may be partially explained by the way dust is modelled in the SKIRT pipeline (see Section \ref{sect:simulations} and \citealt{Rodriguez-Gomez+19}). However, since the simulations are processed in exactly the same way, the relative trends seen (i.e.
IllustrisTNG is overall better than Illustris and that a higher resolution improves performance for star forming galaxies) are robust.
Yet, it is entirely possible that the performance of simulations
is underestimated in this instance, since we expect simulations to reach a better (worse) agreement with observations, i.e. a higher (lower) $\Delta \langle LLR \rangle$, with an optimal (non optimal) treatment of dust. We tested this directly by using mock-observed galaxies from TNG50 where dust radiative transfer was not included. The higher $\Delta \langle LLR \rangle$ achieved in the dust-less case (see Figure \ref{fig:dust}) supports the idea that dust modelling is a non trivial task, and that it can lead to worse agreement with the small-scale light distribution of observed galaxies. 

As for passive galaxies, their dust content is a topic widely discussed the literature (e.g., \citealt{Goudfrooij+94_dust_ellipticals,Temi+04_dust_ETGs, Smith+12_dust_ETGs_Herschel, Yildiz+20_dust_ETGs} amongst many others). In this work the full dust-inclusive radiative transfer is run only if the fraction of star forming gas exceeds 1\% of the total baryonic mass. Therefore, the very low star forming gas content of our passive simulated galaxies implies, at given gas metallicity, that they are essentially dust-free in our model, which may affect the comparison to SDSS observations. While we have not explicitly tested the impact of such small amounts of dust on the detailed structural morphologies of simulated quiescent galaxies in terms of the LLR, we find no discernible differences in the population average stellar size, Gini coefficient, Asymmetry, $M_{20}$ and S\'ersic index of very gas-poor galaxies with and without explicit treatment of dust in SKIRT. Hence, we speculate that the fact that the morphology of quiescent galaxies does not seem to compare well to that observed for SDSS is unlikely to be related to the dust modelling in simulated passive galaxies. It would be interesting to test our framework directly on other simulations where dust is explicitly created and destroyed by detailed physical mechanisms (e.g., SIMBA, \citealt{Dave+19_SIMBA,Li+19_SIMBA_Dust}), and no a-posteriori modelling of dust is required.

We conclude this Section with one last caveat. Given the relatively low amount of star forming gas in the simulated passive galaxies, most objects in this population are modelled using simple stellar populations evolving on the 'Padova 1994' evolutionary tracks and a \citet{Chabrier2003} Initial Mass Function (IMF, see \citealt{Rodriguez-Gomez+19} for more details). However, several observational studies have also reported IMF gradients in passive galaxies (e.g., \citealt{LaBarbera+16_IMF,Conroy+17_IMF,Dominguez-Sanchez+18_IMF_ellipt} only to name a few), which are not modelled here. Since  all stars are formed according to a \citet{Chabrier2003} IMF in our simulations \citep{Vogelsberger+13, Pillepich+18_TNGdescription}, we are unable to quantify how the assumption of a universal IMF affects our results.

\begin{figure*}
    \centering
    \includegraphics[width=0.95 \textwidth]{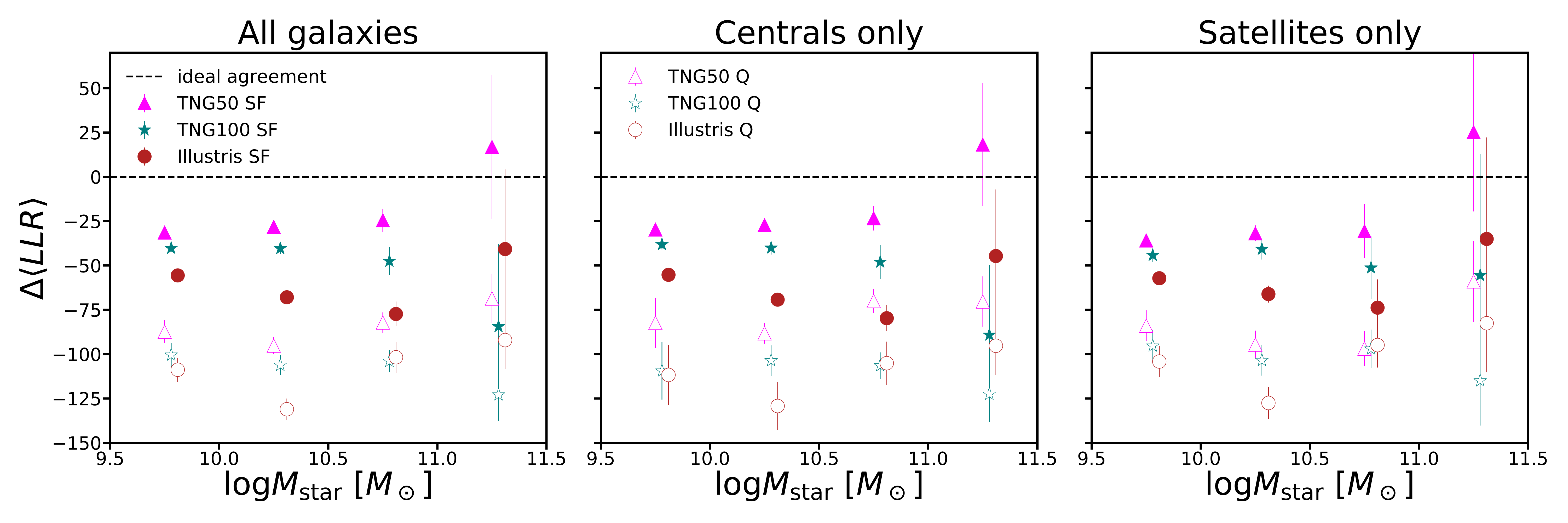}
    \caption{The $\Delta \langle LLR\rangle$ as a function of galaxy stellar mass for TNG50 (magenta triangles), TNG100 (teal stars) and Illustris (red dots).  Star forming and quenched galaxies are shown with filled and empty markers respectively. The left panel shows our results for all galaxies, while the central and right panels are for central galaxies and satellite galaxies respectively. The data points for different simulations are offset for clarity. The error bars represent the 1$\sigma$ uncertainty of 100 bootstrapped realizations of the datasets, of the size of TNG50. See Section \ref{sec:summary} for more details.}
    \label{fig:summary_fig}
\end{figure*}
\subsection{Summary of mass, star formation activity and environmental dependence}
\label{sec:summary}

In this paper we have exploited the $\Delta \langle LLR \rangle$ to quantify the agreement between the detailed light structure of galaxies from SDSS and the TNG50, TNG100 and Illustris simulations. In particular,  we have presented how the $\Delta \langle LLR \rangle$ depends on galaxy stellar mass, star formation activity and environment in Section \ref{sec:quiescent_not_reproduced_LLR}. A comprehensive view of the trends found therein is displayed in Figure \ref{fig:summary_fig}, and is briefly summarised below:
\begin{itemize}
    \item The morphology of star forming galaxies (solid markers) is always better reproduced by simulations compared to quiescent galaxies (empty markers), irrespective of the environment or the stellar mass bin considered;
    \item  At fixed stellar mass and star formation activity, TNG50 provides the highest level of agreement between the small-scale morphological details of simulated and observed galaxies, while TNG100 achieves the second-best $\Delta \langle LLR \rangle$ scores. Illustris features the lowest $\Delta \langle LLR \rangle$, as sign that the disagreement with SDSS is the strongest for this simulation. In the highest stellar mass bin the trend for Illustris and TNG100 are reversed, something that may be due to a combination of different implementations of AGN feedback and the effects of major mergers, as discussed in Section \ref{sec:LLR_mass_SFQ}.
    \item For any given simulation, at fixed star formation activity, it is hard to identify clear trends in the relationship between $\Delta \langle LLR \rangle$ and stellar mass. Perhaps the only significant trend is that, irrespective of a galaxy being central or satellite, for TNG100 and Illustris star forming galaxies the $\Delta \langle LLR \rangle$ declines steadily from $M_{\rm star} \sim 10^{9.5} M_\odot$  to $M_{\rm star} \sim 10^{11} M_\odot$ while in TNG50 the trend is stable. This finding is actually quite puzzling: it would be expected that better sampled galaxies (i.e. higher mass galaxies with larger particle numbers) should be in better agreement with SDSS than lower-mass galaxies. While this is true for TNG50, it is exactly the opposite for TNG100 and Illustris. A possible explanation for this peculiar behaviour is that observed higher mass galaxies may display comparatively more subtle features than low mass galaxies: the number of particles per galaxy at the resolution of TNG100 and Illustris may still not be enough to properly capture them well, as opposed to the higher resolution of TNG50.
    \item Figure \ref{fig:summary_fig} also remarks the little difference in the $\Delta \langle LLR \rangle$ of central and satellite galaxies. While in Section \ref{sec:environment} and Figure \ref{fig:LLR_SFQ_censat} we have shown this for galaxies of all stellar masses, here we further observe that the broad independence on environment applies to all mass scales.
\end{itemize}
We also note that TNG50 and Illustris star forming galaxies seem to have a $\Delta \langle LLR \rangle >0$ at the highest masses, which seems counter-intuitive given that we expect the LLR to be the highest for SDSS (see Section \ref{sec:strategy}). This may be because there are very few SF galaxies in SDSS with stellar mass above $10^{11} M_\odot$. Indeed, the large bootstrapped resampling variance at these masses for star forming galaxies is indicative of a poorly represented and potentially biased population. uture work will be dedicated to bypass this specific limitation of our framework.

\subsection{Convergence study}
\label{sec:convergence}
\begin{figure}
    \centering
    \includegraphics[width=0.5\textwidth]{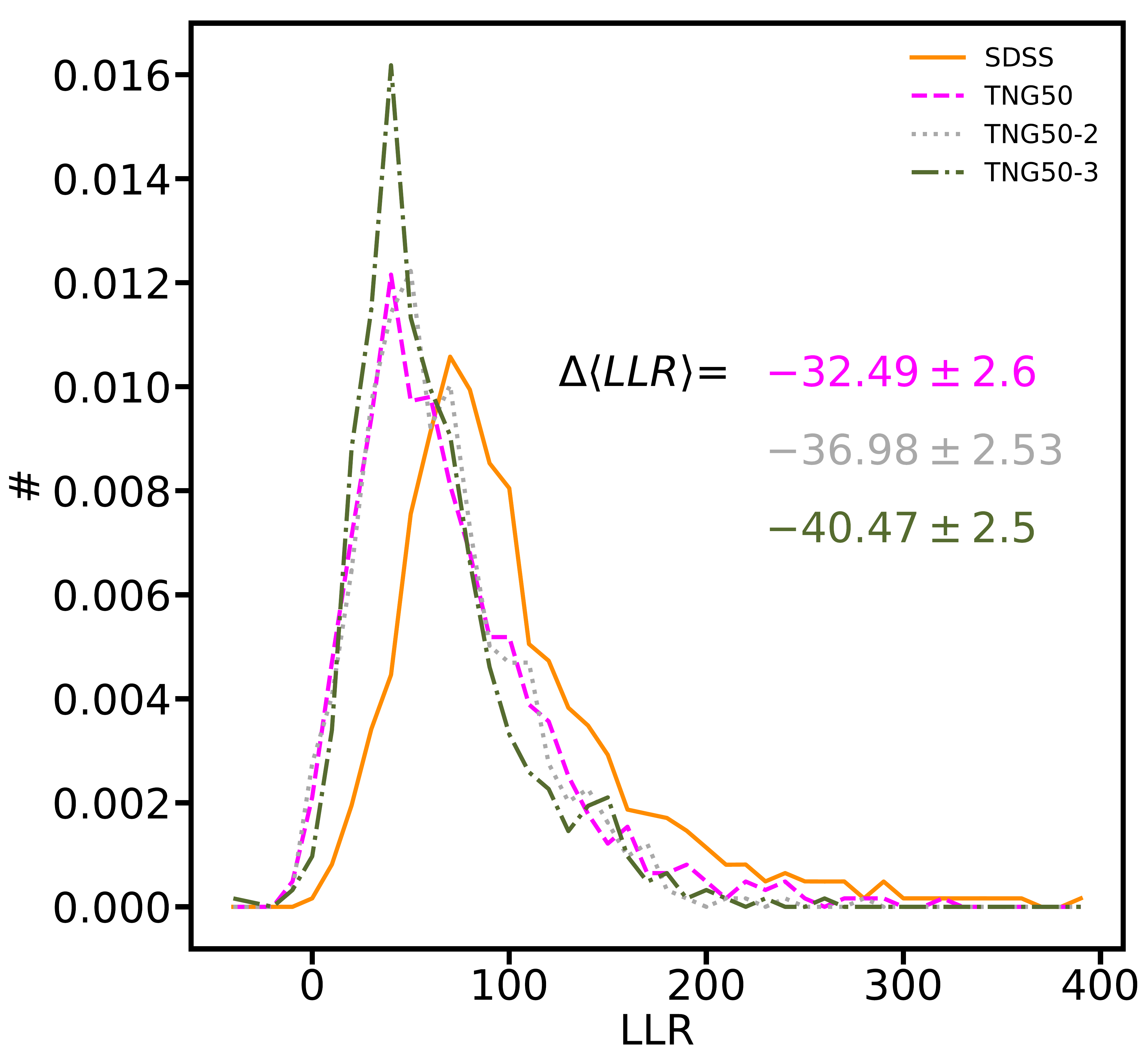}
    \caption{ The LLR distributions of SDSS (orange solid line), TNG50 (dashed magenta line), TNG50-2 (dotted gray line) and TNG50-3 (dot-sashed green line). The $\Delta  \langle LLR \rangle$ increases with improved resolutions, a sign that simulations are converging. Future higher-resolution simulations are likely to be in even better agreement with SDSS. Note that the value of the $\Delta  \langle LLR \rangle$ for the highest resolution run of TNG50 is not comparable to those for TNG50 found in the paper, as we are only considering subsets of the TNG50 simulations and SDSS to match the joint magnitude-S\'ersic index-$R_e$ distribution.}
    \label{fig:LLR_matched_TNG50s}
\end{figure}

As the TNG simulations were run at different resolutions, we are able to perform a direct convergence test. Specifically, we produced mock-observations of two lower-resolution runs of TNG50, TNG50-2 (medium resolution) and TNG50-3 (low resolution, see \citealt{Pillepich+19_TNG50}), as described in Section \ref{sec:data} and in detail in \citet{Rodriguez-Gomez+19}. 
Resolution is known to generate non-trivial changes in the physical properties of simulated galaxies (see for example also Appendix B1 of \citealt{Pillepich+19_TNG50}, and, e.g., \citealt{Sparre+16, Chabanier+20_compactsExtremeHorizonSim}). This is something that we wish to marginalise on, since our aim is to test to what extent an improved resolution brings the small-scale morphology of simulated galaxies into better agreement with that of real galaxies, regardless of the overall structure. Therefore, we match the three TNG50 simulations and SDSS to obtain an identical joint distribution of their global properties, i.e. size, magnitude and S\'ersic index, which are the key observables learned by the $p_{\theta_{sersic}}$ model. This allows us to isolate the effect of resolution on the relationship between the global and local properties of galaxies, as quantified by the LLR.

The $\Delta \langle LLR \rangle$ (see Figure \ref{fig:LLR_matched_TNG50s}) is the highest for the highest resolution run, TNG50, which is followed by TNG50-2, and TNG50-3, the run with the lowest resolution. This result shows that the small-scale morphology of simulated galaxies is converging for progressively improved resolutions, and it is likely that a further improvement in resolution would result in an even better agreement with SDSS. We stress that with our methodology we are able to quantify with just one number, for the first time, the effects of resolution on the detailed morphology of simulated galaxies. 

Note that both the spatial and mass resolution decrease in TNG50-2 and TNG50-3 (see \citealt{Pillepich+19_TNG50} for details), and therefore we are not able to disentangle the contributions of the two here. We will speculate on this matter in the next Section.
 
\subsection{Possible shortcomings of the numerical simulations}
\label{sec:shortcomings}

\begin{figure*}
    \centering
    \includegraphics[width=0.9\textwidth]{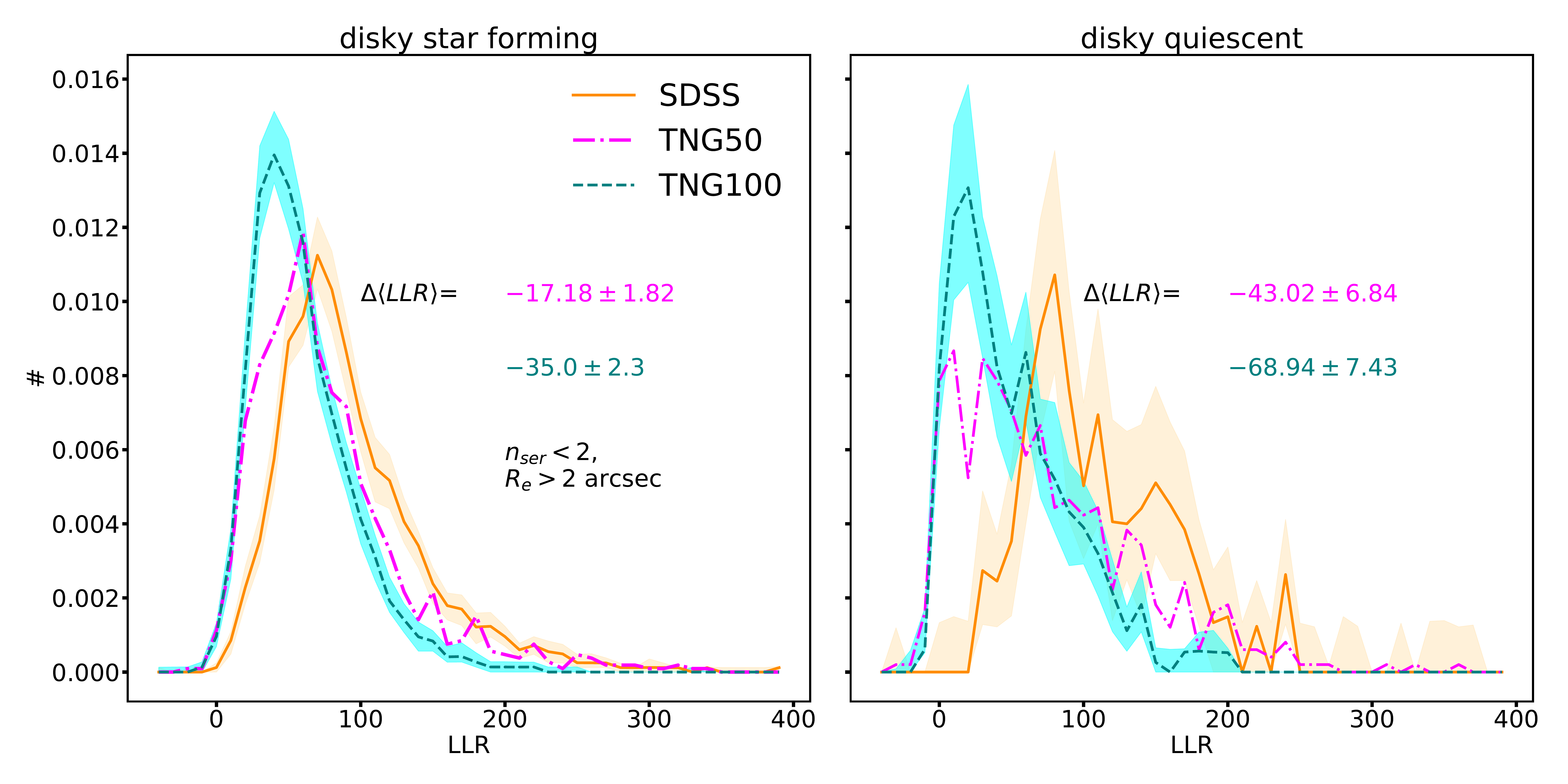}
    \caption{The LLR distributions of star forming (left) and quiescent (right) disks for TNG50 (magenta dot-dashed lines), TNG100 (teal dashed lines) and SDSS (orange solid lines). The cyan and light orange coloured regions indicate the 1$\sigma$ uncertainty of 100 realizations of TNG100 and SDSS with the same sample size of TNG50. Disky galaxies are selected in SDSS and in simulations using the thresholds $n_{ser}<2$ and $R_e>2$arcsec$\approx 2$kpc at z=0.05. The lower $\Delta \langle LLr \rangle$ featured by quiescent disky galaxies is indicative that the processes that lead to quiescence without affecting the stellar morphology (and hence dynamics) still produce a worse agreement with data compared to star forming disks.}
    \label{fig:disks_SFQ}
\end{figure*}
\label{sec:simulations_shortcomings}

In this study we have quantitatively assessed the agreement between SDSS and simulations of galaxy formation on the relationship between the higher-level details and the global morphology. In particular, although the IllustrisTNG simulations agree extremely well with the SDSS structural scaling relations (see also \citealt{Huertas-Company+19} and \citealt{Genel+18}), our findings show that the IllustrisTNG model cannot yet reproduce the detailed distribution of stellar light in comparison to SDSS, particularly for quenched galaxies and regardless of whether quenching is the result of environmental processes like ram-pressure stripping or BH feedback.
Earlier in this Section we have also discussed the results of other deep learning studies that reached similar conclusions for TNG100 and the Horizon-AGN simulation using supervised Bayesian Neural Networks or the anomaly scores of a GAN.  The more classical approach used in \citet{Rodriguez-Gomez+19} also highlights similar tensions in TNG100. Therefore, there are now multiple independent indications that the detailed morphology of quiescent galaxies in cosmological hydrodynamical simulations of galaxy formation is in tension with that of galaxies in our Universe. We speculate below on the possible reasons of this discrepancy, by focusing on the case of the IllustrisTNG simulations.

\subsubsection{The difficulty of reproducing highly-concentrated stellar distributions}

Quenched galaxies in TNG100 (and Illustris) systematically fail at populating the region of high S\'ersic index and small size in the $n_{ser}-R_e$ plane. TNG50 can produce compact quiescent galaxies, and yet small-size and high-S\'ersic index quiescent objects exhibit the worst disagreement with SDSS, also in TNG50. To attempt to disentangle the effects of quenching with possible issues related to the global stellar morphology, in Figure~\ref{fig:disks_SFQ} we contrast TNG100 and TNG50 simulated {\it disky} galaxies to SDSS, divided according to their star formation state: star-forming disks on the left, quiescent disks on the right. 

For disky quiescent galaxies TNG100 features $\Delta \langle LLR \rangle \sim -65$, while TNG50 has $\Delta \langle LLR \rangle \sim -45$.  In comparison to the differences between star-forming and quiescent galaxies of Figure~\ref{fig:LLR_SFQ} without any ``diskiness selection''  (i.e. $\Delta \langle LLR \rangle \sim -101$ and $\sim-83$ for TNG100 and TNG50 respectively), we can see that the disagreement between the real and simulated populations of quenched disks is much less dramatic than that featured by the overall population of quiescent galaxies, which is dominated by smaller spheroids (\citealt{Huertas-Company+19, Joshi+20_disksTNG}). In other words, the TNG simulations return more realistic quenched disk galaxies than quenched galaxies in general: in fact, the TNG model always produces more realistic disk galaxies, whether they are quenched or not. So this suggests that what seems to mostly drive the discrepancy between the TNG and SDSS quiescent populations is not the property of being quenched but rather the fact of being non-disky, with stellar particles mostly in non-rotationally supported orbits.

\subsubsection{The limits of resolution at reproducing high stellar densities}

Since lower-mass galaxies are represented by a lower number of stellar particles in simulations, it could be argued that resolution plays a key role in making quiescent (mostly spheroidal) galaxies look less realistic than the more extended (mostly disky) star forming galaxies.  The fact that quiescent galaxies are better reproduced by TNG50, which offers a higher resolution, seems to support this argument. We also note that, even within the star forming population, smaller galaxies have a lower $\Delta \langle LLR \rangle$ in TNG50 and TNG100.

As highlighted above, differently than TNG100, TNG50 is able to produce compact quiescent galaxies. Moreover, TNG50 features a higher $\Delta \langle LLR \rangle$ for quenched extended galaxies with an intermediate $n_{ser}$ compared to TNG100. This further evidence suggests, again, that an improved resolution is able to better capture the small details of the stellar structure of quenched galaxies. We briefly speculate on the possible physical reason for this.

Quiescent galaxies in the TNG simulations tend to be smaller at fixed stellar mass compared to star forming galaxies, in good agreement with observations (see Figure \ref{fig:sizemass_SFQ}). Furthermore, the quiescent TNG population tends to be dominated by spheroidal galaxies \citep{Huertas-Company+19,Joshi+20_disksTNG}, with the stellar orbits mostly dominated by random motions.
Thus, the finite resolution of the simulations may not reproduce these orbits faithfully.
However, because the levels of (dis)agreement with SDSS do not seem to correlate strongly with a galaxy stellar mass once TNG50 or TNG100 are considered separately (see Figure~\ref{fig:summary_fig}), the issue may be more related to the spatial, rather than the mass, resolution underlying the numerical models we have considered in this work.

Lastly, the fact that the central densities of quenched galaxies appear problematic may be related to an issue that was already identified in the Illustris
simulation by \citet{Sparre+15}, where it was found that the
simulation did not reproduce well the number of starbursting galaxies at the Illustris and TNG100 resolution. If at least some quenched elliptical galaxies formed through gas-rich mergers which drove large amounts of gas into the centers of the merger remnants, the resulting high central densities,  may not be resolved in most of the cosmological simulations studied here. Indeed, the departure from pure S\'ersic profiles in the form of power-law ``cusps" observed in high-resolution imaging (e.g., \citealt{Lauer+95,Faber+97_centralRegionsETGs, Kormendy+99}) has been interpreted of a signature of previous dissipational mergers \citep{Hopkins+09_cusps}, and \citet{Sparre+16} showed that
higher-resolution zoom-in resimulations of selected major mergers
in Illustris are able to produce denser starbursts compared to the
lower resolution Illustris run. \citet{Hopkins+09_cores} have also proposed that the inner stellar ``cores" observed in some elliptical galaxies (e.g., \citealt{Lauer+95}) are the result of dry mergers involving previously formed ``cuspy" ellipticals. If the resolution of TNG50 and TNG100, as well as Illustris, is not able to capture the  formation of ``cusps", as indirectly suggested by \citet{Sparre+16}, then also the formation of stellar ``cores" in these simulations may be unresolved.

\subsubsection{Quenching may affect the small-scale morphology by modifying the underlying gas distribution}
We conclude with a final remark. Figure \ref{fig:sizemass_SFQ} shows that even quenched galaxies with relatively large sizes are not fully reproduced by simulations. In particular, at $M_{\rm star}\lesssim 10^{11}M_\odot$, some of the larger galaxies where star formation has been halted belong to the population of quenched disks (e.g., \citealt{Zhang+19_quenchedDisks}). Furthermore, as shown in Figure \ref{fig:disks_SFQ}, TNG quenched disks are still in worse agreement with SDSS than star-forming disk, the $\Delta \langle LLR \rangle$ of quenched disks being twice as lower than that of star forming disks. This is somewhat unexpected, as the quenching mechanism that operates on them has preserved the bulk of the ordered stellar motions proper of disk galaxies. A possible explanation for this is that the mechanisms that quench disks may displace the distribution of gas within the galaxy, thus affecting the distribution of dust and hence the small-scale light distribution.

\section{Conclusions and future outlook}
\label{sec:conclusions}
Since the time of \citet{Hubble1926}, the astronomical community has strived to understand the physical origin of the variety of morphologies that galaxies display in our Universe. The simulations of galaxy evolution available to date have achieved an unprecedented accuracy in reproducing galaxy properties, and, with them, a plethora of galaxy morphologies. Assessing how \emph{exactly} the small-scale morphological details of simulated galaxies agree with the real ones is a crucial test for models of galaxy formation and evolution. Our contributions to this topic are summarized as follows:
\begin{itemize}
    \item We have introduced an unsupervised deep learning method to accurately and quantitatively compare the small-scale stellar morphology of galaxies produced by cosmological hydrodynamical simulations with that of real galaxies (Section \ref{sec:PixelCNN}). This assessment is based on a single-valued metric which is the combination of the likelihood of two deep generative models, the log-likelihood ratio, LLR (Section \ref{sec:strategy}). We demonstrate that the LLR is broadly independent from the sky background statistically, and specifically is mostly sensitive to internal, small-scale morphological structure. The behaviour of the LLR indeed follows these expectations, as shown in Appendix \ref{app:robustness}. We also prove that the LLR is a metric that can be used to assess the similarity of two datasets based on the mean value of its distribution, and we adopt the $\Delta \langle LLR \rangle \equiv \langle LLR \rangle - \langle LLR_{\rm SDSS} \rangle$ to assess the quality of the small scale light structure of fully realistic mock observations of galaxies from the Illustris, TNG50 and TNG100 simulations against observations from the Sloan Digital Sky Survey. 
    
    \item  In Figure \ref{fig:LLR} we show that our approach can identify TNG50 as the simulation that is able to produce galaxies with small-scale morphological features that most closely resemble observations, followed by TNG100 and the original Illustris implementation, which performs the worst. This can be interpreted as a sign that the improvement in the modelling of galaxy formation physics featured by the more recent IllustrisTNG simulations is more effective than that implemented in Illustris. \citet{Rodriguez-Gomez+19} reached similar conclusions using non-parametric morphologies. Moreover, we find that the improved resolution of TNG50 results in an even better match to SDSS morphologies. 
    
    \item We split our data sets in star forming ($\rm{sSFR/yr}^{-1}>-11$) and quiescent ($\rm{sSFR/yr}^{-1}<-11$) galaxies and show the respective LLR distributions in the upper panel of Figure \ref{fig:LLR_SFQ}. We find a marked improvement in the morphology of star forming galaxies from Illustris to TNG100 and from the latter to TNG50, which indicates that a better treatment of star formation regulation and an improved resolution are key to accurately reproduce the morphology of star forming galaxies. On the other hand, we see only a marginal improvement for quiescent galaxies from Illustris to its successor IllustrisTNG, and we note that the better resolution offered by TNG50 over TNG100 does not lead to a significantly better agreement with SDSS. 
    
     \item We find the trends with star formation activity to be weakly dependent on stellar mass (middle and lower panels of Figure \ref{fig:LLR_SFQ}) and environment (Figure \ref{fig:LLR_SFQ_censat}), so that simulated quenched galaxies are in similar disagreement with SDSS regardless of the nature of the quenching mechanism i.e. regardless of whether quenching is driven by e.g. ram-pressure stripping or AGN feedback. This information is displayed in a more self-contained way in Figure \ref{fig:summary_fig}.
     
    \item We study how well simulated galaxies are reproduced across scaling relations of galaxy size, star formation rate and S\'ersic index in Figures \ref{fig:scalingrelations}, \ref{fig:sizemass_SFQ} and \ref{fig:n_R_SFQ}. We note a significant change in the quality of simulated galaxies, whereby large, star forming, disky objects are the most similar to SDSS, while the smaller, high-S\'ersic index, quenched galaxies are found less realistic by our deep learning framework. We also note that even within the structural scaling relations of star forming and quiescent galaxies some trends are appreciable. More massive, extended galaxies are more realistic in both quenched and star forming TNG50 galaxies, while the same is true of TNG100 star forming galaxies only.
    
    \item Our main finding is that reproducing the \emph{small-scale morphological features} of quiescent, small and/or concentrated galaxies remains a challenge for state-of-the-art hydrodynamical cosmological simulations of galaxy formation. We show that this kind of evidence has started to emerge in the literature in Sections \ref{sec:nonparam_discussion} and \ref{sec:other_deep_L}. We speculate that a limited resolution may be at the origin of these findings. First, we carry out a specific convergence test in Section \ref{sec:convergence}, where we show that the lower-resolution runs TNG50-2 and TNG50-3 perform worse than the flagship, better-resolution TNG50 simulation. Secondly, in Section \ref{sec:simulations_shortcomings} we also argue that the high density of stellar particles in the central regions of quenched galaxies may not be properly captured by the finite resolution of simulations, as also shown by the "LLR maps" in Figure \ref{fig:LLR_interpret}. This argument is also supported by the similar level of (dis)agreement with SDSS observations reached by both Illustris and TNG100 for massive quenched galaxies  (see Fig. \ref{fig:LLR_SFQ}), despite the AGN feedback mechanism implemented in the two simulations is substantially different. In fact, the formation histories of these galaxies are affected by similar rates of major mergers that cause a similar change in the stellar dynamics, since the resolution of the two simulations is comparable.  We also speculate that the displacement of gas, and the consequent dust obscuration patterns, that quenching mechanisms cause within a galaxy, may also partially explain the lesser agreement between simulated and real quenched galaxies.
    \item Finally, we remark that the results listed above have been obtained at the seeing-limited resolution of SDSS, i.e. $\approx 1$kpc, which means that the some of small-scale details of the stellar light structure that characterizes galaxies have been lost. Future work that will exploit higher resolution images may be able to unveil some trends that are  not found in this exploratory study.
    
\end{itemize}

The deep learning framework outlined here provides a useful tool to evaluate the performance of hydrodynamical simulations of galaxy formation, that generalizes over the parametric and non-parametric approaches taken in the past. With our strategy, we can identify meaningful physical information encoded in the galaxy structure, which proves key in identifying the shortcomings and successes of simulations. Our methodology still works only in a statistical sense, given the not completely null contribution of the sky background to the metric that we use (see Appendix \ref{sec:LLR_variance}). However, future work in this direction will make it possible to evaluate the morphology of simulated galaxies at the time of calibrating the next generation of simulations of galaxy formation and evolution.

 Lastly, Out of Distribution detection tasks are of paramount importance in Astronomy since they are able to unearth the potentially most interesting objects in a dataset, and will be even more important when the next observing facilities such as EUCLID and JWST will come online and collect an unprecedented load of data. Our framework may be applied also in this context, similarly to \citet{Margalef-Bentabol+20} and Storey-Fisher et al. in prep.

 \section*{ACKNOWLEDGEMENTS}
We thank the anonymous referee for a constructive report.
L.Z. and C.B. warmly thank the organizers of the Kavli Summer Program in Astrophysics 2019 and UCSC for hosting the school. We thank the Kavli Foundation and the National Science Foundation for their generous support. L.Z. also wishes to thank all the participants for their help and support during the school and, in particular, Mike Walmsley and Dezso Ribli for very helpful comments and suggestions. L.Z. acknowledges funding from the Science and Technology Facilities Council (STFC) through the Data Intensive Science Centre for Doctoral Training (DISCnet). C.B. is grateful for support from the National Sciences and Engineering Council of Canada (NSERC). The computations in this research were enabled in part by the resources provided by Compute Canada (\url{www.computecanada.ca}). F.S. acknowledges partial support from a Leverhulme Trust Research Fellowship. 

\section*{ Data availability}
The codes used to produce the results presented in this work are available at \url{https://github.com/lorenzozanisi/pixelcnn-ood}. Illustris and TNG100 data products are publicly available at \url{https://www.illustris-project.org} and \url{https://www.tng-project.org} respectively. Data for the TNG50 simulation, in particular, are expected to be made publicly available within some months from this publication, at the same IllustrisTNG repository. SDSS images can be downloaded from \url{http://skyserver.sdss.org}. The \citet{Meert+15} SDSS dataset is located at \url{http://alan-meert-website-aws.s3-website-us-east-1.amazonaws.com/fit_catalog/download/index.html}.

\appendix 



\section{Training}
\label{app:training}
\begin{figure}
    \centering
    \includegraphics[width=0.5\textwidth]{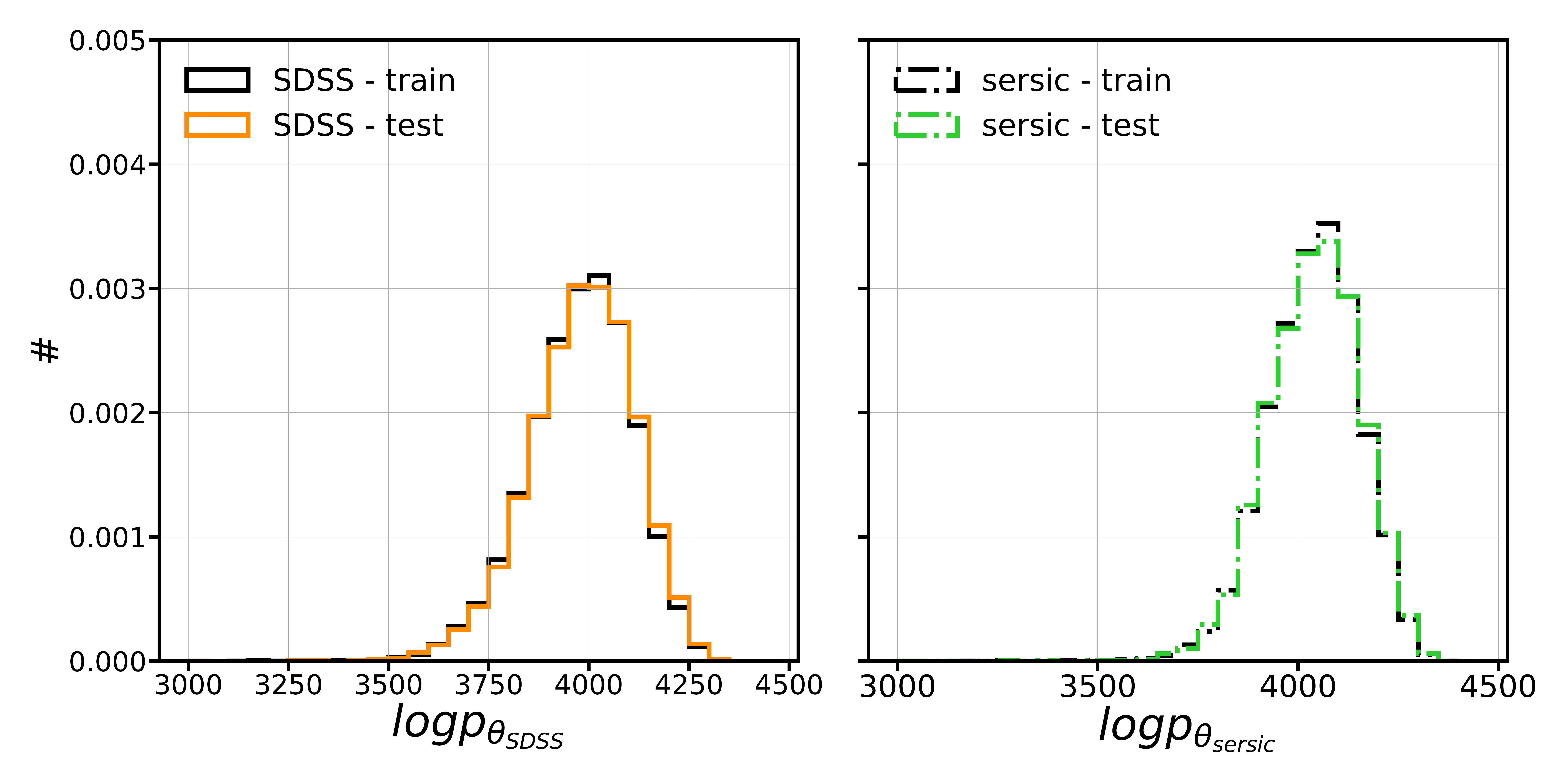}
    \caption{\emph{Left:} The likelihood distribution of the SDSS training set (black thin line) and test set (orange thick line). \emph{Right:} The likelihood distribution of the training (black) and test (green) sets for the best S\'ersic models. The overlap between the distributions shows that the model has converged.}
    \label{fig:traintest}
\end{figure}

The likelihood distributions of training and test sets for both models are shown in Figure \ref{fig:traintest}. The good agreement between the training and test sets is indicative of the convergence of the models.

\begin{figure*}
    \centering
    \includegraphics[width=0.8\textwidth]{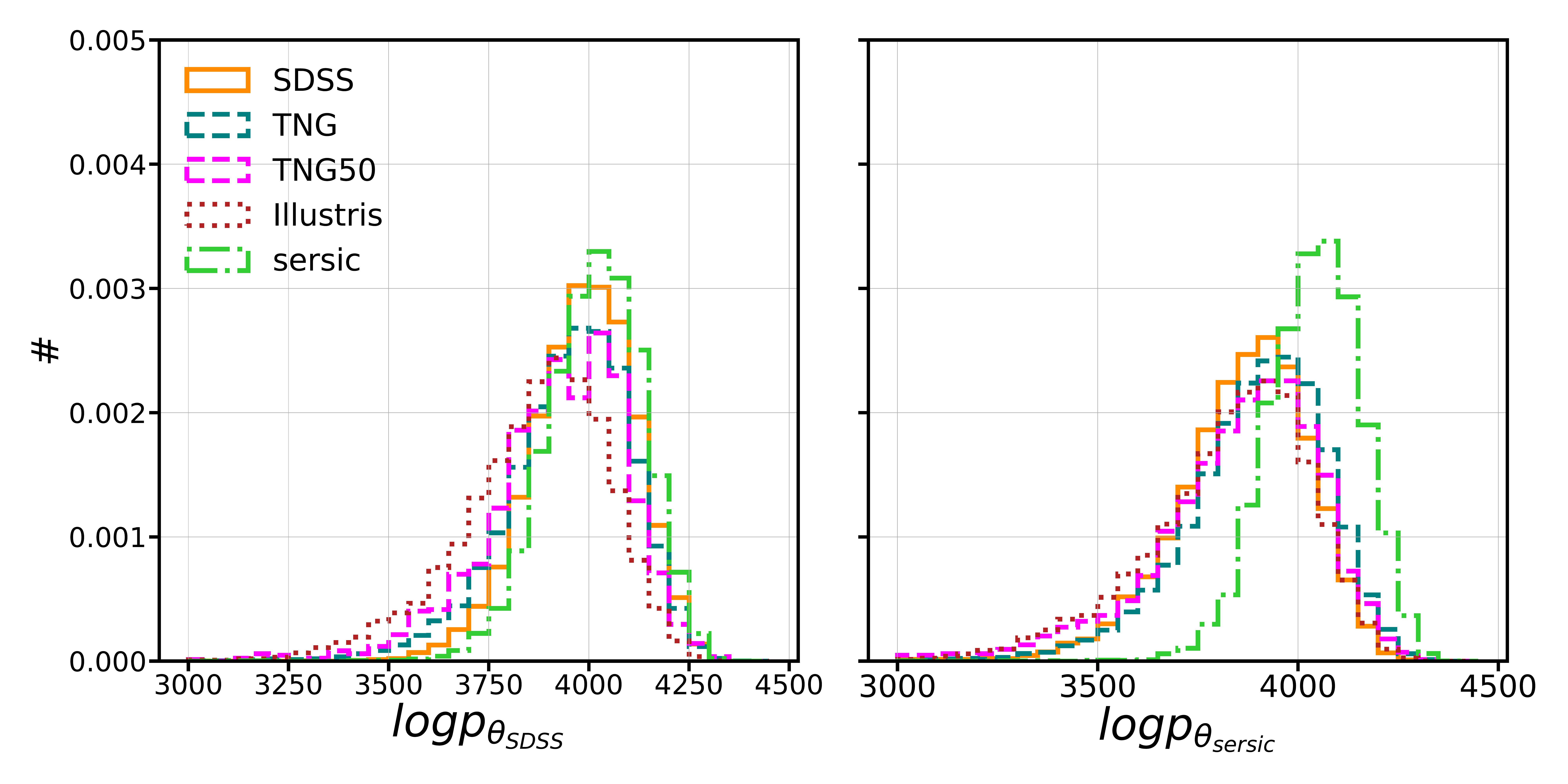}
    \caption{ Likelihood distributions of SDSS (solid orange lines), TNG100 (teal dashed lines), TNG50 (magenta lines), Illustris (dotted red lines) and the best S\'ersic fits (dot-dashed green lines) according to the $p_{\theta_{\rm SDSS}}$ (\emph{left}) and the $p_{\theta_{\rm sersic}}$ (\emph{right}) models. }
    \label{fig:likelihoods_distribs}
\end{figure*}

\begin{figure*}
    \centering
    \subfloat[]{\label{fig:low_likelihood}\includegraphics[height=0.3\textwidth]{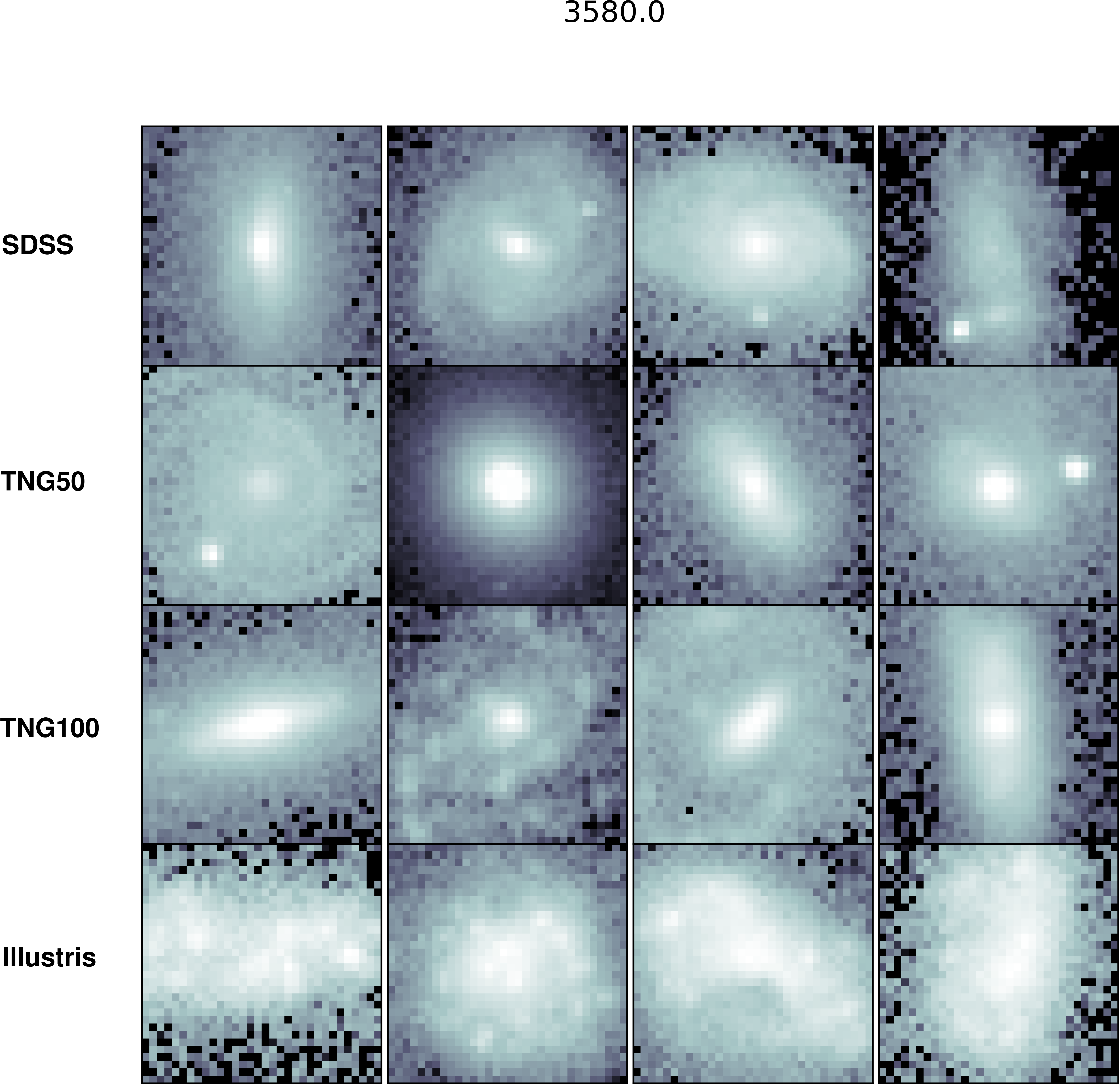} }
    \subfloat[]{\label{fig:med_likelihood}\includegraphics[height=0.3\textwidth]{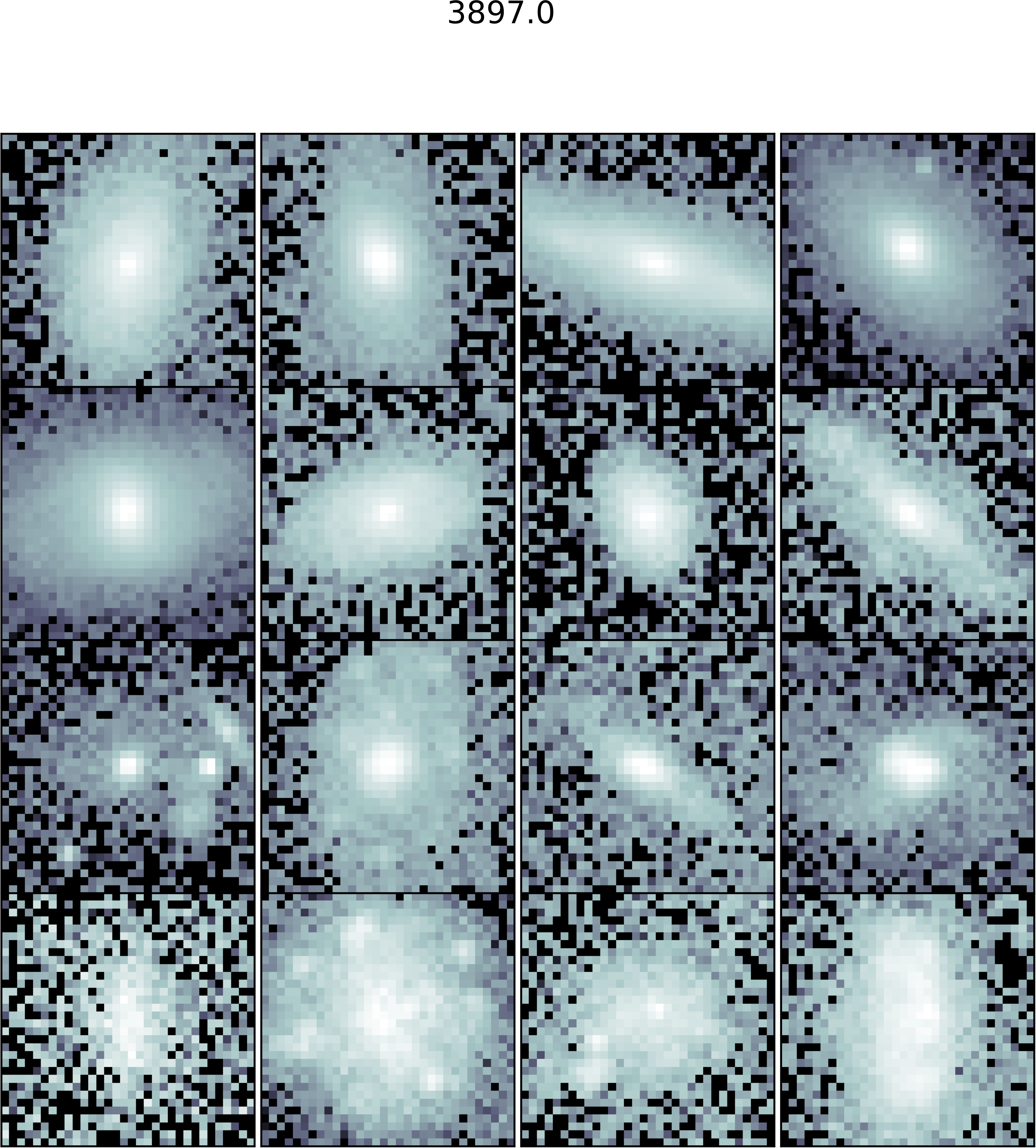} }
     \subfloat[]{\label{fig:high_likelihood}\includegraphics[height=0.3\textwidth]{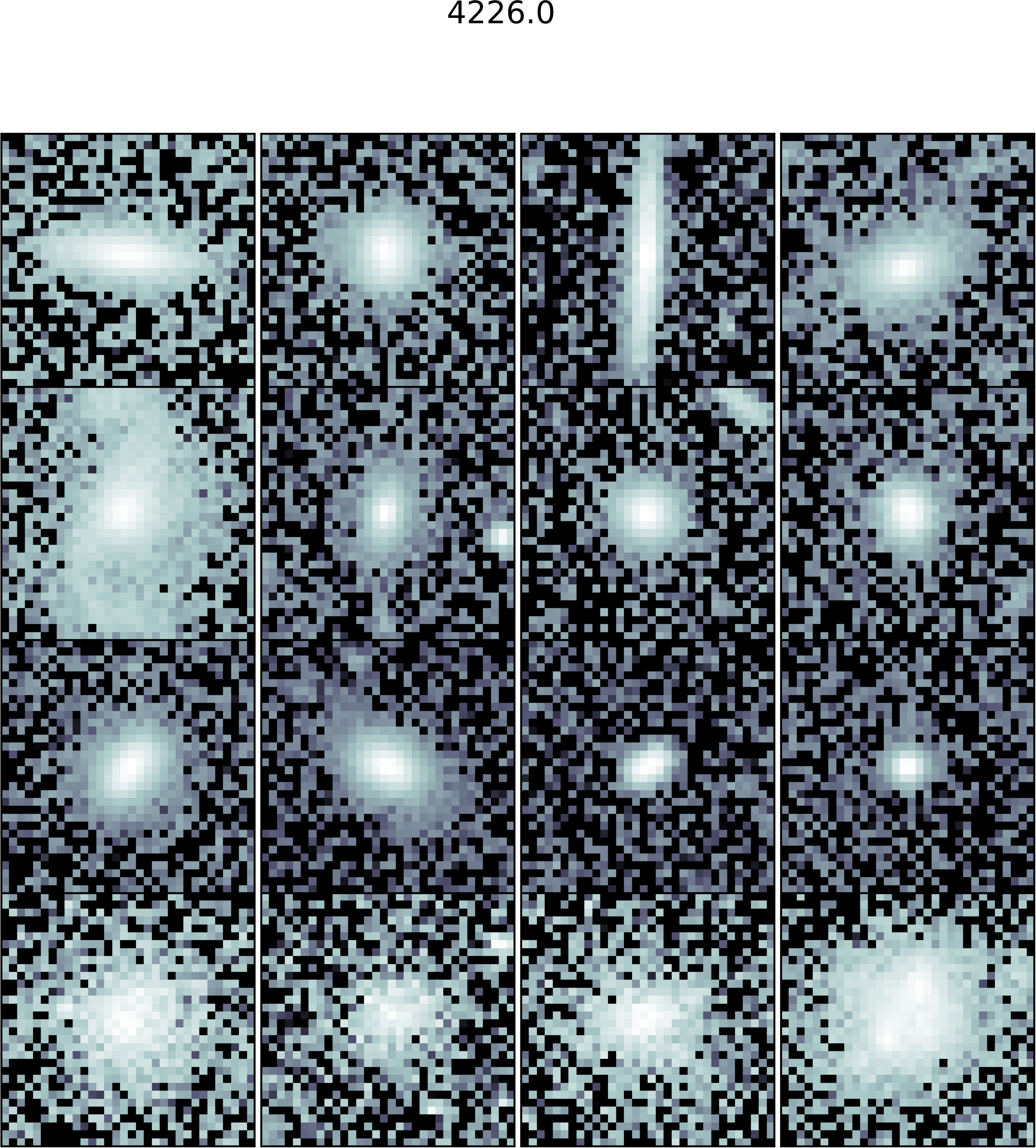} }
    \caption{Typical galaxies at low (left panel) medium (central panel) and high (right panel) likelihood. The values of the likelihood are reported in the title of each panel. The first row of each panel shows SDSS galaxies, the second TNG50 galaxies, and the third and fourth TNG100 and Illustris galaxies. It can be seen that images with a lower likelihood tend to be those of more complex, larger galaxies, while smaller galaxies have the highest contribution to the likelihood from the sky background.}
    \label{fig:likelihoods_summary}
\end{figure*}

\begin{figure*}
   \centering
    \subfloat[(a)]{{\includegraphics[height=0.15\textwidth]{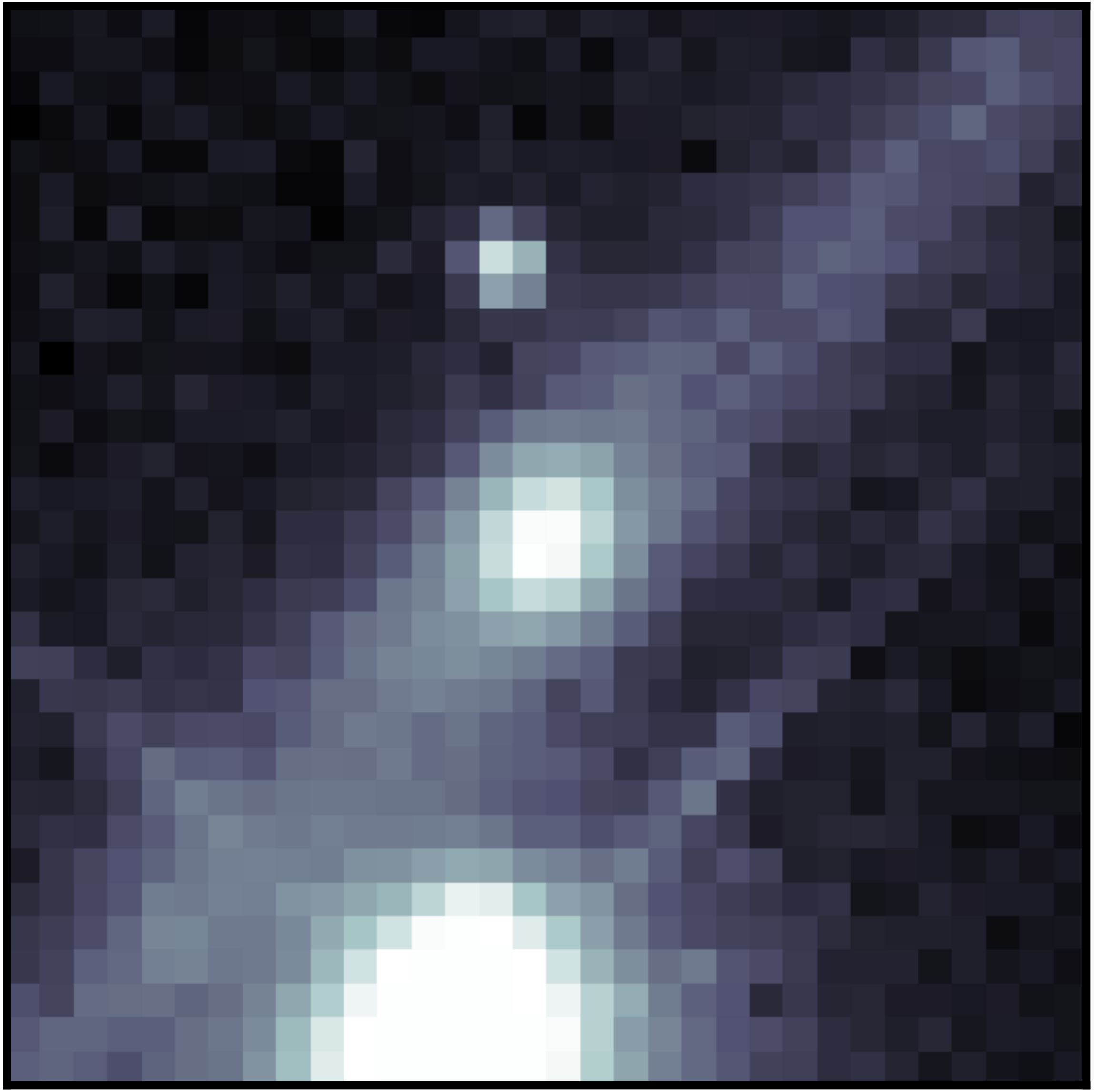} }} \hspace{0.2cm}
    \subfloat[(b)]{{\includegraphics[height=0.15\textwidth]{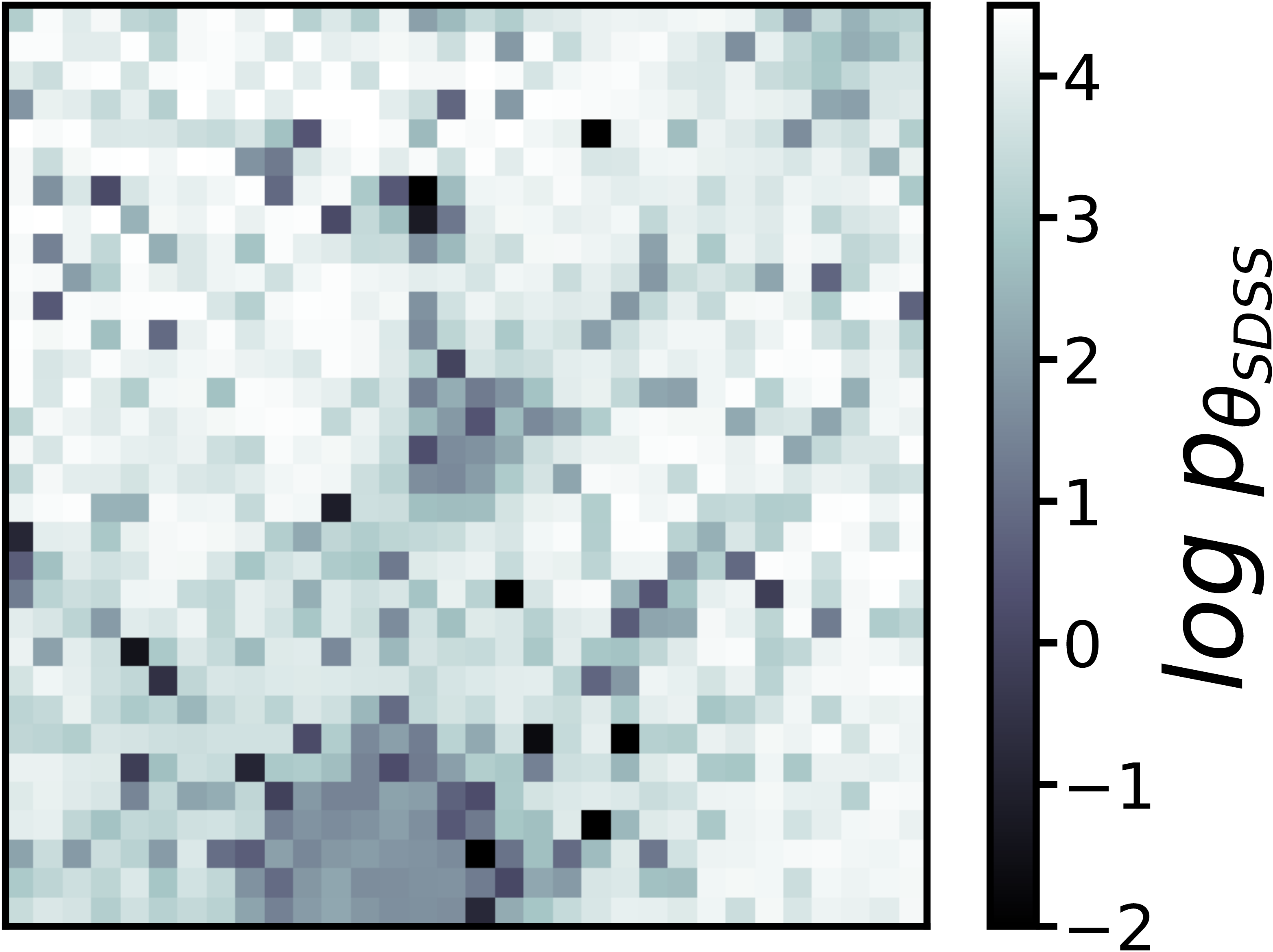} }}\hspace{0.07cm}
    \subfloat[(c)]{{\includegraphics[height=0.15\textwidth]{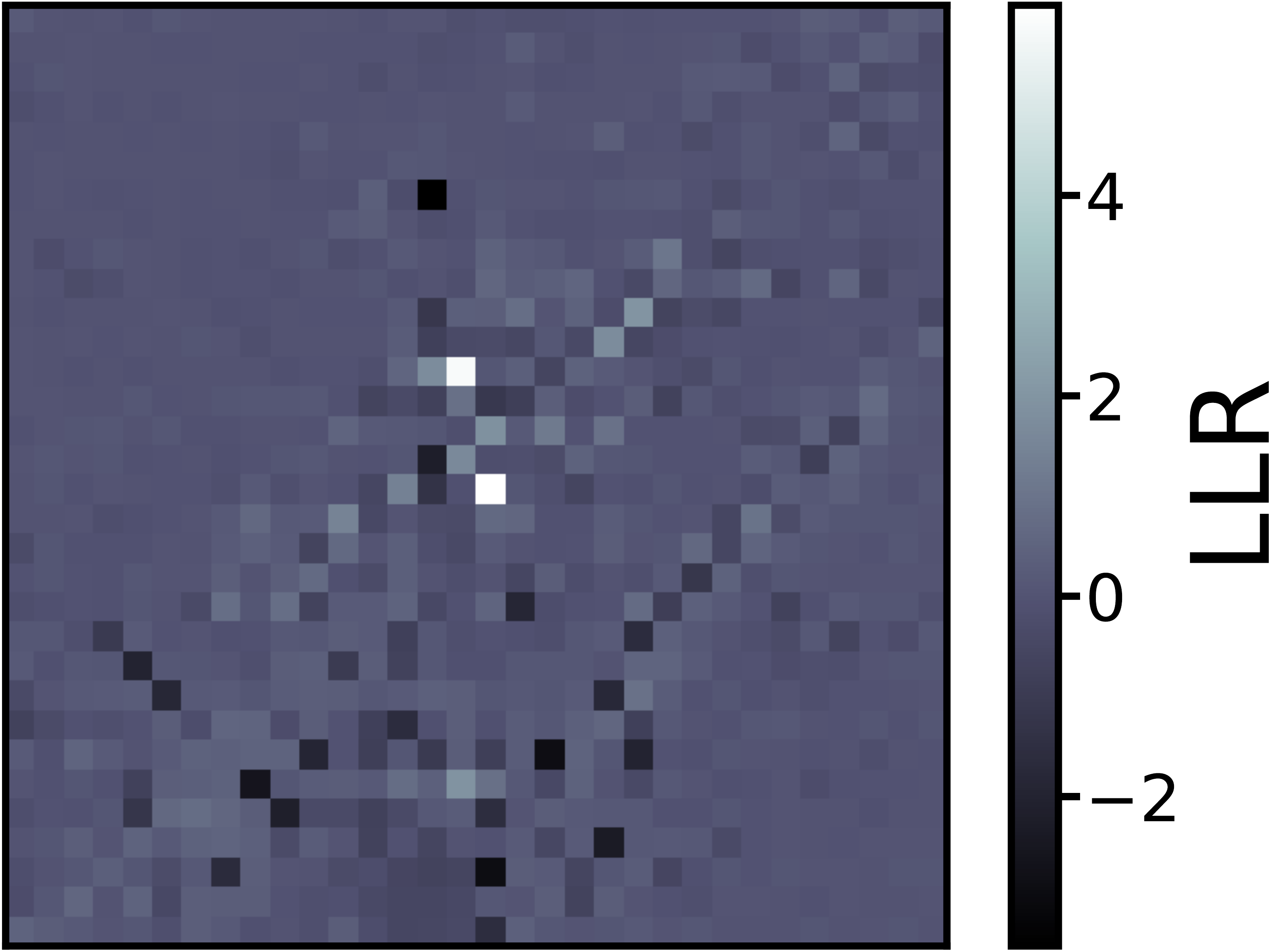} }}\\
    \subfloat[(d)]{{\includegraphics[height=0.15\textwidth]{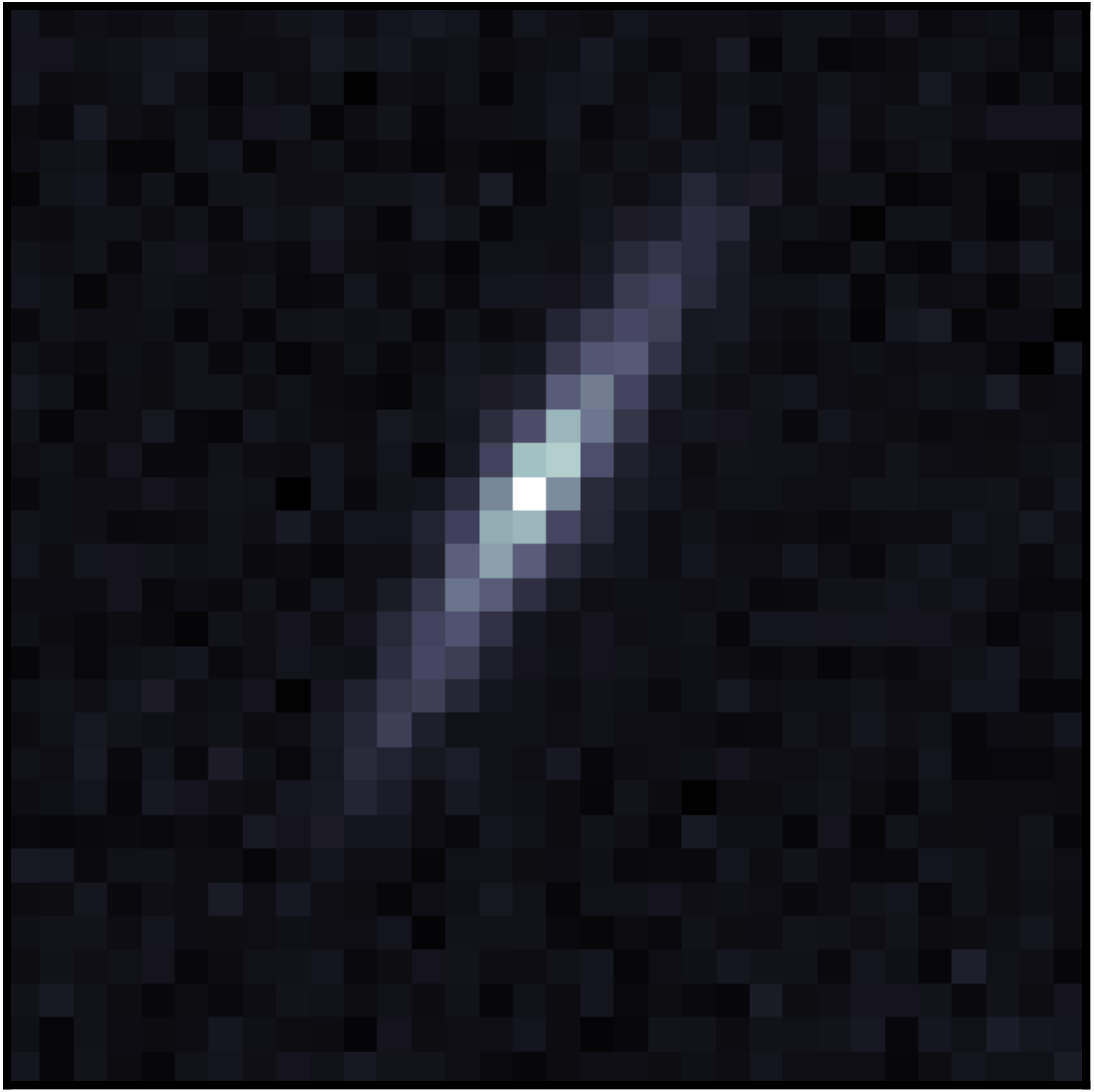} }}\hspace{0.2cm}
    \subfloat[(e)]{{\includegraphics[height=0.15\textwidth]{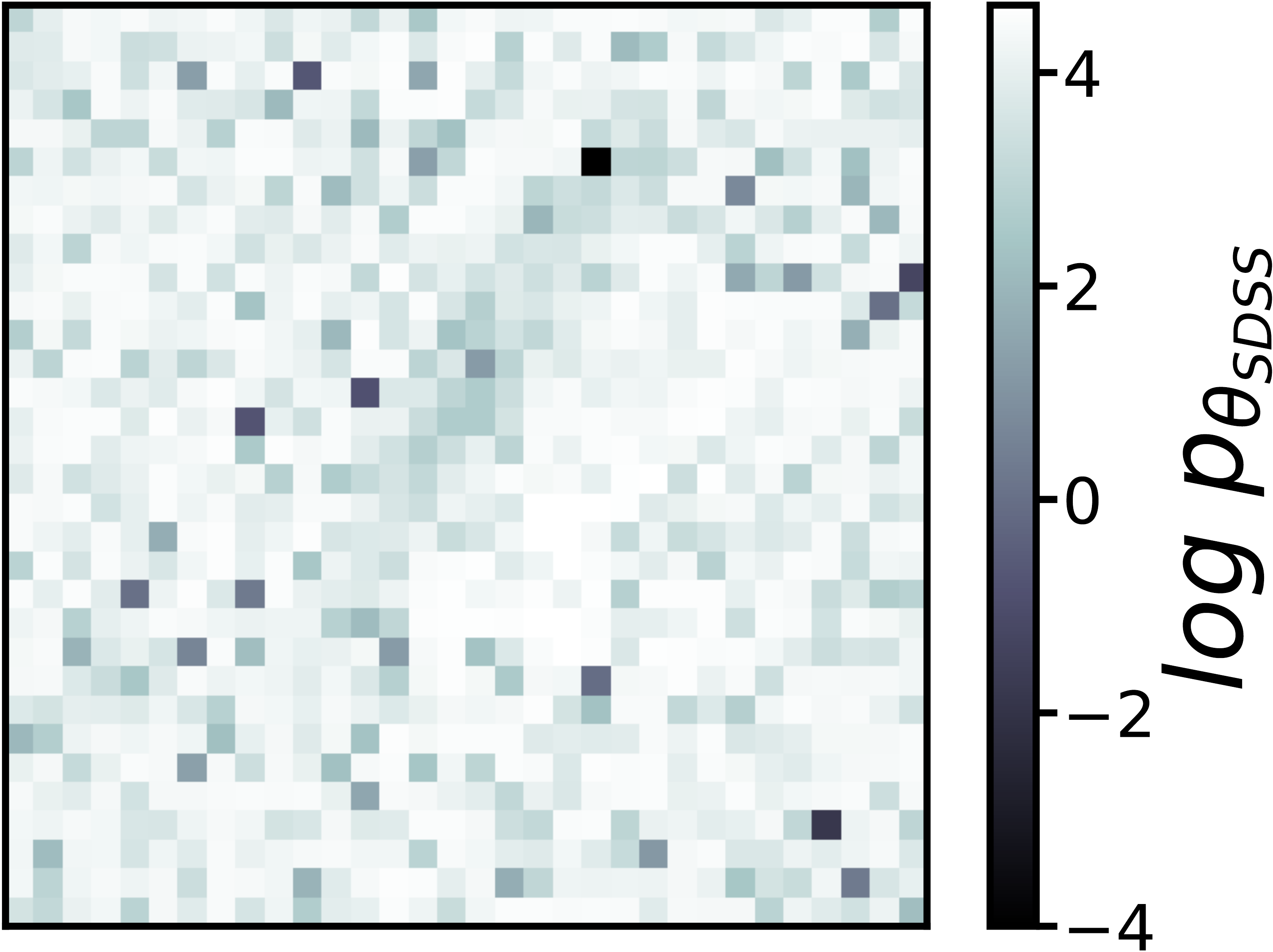} }}\hspace{0.07cm}
    \subfloat[(f)]{{\includegraphics[height=0.15\textwidth]{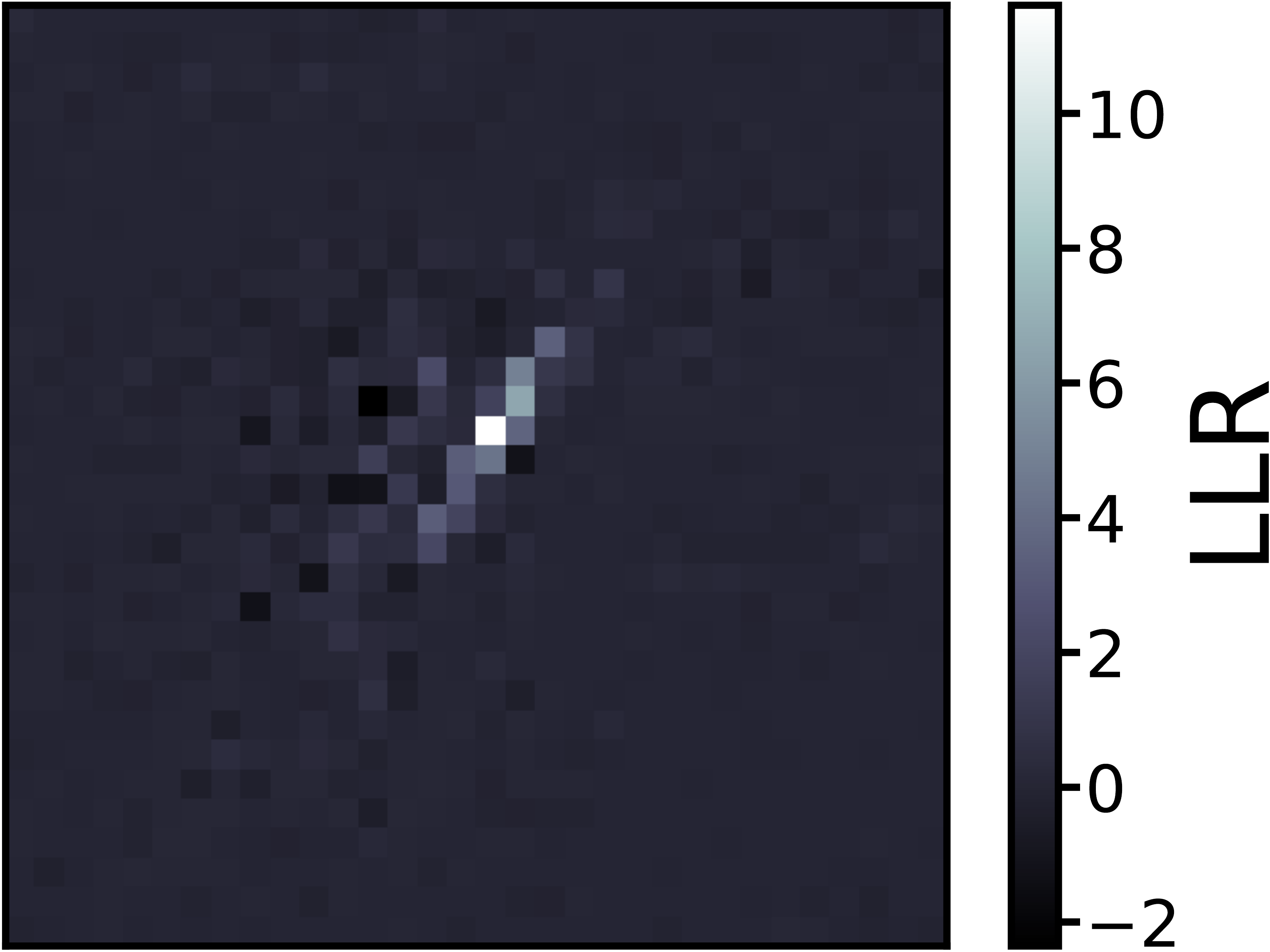} }}
\caption{\small \emph{(a):}A galaxy from IllustrisTNG  that was assigned a sky patch with both a large and a small Milky Way star by \texttt{RealSim}.  
    \emph{(b):} The pixel-wise likelihood of the $p_{\boldsymbol{\theta_{\rm SDSS}}}$ model.
    \emph{(c):} The pixel-wise LLR for (a). The net contribution of the pure sky noise is zero, while the galaxy contributes positively to the LLR. The spikes and edges of the larger star as well as the smaller star contribute negatively. The contribution of the larger star itself is mostly null. \emph{(d):} An SDSS galaxy in an empty background. \emph{(e): } This galaxy has a lower likelihood compared to the rest of the sky. \emph{(f):} The galaxy gives the largest positive contribution to the LLR. }
    \label{fig:llrmap}
\end{figure*}

\section{Robustness of the methodology}
\label{app:robustness}
In the main text we have extensively used the fact that the LLR is a good metric to compare the morphology of observed and simulated galaxies. Conversely, in Section \ref{sec:strategy} we argued that the likelihood alone may not be as good as a metric. In the following we outline the rationale behind this statement.

We start by showing  the likelihood distributions of our datasets for both the $p_{\theta_{\rm SDSS}}$ and the $p_{\theta_{\rm sersic}}$ models in Figure \ref{fig:likelihoods_distribs} . It can be seen that in the former case the distributions of simulations are displaced at slightly lower  likelihoods and feature a higher variance  compared to SDSS. This is less severe for both the IllustrisTNG realizations, and more substantial for Illustris. Interestingly, the best S\'ersic fits appear to peak at a higher likelihood than all the other datasets, including SDSS. 
This fact is suggestive that simpler images have a higher likelihood compared to more complex samples, including the training set (SDSS in this case), since the best S\'ersic fits are simple, smooth objects. 
To further explore this hypothesis, already formulated in \citet{serra+19}, in  Figure \ref{fig:likelihoods_summary} we show random samples of SDSS and simulated galaxies in three narrow bins of likelihood. It is readily appreciable that indeed more extended galaxies with a complex structure and the presence of interlopers dominate the low likelihood tail of the distributions, while smaller, smoother objects are located at very high likelihood values. Figure \ref{fig:likelihoods_summary} raises two important issues that undermine the use of the likelihood alone to compare simulations and observations. We discuss them in the following discussed below.

\subsection{The role of the sky background}
First of all, the fact that large and small galaxies are at the opposite ends of the likelihood spectrum is suggestive that the number of sky pixels in an image is an important predictor of the likelihood. This is not really a surprise, since the overall likelihood of an image is the sum of that of all pixels\footnote{Recall that we are actually using the \emph{log} likelihood.}, but it is certainly not desirable that the sky background plays such an important role, given that what we are really interested in is, of course, only the structure of the galaxy. How to solve this issue?

In Section \ref{sec:strategy} we have hypothesized that the $p_{\theta_{\rm SDSS}}$ and the $p_{\theta_{\rm sersic}}$ models are able to capture the background equally well, and therefore their LLR should isolate the contribution of the galaxy alone. We show that this is indeed the case in the third column of Figure \ref{fig:llrmap}, where most of the sky pixels have an LLR close to zero, whereas  in the middle panels of Figure \ref{fig:llrmap} is shown that the sky background gives the most positive contribution to the likelihood. 

\subsubsection{The sky generates variance in the LLR}
\label{sec:LLR_variance}
Figure \ref{fig:llrmap} reveals also that bright interlopers (first row) may still contribute significantly to the LLR. It is important to recall that we implement observational realism on simulations by assigning a simulated galaxy to a random SDSS field. Given the potential presence of interlopers in that field, we expect this to be a process that generates some variance in the LLR of a given galaxy cutout. Therefore, the LLR of any single object should not be strictly interpreted as a measure of its quality compared to observations. However, the mean LLR of selected subpopulations can still be robustly compared.

\subsection{The role of image complexity}
Secondly, it is clear from Figure \ref{fig:likelihoods_summary} that some of the simulated galaxies, especially in the low likelihood bin, are all but realistic. Hence the second question: is the substantial overlap in the likelihood distributions of Figure \ref{fig:likelihoods_distribs} really meaningful to assess the quality of simulated images? We discuss how the LLR may be a more meaningful metric below.

We start by discussing the right panel of Figure \ref{fig:likelihoods_distribs}, where the likelihood distributions of our datasets evaluated through the $p_{\theta_{\rm sersic}}$ model are shown. The most significant feature is certainly that in this case the likelihood of the best S\'ersic fits is markedly higher compared to that of all the other datasets, and not only slightly larger as it was the case for the $p_{\theta_{\rm SDSS}}$ model. We interpret this as a sign that a model trained on the smooth archetypes (i.e. the best S\'ersic fits) of more irregular objects (i.e. the real SDSS galaxies) is able to identify the complexity of the latter and other non-smooth datasets (i.e. the simulations). We have already discussed above that in Figure \ref{fig:likelihoods_summary} the likelihood of a galaxy is heavily dependent on the complexity of its internal patterns. Although the discussion referred only to the $p_{\theta_{\rm SDSS}}$ model (the left panel of Figure \ref{fig:likelihoods_distribs}) a similar behaviour is found in the $p_{\theta_{\rm sersic}}$ model.  Crucially this dependence is not the same in the $p_{\theta_{\rm SDSS}}$ and the $p_{\theta_{\rm sersic}}$ models, as shown by the fact that the two models produce a different likelihood distribution for the same dataset. Therefore the hope is that by combining the likelihood of the same object evaluated by both models some trends will arise which will depend only on the galaxy's internal structure, as derived mathematically in Section \ref{sec:strategy}. Let us now come back once again to Figure \ref{fig:llrmap}, which already helped us prove that the LLR is able to factor out most of the sky contribution. What is really interesting is that not all the pixels of a galaxy show up in the LLR, but only a few of them carry a high LLR value. These may be the fine morphological details that deviate from the smoothness of the S\'ersic profile. In short, this corroborates our hypothesis that the LLR of the the $p_{\theta_{\rm SDSS}}$ and the $p_{\theta_{\rm sersic}}$ models is able to isolate the subtle patterns that are present in galaxy structure (eq. \ref{eq:LLR_final}).

The issues discussed in this Section are known and have been addressed in previous studies \citep{Shafaei+18,Nalisnick+18_do_models_dont_know}, which however made use of toy datasets popular in the machine learning community\footnote{ These datasets included MNIST and  FashionMNIST (pictures of numbers and clothing in a monochromatic background respectively) and ImageNet (natural images such as dogs and boats) amongst others.}. We have shown that similar considerations can be made for astronomical images, which is not obvious in principle.

\bibliographystyle{mnras}
\interlinepenalty=1000000
\bibliography{main.bib} 




\bsp	
\label{lastpage}
\end{document}